\pdfoutput=1

\documentclass[12pt,a4paper]{article}

\usepackage{ifthen} 
\newboolean{pdflatex}
\setboolean{pdflatex}{true} 

\newboolean{articletitles}
\setboolean{articletitles}{true} 

\newboolean{uprightparticles}
\setboolean{uprightparticles}{false} 

\usepackage[top=1in, bottom=1.25in, left=1in, right=1in]{geometry}

\usepackage{microtype}
\usepackage{lineno}  
\usepackage{xspace} 
\usepackage{caption} 

\usepackage{graphicx}  
\usepackage{color}
\usepackage{colortbl}
\graphicspath{{./}} 

\usepackage{amsmath} 
\usepackage{amssymb}
\usepackage{amsfonts}
\usepackage{upgreek} 
\usepackage{multirow}
\newcommand*\patchAmsMathEnvironmentForLineno[1]{%
\expandafter\let\csname old#1\expandafter\endcsname\csname #1\endcsname
\expandafter\let\csname oldend#1\expandafter\endcsname\csname
end#1\endcsname
 \renewenvironment{#1}%
   {\linenomath\csname old#1\endcsname}%
   {\csname oldend#1\endcsname\endlinenomath}%
}
\newcommand*\patchBothAmsMathEnvironmentsForLineno[1]{%
  \patchAmsMathEnvironmentForLineno{#1}%
  \patchAmsMathEnvironmentForLineno{#1*}%
}
\AtBeginDocument{%
\patchBothAmsMathEnvironmentsForLineno{equation}%
\patchBothAmsMathEnvironmentsForLineno{align}%
\patchBothAmsMathEnvironmentsForLineno{flalign}%
\patchBothAmsMathEnvironmentsForLineno{alignat}%
\patchBothAmsMathEnvironmentsForLineno{gather}%
\patchBothAmsMathEnvironmentsForLineno{multline}%
\patchBothAmsMathEnvironmentsForLineno{eqnarray}%
}

\usepackage[bookmarks=false]{hyperref}
\usepackage[all]{hypcap} 

\usepackage{xspace} 
\usepackage{upgreek}

\def\lhcb {\mbox{LHCb}\xspace}

\def\MagUp {\mbox{\em Mag\kern -0.05em Up}\xspace}

\ifthenelse{\boolean{uprightparticles}}%
{

 \def\Ppsi        {\ensuremath{\uppsi}\xspace}

 \def\PDelta      {\ensuremath{\Delta}\xspace}                 
 \def\PXi      {\ensuremath{\Xi}\xspace}                 
 \def\PLambda      {\ensuremath{\Lambda}\xspace}                 
 \def\PSigma      {\ensuremath{\Sigma}\xspace}                 
 \def\POmega      {\ensuremath{\Omega}\xspace}                 
 \def\PUpsilon      {\ensuremath{\Upsilon}\xspace}                 
 
 \def\PB      {\ensuremath{\mathrm{B}}\xspace}                 
                  
 \def\PD      {\ensuremath{\mathrm{D}}\xspace}

 \def\PJ      {\ensuremath{\mathrm{J}}\xspace}                 
 \def\PK      {\ensuremath{\mathrm{K}}\xspace}

 \def\Pc      {\ensuremath{\mathrm{c}}\xspace}

 \def\Pi      {\ensuremath{\mathrm{i}}\xspace}

 \def\Ps      {\ensuremath{\mathrm{s}}\xspace}

}
{

 \def\Ppsi        {\ensuremath{\psi}\xspace}                 
                  
 \mathchardef\PDelta="7101
 \mathchardef\PXi="7104
 \mathchardef\PLambda="7103
 \mathchardef\PSigma="7106
 \mathchardef\POmega="710A
 \mathchardef\PUpsilon="7107
                  
 \def\PB      {\ensuremath{B}\xspace}                 
                  
 \def\PD      {\ensuremath{D}\xspace}

 \def\PJ      {\ensuremath{J}\xspace}                 
 \def\PK      {\ensuremath{K}\xspace}

 \def\Pc      {\ensuremath{c}\xspace}

 \def\Pi      {\ensuremath{i}\xspace}

 \def\Ps      {\ensuremath{s}\xspace}

}

\makeatletter
\ifcase \@ptsize \relax
  \newcommand{\miniscule}{\@setfontsize\miniscule{4}{5}}
\or
  \newcommand{\miniscule}{\@setfontsize\miniscule{5}{6}}
\or
  \newcommand{\miniscule}{\@setfontsize\miniscule{5}{6}}
\fi
\makeatother

\DeclareRobustCommand{\optbar}[1]{\shortstack{{\miniscule (\rule[.5ex]{1.25em}{.18mm})}
  \\ [-.7ex] $#1$}}

\def\squark    {{\ensuremath{\Ps}}\xspace}

\def\cquark    {{\ensuremath{\Pc}}\xspace}

  \def\Kbar    {{\kern 0.2em\overline{\kern -0.2em \PK}{}}\xspace}

\def\KorKbar    {\kern 0.18em\optbar{\kern -0.18em K}{}\xspace}

  \def\Dbar    {{\kern 0.2em\overline{\kern -0.2em \PD}{}}\xspace}
\def\D       {{\ensuremath{\PD}}\xspace}

\def\DorDbar    {\kern 0.18em\optbar{\kern -0.18em D}{}\xspace}
\def\Dz      {{\ensuremath{\D^0}}\xspace}

\def\Dp      {{\ensuremath{\D^+}}\xspace}

\def\Ds      {{\ensuremath{\D^+_\squark}}\xspace}

\def\Bbar    {{\ensuremath{\kern 0.18em\overline{\kern -0.18em \PB}{}}}\xspace}

\def\BorBbar    {\kern 0.18em\optbar{\kern -0.18em B}{}\xspace}

\def\jpsi     {{\ensuremath{{\PJ\mskip -3mu/\mskip -2mu\Ppsi\mskip 2mu}}}\xspace}
\def\psitwos  {{\ensuremath{\Ppsi{(2S)}}}\xspace}

  \def\Y#1S{\ensuremath{\PUpsilon{(#1S)}}\xspace}

\def\Lz          {{\ensuremath{\PLambda}}\xspace}
\def\Lbar        {{\ensuremath{\kern 0.1em\overline{\kern -0.1em\PLambda}}}\xspace}
\def\LorLbar    {\kern 0.18em\optbar{\kern -0.18em \PLambda}{}\xspace}

\def\Lc      {{\ensuremath{\Lz^+_\cquark}}\xspace}

\def\BF         {{\ensuremath{\mathcal{B}}}\xspace}

\def\BR         {\BF}

\def\to                 {\ensuremath{\rightarrow}\xspace}

\newcommand{\etot}{{\ensuremath{\varepsilon_{\mathrm{ tot}}}}\xspace}

\def\AT#1     {\ensuremath{A_{\mathrm{T}}^{#1}}\xspace}           

\def\C#1      {\ensuremath{\mathcal{C}_{#1}}\xspace}                       
\def\Cp#1     {\ensuremath{\mathcal{C}_{#1}^{'}}\xspace}                    
\def\Ceff#1   {\ensuremath{\mathcal{C}_{#1}^{\mathrm{(eff)}}}\xspace}        
\def\Cpeff#1  {\ensuremath{\mathcal{C}_{#1}^{'\mathrm{(eff)}}}\xspace}       
\def\Ope#1    {\ensuremath{\mathcal{O}_{#1}}\xspace}                       
\def\Opep#1   {\ensuremath{\mathcal{O}_{#1}^{'}}\xspace}                    

\newcommand{\tev}{\ifthenelse{\boolean{inbibliography}}{\ensuremath{~T\kern -0.05em eV}\xspace}{\ensuremath{\mathrm{\,Te\kern -0.1em V}}}\xspace}
\newcommand{\tevv}{\ensuremath{\mathrm{\,Te\kern -0.1em V}}\xspace}
\newcommand{\gev}{\ensuremath{\mathrm{\,Ge\kern -0.1em V}}\xspace}
\newcommand{\mev}{\ensuremath{\mathrm{\,Me\kern -0.1em V}}\xspace}
\newcommand{\kev}{\ensuremath{\mathrm{\,ke\kern -0.1em V}}\xspace}
\newcommand{\ev}{\ensuremath{\mathrm{\,e\kern -0.1em V}}\xspace}
\newcommand{\gevc}{\ensuremath{{\mathrm{\,Ge\kern -0.1em V\!/}c}}\xspace}
\newcommand{\mevc}{\ensuremath{{\mathrm{\,Me\kern -0.1em V\!/}c}}\xspace}
\newcommand{\gevcc}{\ensuremath{{\mathrm{\,Ge\kern -0.1em V\!/}c^2}}\xspace}
\newcommand{\gevgevcccc}{\ensuremath{{\mathrm{\,Ge\kern -0.1em V^2\!/}c^4}}\xspace}
\newcommand{\mevcc}{\ensuremath{{\mathrm{\,Me\kern -0.1em V\!/}c^2}}\xspace}

\def\mum  {\ensuremath{{\,\upmu\mathrm{m}}}\xspace}

\def\mbarn{\ensuremath{\mathrm{ \,mb}}\xspace}

\def\invnb {\ensuremath{\mbox{\,nb}^{-1}}\xspace}

\newcommand{\chisqip}{\ensuremath{\chi^2_{\text{\ensuremath{\rm IP}}}}\xspace}

\def\deriv {\ensuremath{\mathrm{d}}}

\def\gsim{{~\raise.15em\hbox{$>$}\kern-.85em
          \lower.35em\hbox{$\sim$}~}\xspace}
\def\lsim{{~\raise.15em\hbox{$<$}\kern-.85em
          \lower.35em\hbox{$\sim$}~}\xspace}

\def\ptot       {\mbox{$p$}\xspace}
\def\pt         {\mbox{$p_{\mathrm{ T}}$}\xspace}

\newcommand{\lum} {\ensuremath{\mathcal{L}}\xspace}

\def\evtgen     {\mbox{\textsc{EvtGen}}\xspace}

\def\geant      {\mbox{\textsc{Geant4}}\xspace}

\def\photos     {\mbox{\textsc{Photos}}\xspace}

\def\pythia     {\mbox{\textsc{Pythia}}\xspace}

\def\tell1  {TELL1\xspace}
\def\ukl1   {UKL1\xspace}

\newcommand{\ie}{\mbox{\itshape i.e.}\xspace}

\newcommand{\pPb}{{\ensuremath{p\mathrm{Pb}}}\xspace}
\newcommand{\Tev}{\ensuremath{\mathrm{\,Te\kern -0.1em V}}\xspace}

\newcommand{\snn}{{\ensuremath{\sqrt{s_{\mathrm{NN}}}}}\xspace}

\newcommand{\lhcborcid}[1]{\href{https://orcid.org/#1}{\hspace*{0.1em}\raisebox{-0.45ex}{\includegraphics[width=1em]{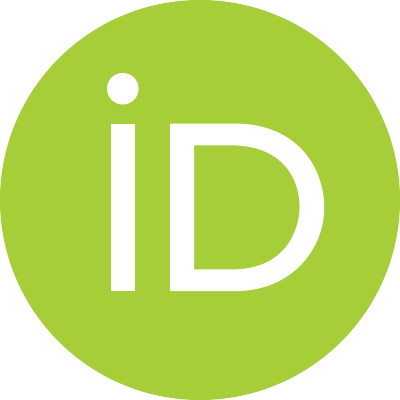}}}}

\usepackage{cite} 
\usepackage{mciteplus}
\usepackage{longtable} 
\usepackage{amsmath}
\usepackage{varwidth}
\usepackage[figuresright]{rotating}
\begin{document}

\renewcommand{\thefootnote}{\fnsymbol{footnote}}
\setcounter{footnote}{1}


\begin{titlepage}
\pagenumbering{roman}

\vspace*{-1.5cm}
\centerline{\large EUROPEAN ORGANIZATION FOR NUCLEAR RESEARCH (CERN)}
\vspace*{1.5cm}
\noindent
\begin{tabular*}{\linewidth}{lc@{\extracolsep{\fill}}r@{\extracolsep{0pt}}}
\ifthenelse{\boolean{pdflatex}}
{\vspace*{-2.7cm}\mbox{\!\!\!\includegraphics[width=.14\textwidth]{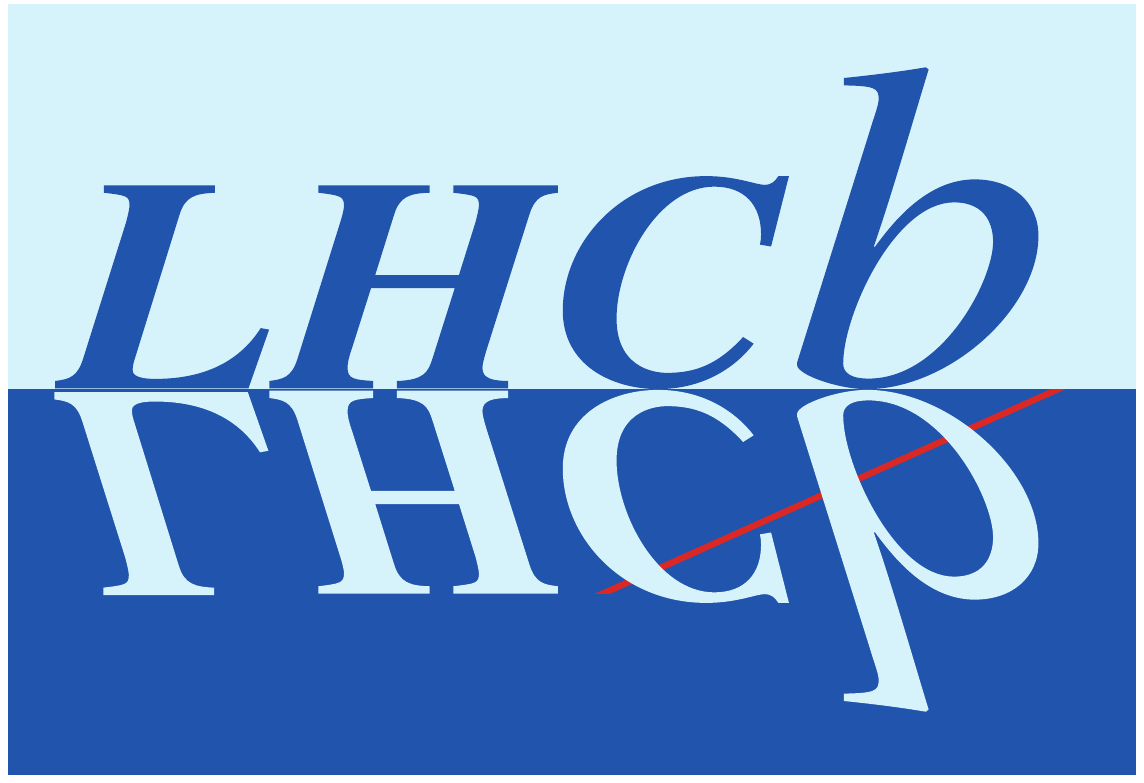}} & &}%
{\vspace*{-1.2cm}\mbox{\!\!\!\includegraphics[width=.12\textwidth]{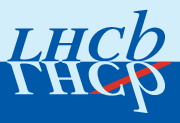}} & &}%
\\
 & & CERN-EP-2023-142 \\  
 & & LHCb-PAPER-2023-006 \\  
 & & September 5, 2023
\end{tabular*}

\vspace*{3.0cm}

{\normalfont\bfseries\boldmath\huge
\begin{center}
 Measurement of prompt $D^+$ and $D^+_{s}$ production in $p\mathrm{Pb}$ collisions at $\sqrt {s_{\mathrm{NN}}}=5.02$$\ensuremath{\mathrm{\,Te\kern -0.1em V}}$
\end{center}
}

\vspace*{2.0cm}

\begin{center}
The LHCb collaboration\footnote{Authors are listed at the end of this paper.}
\end{center}

\vspace{\fill}

\begin{abstract}
  \noindent
  The production of prompt $D^+$ and $D^+_{s}$ mesons is studied in proton-lead collisions at a centre-of-mass energy of $\sqrt {s_{\mathrm{NN}}}=5.02$$\ensuremath{\mathrm{\,Te\kern -0.1em V}}$.
  The data sample corresponding to an integrated luminosity of $(1.58\pm0.02)\mathrm{nb}^{-1}$ is collected by the LHCb experiment at the LHC.
  The differential production cross-sections are measured using $D^+$ and $D^+_{s}$ candidates with transverse momentum in the range of $0<p_{\mathrm{T}} <14\,\mathrm{GeV}/c$ and rapidities in the ranges of $1.5<y^*<4.0$ and $-5.0<y^*<-2.5$ in the nucleon-nucleon centre-of-mass system.
  For both particles, the nuclear modification factor and the forward-backward production ratio are determined.
  These results are compared with theoretical models that include initial-state nuclear effects.
  In addition, measurements of the cross-section ratios between $D^+$, $D^+_{s}$ and $D^0$ mesons are presented, providing a baseline for studying the charm hadronization in lead-lead collisions at LHC energies.

\end{abstract}

\vspace*{2.0cm}

\begin{center}
  Published in JHEP 01 (2024) 070
\end{center}

\vspace{\fill}

{\footnotesize 
\centerline{\copyright~CERN on behalf of the \lhcb collaboration, licence \href{http://creativecommons.org/licenses/by/4.0/}{CC-BY-4.0}.}}
\vspace*{2mm}

\end{titlepage}


\newpage
\setcounter{page}{2}
\mbox{~}
%
\cleardoublepage


\renewcommand{\thefootnote}{\arabic{footnote}}
\setcounter{footnote}{0}



\pagestyle{plain} 
\setcounter{page}{1}
\pagenumbering{arabic}

\section{Introduction}
\label{sec:Introduction}

Quark gluon plasma (QGP) is a new form of matter consisting of deconfined quarks and gluons as basic components at high energy densities~\cite{Shuryak:1980tp}.
The production of QGP and investigation of its properties are among the primary goals of high energy heavy-ion collision experiments.
Heavy quarks are sensitive and effective probes for QGP transport properties, as they are produced in pairs in the early stage of heavy-ion collisions by hard scatterings and experience the entire evolution of the fireball prior to the hadronization process.
The measured $D$-meson nuclear modification factors ($R_{\mathrm{AA}}$) in nucleus-nucleus (AA) collisions at the RHIC~\cite{STAR:2014wif,STAR:2018zdy} and the LHC~\cite{ALICE:2015vxz,ALICE:2015dry,ALICE:2021rxa} colliders show a strong suppression at high transverse momentum ($\pt$) and demonstrate a significant charm-quark energy loss in the medium~\cite{Wicks:2005gt,Uphoff:2012gb,He:2014cla,Horowitz:2011wm}.
This modification factor is calculated as the ratio of $D$-meson cross-sections from AA to proton-proton ($pp$) collisions, and is scaled by the binary collision number.
The production ratio of strange \Ds to non-strange \Dz mesons is found to be enhanced in $\snn\,=\,$200$\gev$ gold-gold collisions~\cite{STAR:2021tte} and in $\snn\,=\,5.02$$\ensuremath{\mathrm{\,Te\kern -0.1em V}}$ lead-lead collisions~\cite{ALICE:2021kfc,ALICE:2018lyv}, compared with the results from $pp$ collisions data~\cite{ALICE:2021mgk,ALICE:2019nxm} and \pythia $pp$ simulation~\cite{Sjostrand:2006za,Bierlich:2015rha}.  
This is attributed to the enhanced production of strange quarks in the hot dense medium and charm quark hadronization {\it via} the coalescence mechanism~\cite{Greco:2003xt,Fries:2003vb,Greco:2003mm,Ravagli:2007xx}.
To fully understand the experimental measurements for nucleus-nucleus collisions, it is essential to characterize the cold nuclear matter (CNM) effects due to the involvement of heavy nuclei like lead in the colliding system.

Multiple CNM effects could modify the heavy-flavor hadron production and kinematic distributions. One way to investigate these effects is to study proton-nucleus ($p$A) collisions, where it is assumed that QGP effects are not dominant.
In the initial state, the nuclear environment influences the parton distribution functions (PDFs) of the bound nucleons, 
and this modification depends on the parton momentum fraction (Bjorken-$x$), the momentum transfer squared ($Q^2$), and the nucleus mass number (A)~\cite{Kharzeev:2005zr,Fujii:2006ab}.
At LHC energies and at forward rapidity (corresponding to Bjorken-$x \approx 10^{-6}-10^{-5}$),
the most relevant effect on PDFs is nuclear shadowing~\cite{Armesto:2006ph}.
If the gluon phase space is saturated then the $D$-meson yield, which may be significantly affected at low $\pt$, can be described by the Colour Glass Condensate (CGC) effective theory~\cite{Fujii:2013yja,Tribedy:2011aa,Albacete:2012xq,Rezaeian:2012ye}.

Measurements of lepton production from semileptonic decays of heavy flavour hadrons in deuteron-gold collisions at the RHIC~\cite{PHENIX:2012hww,PHENIX:2013txu} and in proton-lead (\pPb) collisions at the LHC~\cite{Adam:2015qda,ALICE:2016uid,Acharya:2017hdv} suggest that CNM effects are present.
These can be further studied with measurements of decays of charm hadrons
in $\pPb$ collisions at the LHC~\cite{LHCb:2013gmv,LHCb:2016vqr,LHCb:2017yua,LHCb:2018weo,Acharya:2019mno,ALICE:2017wet,Adam:2016ich,Adam:2016ohd,Adam:2016mkz, Adam:2015jsa, Abelev:2014hha,Abelev:2013yxa,Sirunyan:2017mzd,CMS:2016wma}.
Recently, the LHCb experiment measured  $\jpsi$~\cite{LHCb:2013gmv}, \psitwos ~\cite{LHCb:2016vqr}, $\Dz$~\cite{LHCb:2017yua} meson and \Lc~\cite{LHCb:2018weo} baryon production at $\snn=5.02$$\ensuremath{\mathrm{\,Te\kern -0.1em V}}$ and $D$-meson production at $\snn=8.16$$\ensuremath{\mathrm{\,Te\kern -0.1em V}}$~\cite{LHCb:2022dmh,LHCb:2023rpm} in \pPb collisions at forward and backward rapidities.
The ALICE collaboration also studied the $D$-meson production in \pPb collisions~\cite{Acharya:2019mno,Adam:2016ich,Adam:2016mkz,Abelev:2014hha} 
at the same centre-of-mass energy in the rapidity interval $- 0.96 < y^* <  0.04$, 
where $y^*$ is the rapidity of the $D$ mesons in the nucleon-nucleon centre-of-mass
frame.
Extensive measurements of charm-quark production at low \pt have greatly constrained in the PDFs~\cite{Eskola:2019bgf,AbdulKhalek:2022fyi}, in particular the \Dz results in LHCb \pPb collisions~\cite{AbdulKhalek:2022fyi}.

This paper reports measurements of production cross-sections, nuclear modification factors and forward-backward production ratios for prompt \Dp and \Ds mesons produced directly in proton-lead interactions or from excited charmed hadron decays.
The measurement is performed at $\snn=5.02$$\ensuremath{\mathrm{\,Te\kern -0.1em V}}$ with the LHCb detector~\cite{Alves:2008zz}, which is able to cover two different acceptance regions in the nucleon-nucleon rest frame. In the ``forward'' (``backward'') configuration, the region is $1.5 < y^* < 4.0$ ($- 5.0 < y^* < - 2.5$), and the positive direction is defined with respect to the direction of the proton beam.
The measurement of the production cross-section is performed over the range of $\Dp$ and $\Ds$ transverse momentum $0<\pt<14\gevc$, in both backward and forward configurations.
Finally, the production ratios between \Dp, \Ds and \Dz mesons in \pPb collisions are presented as a function of \pt and $y^*$, and compared with the measurements in $pp$~\cite{LHCb:2016ikn,ALICE:2019nxm} and \pPb collisions~\cite{Acharya:2019mno}
at same nucleon-nucleon centre-of-mass energy at the LHC.

\section{Detector and data samples}
\label{sec:Detector}
The \lhcb detector~\cite{Alves:2008zz,LHCb:2014set} is a single-arm forward
spectrometer covering the \mbox{pseudorapidity} range $2<\eta <5$,
designed for the study of particles containing charm or beauty
quarks. The detector includes a high-precision tracking system
consisting of a silicon-strip vertex detector surrounding the $pp$
interaction region (VELO), a large-area silicon-strip detector (TT) located
upstream of a dipole magnet with a bending power of about
$4{\mathrm{\,Tm}}$, and three stations of silicon-strip detectors (IT) and straw
drift tubes (OT) placed downstream of the magnet.
The tracking system provides a measurement of momentum, \ptot, of charged particles with
a relative uncertainty that varies from 0.5\% at low momentum to 1.0\% at 200\gevc.
The minimum distance of a track to a primary interaction vertex (PV), the impact parameter (IP), is measured with a resolution of $(15+29/\pt)\mum$.  
Different types of charged hadrons are distinguished using information
from two ring-imaging Cherenkov detectors. 
Photons, electrons and hadrons are identified by a calorimeter system consisting of
scintillating-pad and preshower detectors, an electromagnetic
calorimeter and a hadronic calorimeter. Muons are identified by a
system composed of alternating layers of iron and multiwire
proportional chambers~\cite{AbellanBeteta:2020amj}.
The online event selection is performed by a trigger~\cite{Aaij:2012me}, 
which consists of a hardware stage, based on information from the calorimeter and muon
systems, followed by a software stage, which applies a full event
reconstruction.

This analysis uses the \pPb data sample collected by the LHCb detector in early 2013, corresponding to integrated luminosities of ($1.06\pm 0.02$)\invnb and ($0.52\pm0.01$)\invnb for the forward and backward collisions, respectively~\cite{LHCb:2013gmv}.
The instantaneous luminosity during the data taking period was about 5$\times 10^{27}$~cm$^{-2}$~s$^{-1}$, resulting in event rates three orders of magnitude lower than for typical \lhcb $pp$ interactions. 
Consequently, the hardware trigger only rejected empty events, while the software trigger accepted all events with at least one track reconstructed in the VELO.

Simulated samples of \pPb collisions at $\snn=5.02$$\ensuremath{\mathrm{\,Te\kern -0.1em V}}$ are used to determine detector efficiencies. 
In the simulation, $D$ mesons in $pp$ collisions are generated using \pythia~\cite{Sjostrand:2006za,Sjostrand:2007gs}, embedded into minimum bias \pPb events from the \mbox{\textsc{EPOS}} generator~\cite{Pierog:2013ria} and calibrated with LHC data~\cite{LHCb-PROC-2010-056}. 
Hadronic decays are generated using \evtgen~\cite{Lange:2001uf}, 
in which final-state radiation is simulated by \photos~\cite{Golonka:2005pn}.
The interaction of the generated particles with the \lhcb detector, and its response, are implemented using the \geant toolkit~\cite{Allison:2006ve, *Agostinelli:2002hh,LHCb-PROC-2011-006}. 
\section{Cross-section determination}\label{sec:xsec}

The prompt $D$-meson double-differential production cross-section in a given $(\pt,y^{*})$ kinematic bin is defined as 

\begin{equation}\label{eq:cross-section}
  \frac{\deriv^2\sigma}{\deriv \pt\deriv y^*} 
  = \frac{N(\pt,y^{*})}
  {\lum\etot\BR \Delta\pt\Delta y^*}.
\end{equation}
In the formula, $N(\pt,y^{*})$ is the signal yield of prompt $D$-meson candidates, $\etot$ is the total detection efficiency in a specific bin of $(\pt,y^{*})$, $\lum$ is the integrated luminosity, $\BR$ is the branching fraction of the corresponding $D$-meson decay and $\Delta \pt$ and $\Delta y^*$ are the bin widths of the \pt and rapidity, respectively.
The $D$-meson candidates are reconstructed through the $\Dp \to K^- \pi^+ \pi^+$ or $\Ds \to K^- K^+ \pi^+$ decay channels\footnote{In this paper charge-conjugated processes are implied unless stated otherwise.}, where the mass of the $K^+K^-$ pair in the \Ds meson candidate is required to fall within $\pm20$\mevcc of the known mass of the $\phi(1020)$ meson.
The corresponding branching fractions are $\BR=(9.38\pm0.15)\%$ for the $\Dp \to K^- \pi^+ \pi^+$ decay
obtained from Ref.~\cite{ParticleDataGroup:2022pth}, and
$\BR=(2.24\pm0.13)\%$ for the $\Ds \to K^- K^+ \pi^+$  decay, obtained from Ref.~\cite{CLEO:2008hzo}.

The total cross-section over a specific kinematic range is determined by 
integrating the double-differential cross-section. The nuclear 
modification factor, $R_{p{\rm Pb}}$, \ie the normalized ratio of the $D$-meson  production cross-section
in \pPb collisions to that in $pp$ interactions at the same nucleon-nucleon centre-of-mass energy is defined as
\begin{equation}
R_{p{\rm Pb}} (\pt,y^*) \equiv \frac{1}{A} \frac{{\rm d}^2 \sigma_{p{\rm Pb}}(\pt,y^*)/{\rm d}\pt{\rm d}y^*}
{{\rm d}^2\sigma_{pp}(\pt,y^*)/{\rm d}\pt{\rm d}y^*},
\end{equation}
where $A$=208 is the mass number of the lead nucleus. The forward-backward production ratio  is defined as
\begin{equation}
R_{\rm FB} (\pt,y^*) \equiv \frac{{\rm d}^2 \sigma_{\rm forward}(\pt,|y^*|;y^*>0)/{\rm d}\pt{\rm d}y^*}{{\rm d}^2 \sigma_{\rm backward}(\pt,|y^*|;y^*<0)/{\rm d}\pt{\rm d}y^*},
\end{equation}
in the common rapidity region $2.5<|y^*|<4.0$. 

The $D$-meson candidates are selected using similar selection criteria to those for the open charm production measurements in $pp$ collisions at $\sqrt{s}=5.02$$\ensuremath{\mathrm{\,Te\kern -0.1em V}}$~\cite{LHCb:2016ikn}, 7$\ensuremath{\mathrm{\,Te\kern -0.1em V}}$~\cite{LHCb:2013xam} and 13$\ensuremath{\mathrm{\,Te\kern -0.1em V}}$~\cite{LHCb:2015swx} at \lhcb.
The trajectories of kaons and pions from the $D$-meson candidates are required to be of good quality and originate from a common vertex. 
In order to improve the signal purity, more stringent particle identification (PID) requirements than in $pp$ interactions are exploited.

\begin{figure}[!tbp]
\centering
\begin{minipage}[t]{0.99\textwidth}
\centering
\begin{tabular}{cc}
\includegraphics[width=0.5\textwidth]{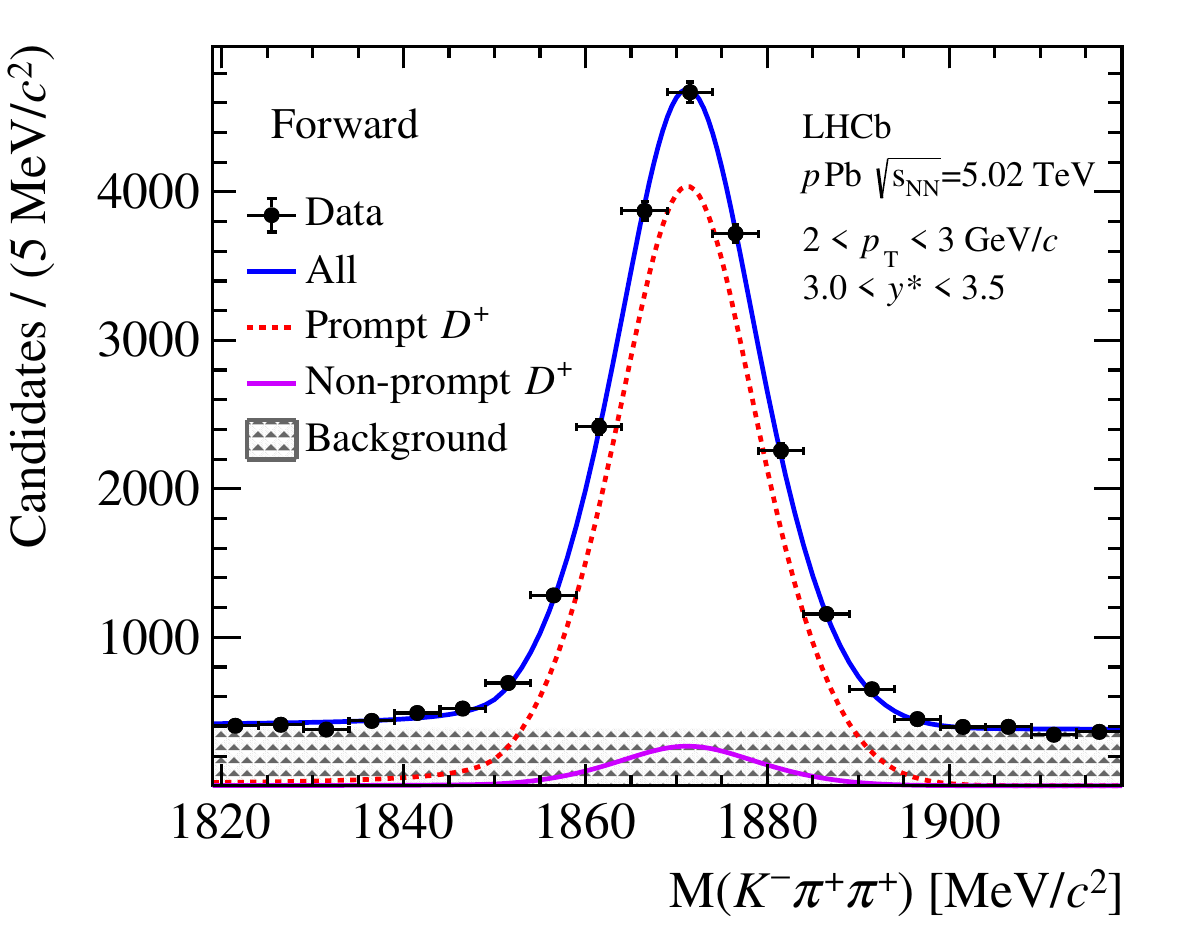}&
\includegraphics[width=0.5\textwidth]{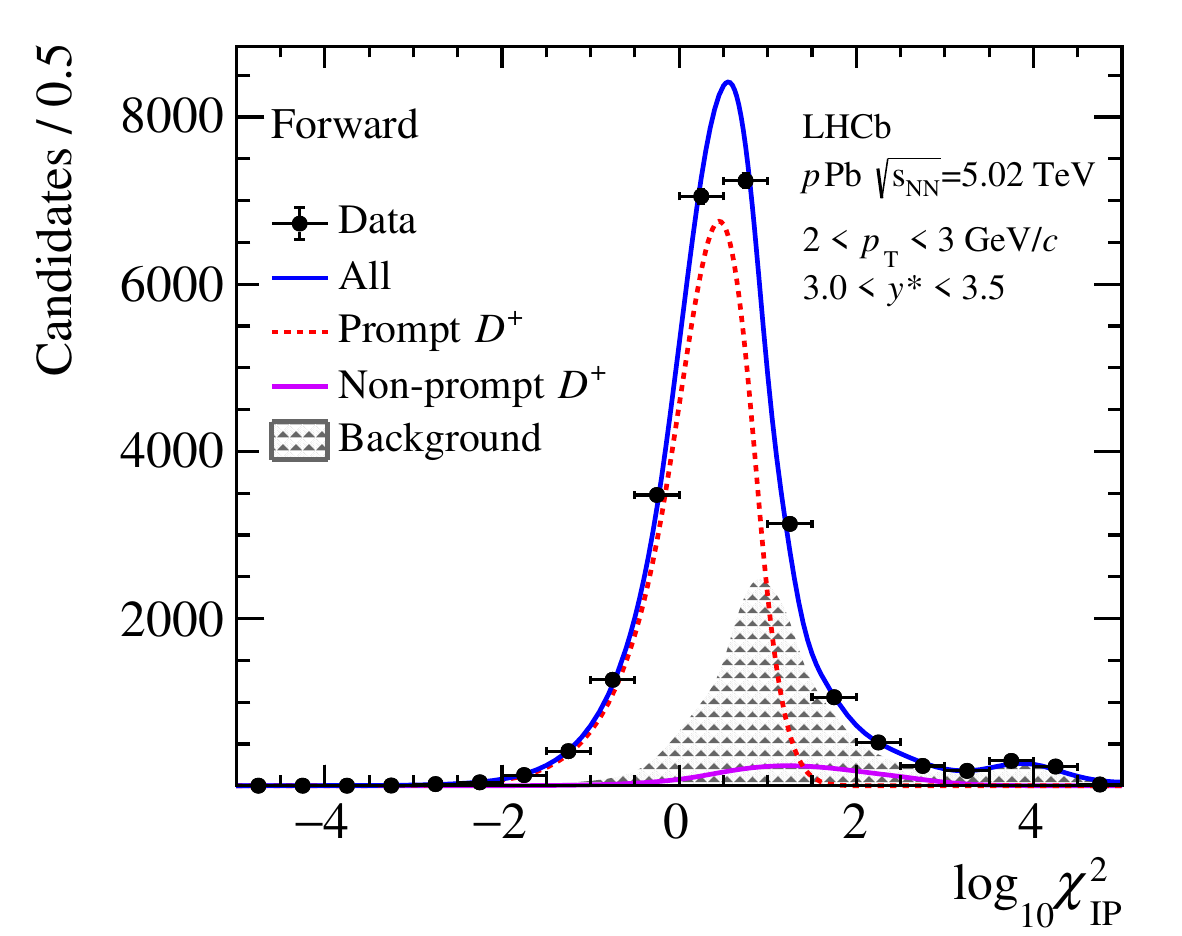}\\
\end{tabular}
\end{minipage}
\caption{Distributions of the simultaneous fits to the (left) $\text{M}(K^- \pi^+ \pi^+)$ and (right) $\log_{10}\chisqip$ for $\Dp$ mesons in the forward data sample in the kinematic bin of $2<\pt<3\gevc$
and $3.0<y^*<3.5$.}
\label{fig:MassIPFittPA}
\end{figure}
\begin{figure}[!tbp]
\centering
\begin{minipage}[t]{0.99\textwidth}
\centering
\begin{tabular}{cc}
\includegraphics[width=0.5\textwidth]{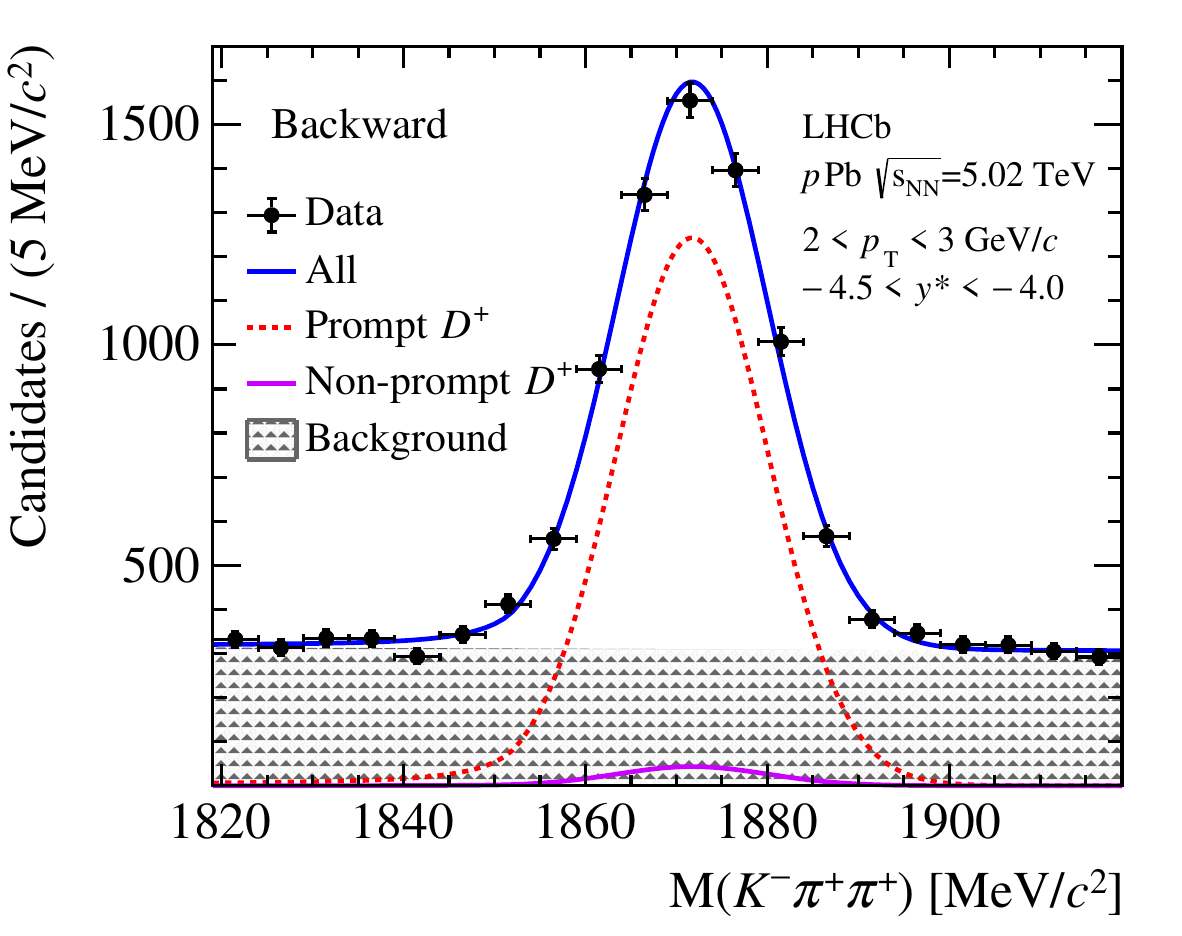}&
\includegraphics[width=0.5\textwidth]{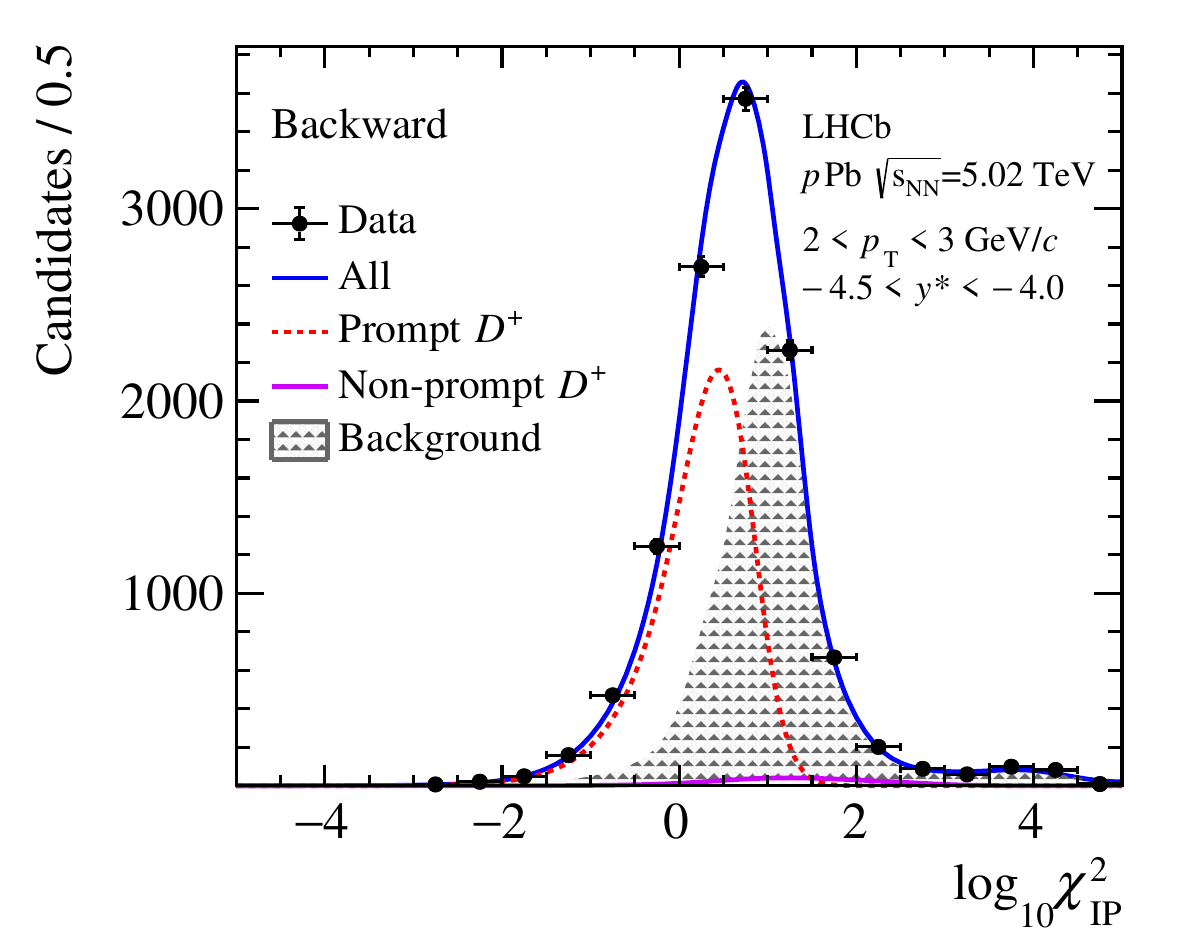}\\
\end{tabular}
\end{minipage}
\caption{Distributions of the simultaneous fits to the (left) $\text{M}(K^- \pi^+ \pi^+)$ and (right) $\log_{10}\chisqip$ for $\Dp$ mesons in the backward data sample in the kinematic bin of $2<\pt<3\gevc$
and $-4.5<y^*<-4.0$.}
\label{fig:MassIPFittAP}
\end{figure}

\begin{figure}[!tbp]
\centering
\begin{minipage}[t]{0.99\textwidth}
\centering
\begin{tabular}{cc}
\includegraphics[width=0.5\textwidth]{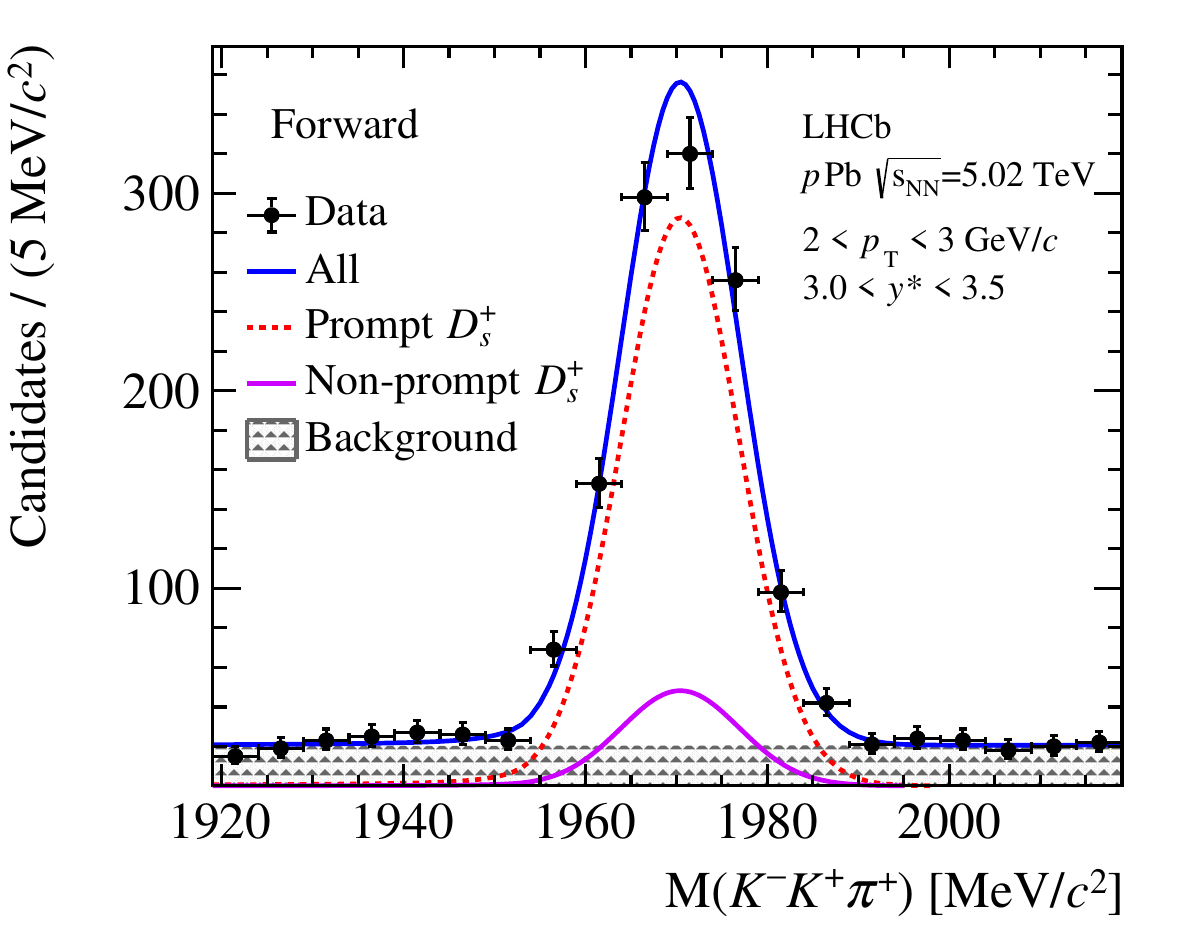}&
\includegraphics[width=0.5\textwidth]{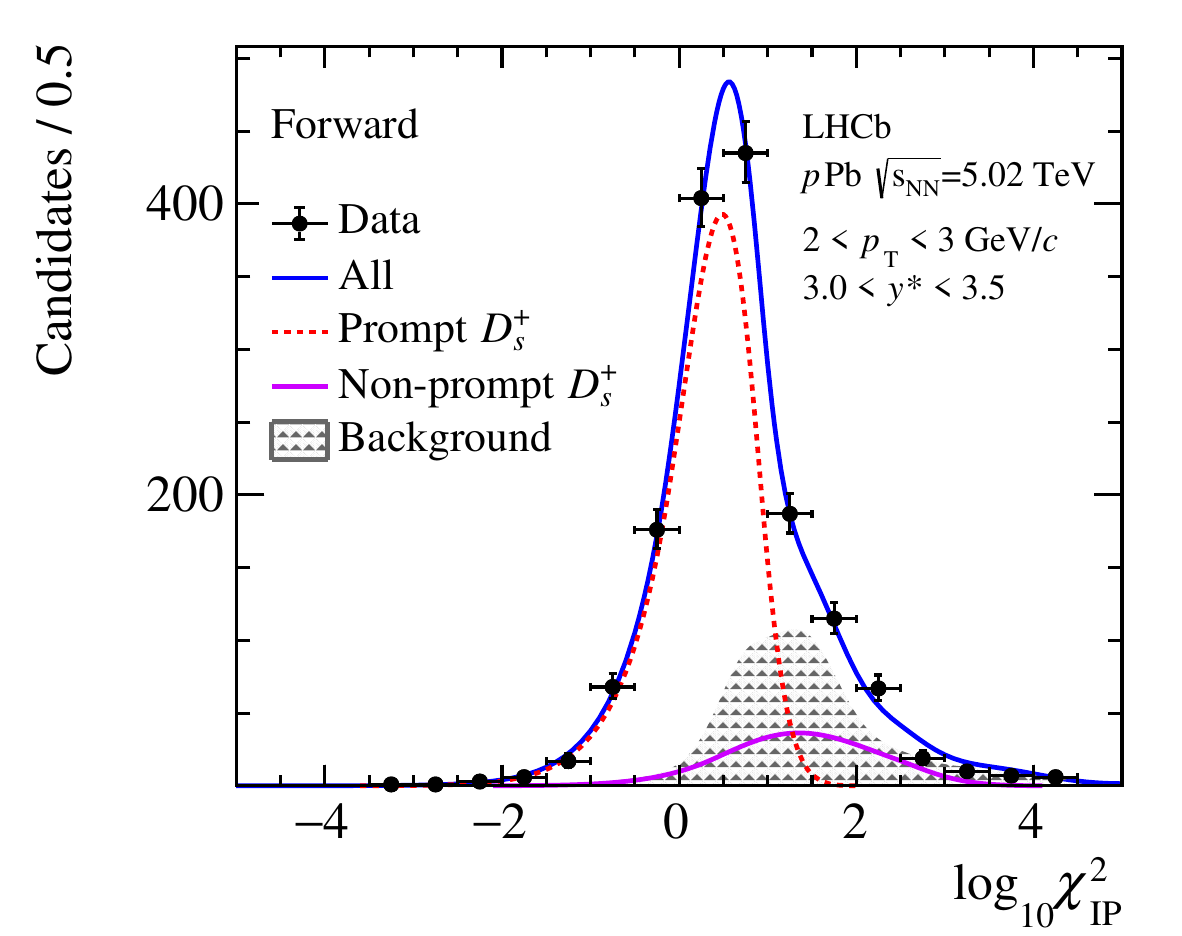}\\
\end{tabular}
\end{minipage}
\caption{Distributions of the simultaneous fits to the (left) $\text{M}(K^-K^+ \pi^+)$ and (right) $\log_{10}\chisqip$ for $\Ds$ mesons in the forward data sample in the kinematic bin of $2<\pt<3\gevc$
and $3.0<y^*<3.5$.}
\label{fig:MassIPFittPA_Ds}
\end{figure}
\begin{figure}[!tbp]
\centering
\begin{minipage}[t]{0.99\textwidth}
\centering
\begin{tabular}{cc}
\includegraphics[width=0.5\textwidth]{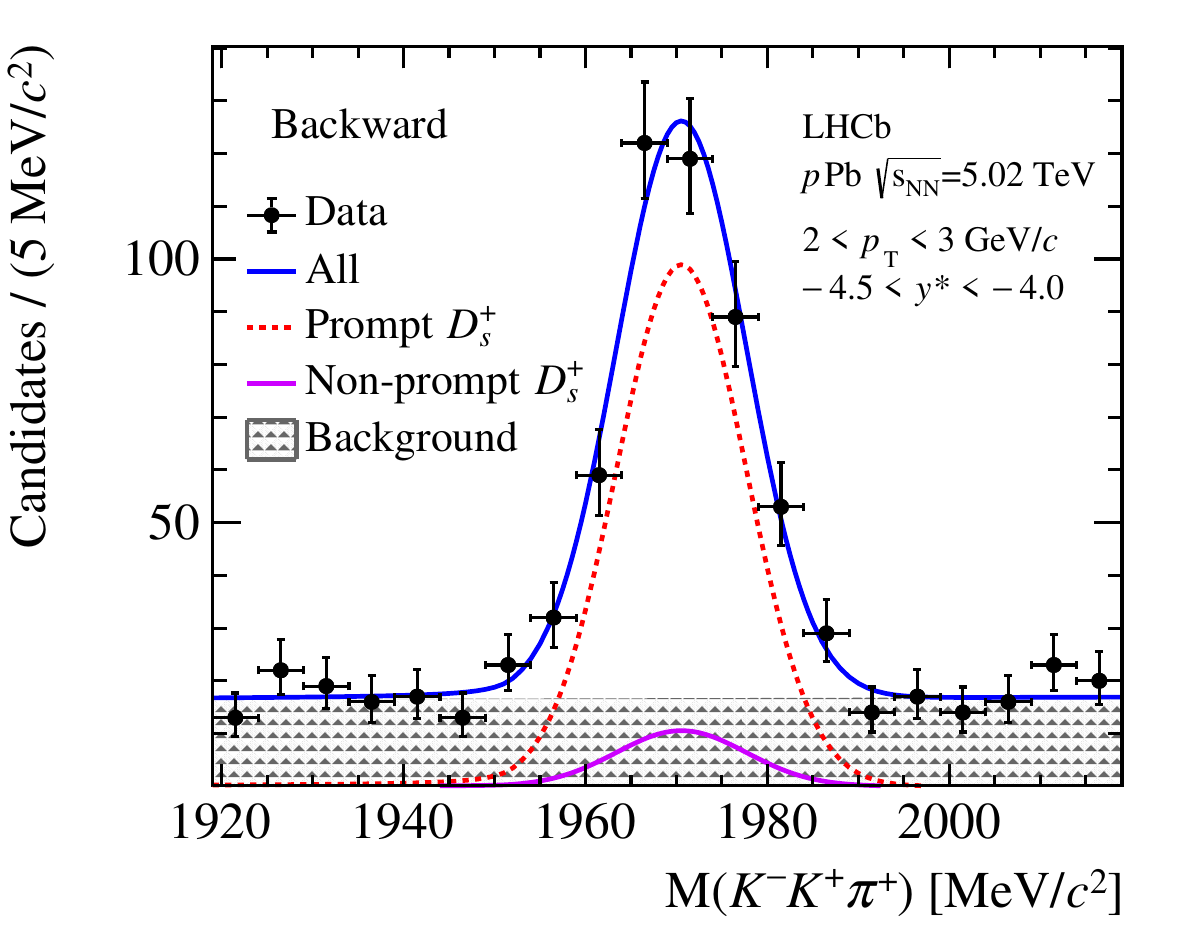}&
\includegraphics[width=0.5\textwidth]{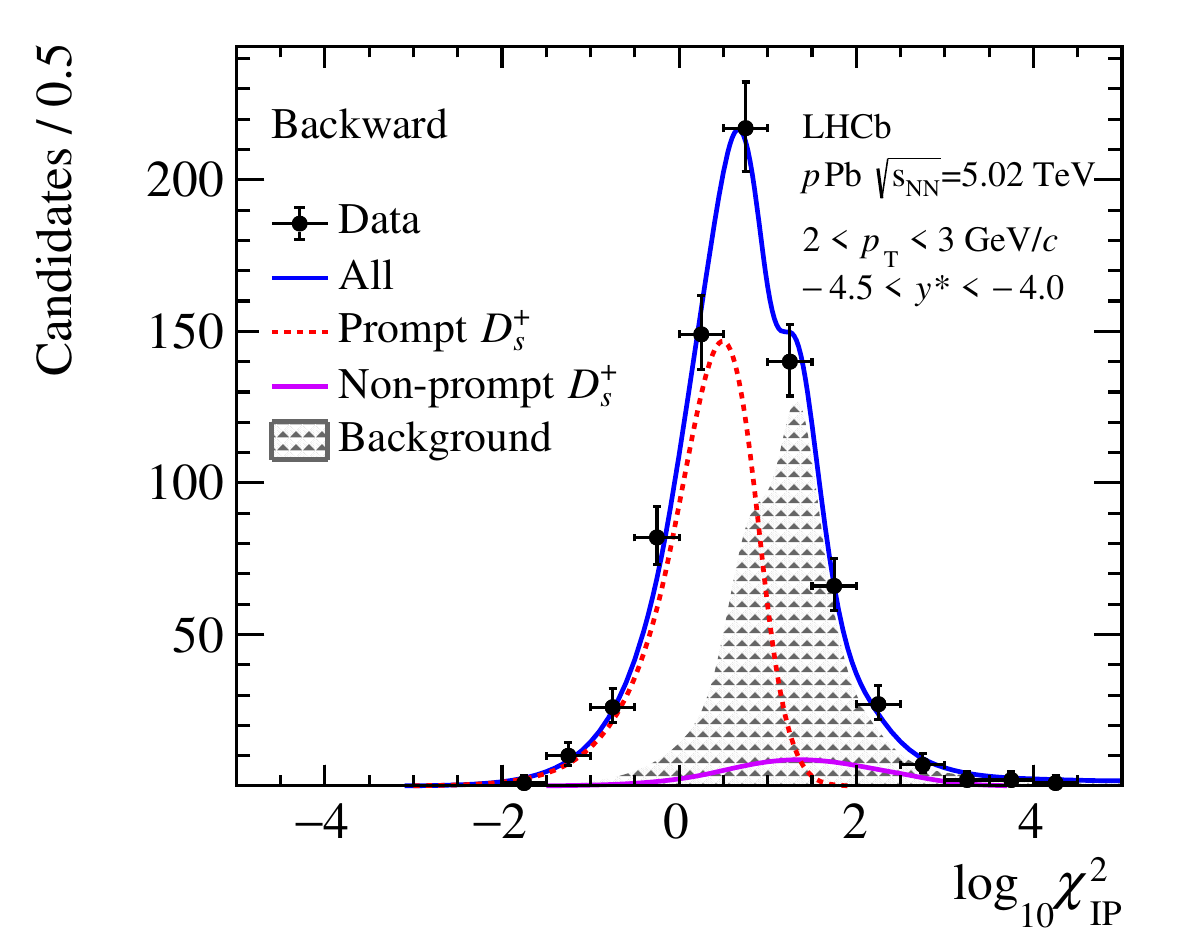}\\
\end{tabular}
\end{minipage}
\caption{Distributions of the simultaneous fits to the (left) $\text{M}(K^- K^+ \pi^+)$ and (right) $\log_{10}\chisqip$ for $\Ds$ mesons in the backward data sample in the kinematic bin of $2<\pt<3\gevc$
and $-4.5<y^*<-4.0$.}
\label{fig:MassIPFittAP_Ds}
\end{figure}

\newcommand{\rhoL}{\ensuremath{\rho_\mathrm{L}}\xspace}
\newcommand{\rhoR}{\ensuremath{\rho_\mathrm{R}}\xspace}

The prompt and non-prompt (\ie from $b$-hadron decays) $D$-meson yields in intervals of the \pt and $y^*$ are determined from simultaneous extended maximum-likelihood fits to the unbinned invariant-mass and $\log_{10}\chisqip$ distributions. Here, $\chisqip$ is the difference in vertex-fit $\chi^2$ of a given PV reconstructed with and without consideration of the signal candidates.
The simultaneous fits are performed to the $D$-meson candidates whose invariant mass lies within 
$\pm50\mevcc$ of their known mass~\cite{ParticleDataGroup:2022pth}.
The inclusive signal shape in the $M(K^- \pi^+ \pi^+)$ or $M(K^- K^+ \pi^+)$ distributions is modeled by the sum of a Crystal Ball (CB) function~\cite{Skwarnicki:1986xj} and a Gaussian with the same mean. The widths of CB function and Gaussian vary freely in the fit, and the fraction and the tail parameters of the CB function are fixed to values obtained from simulation. 
The combinatorial background is described by a linear function. 
The $\log_{10}\chisqip$ shapes for the prompt and non-prompt $D$-meson signal candidates are estimated using simulated events and modeled with an asymmetric Bukin curve~\cite{Bukin:2007zha} with tails described by Gaussian functions. 
In the forward and backward cases for both prompt and non-prompt components,
the width, asymmetry and tail coefficients of the Bukin function are fixed to the values obtained from simulated events. Additionally, the Bukin peak location for non-prompt $D$ mesons is fixed to the value obtained from simulated events.
The Bukin parameters are determined from the fit to the whole forward or backward simulation sample.
The combinatorial background $\log_{10}\chisqip$ distribution is modeled by a kernel density estimate function~\cite{Cranmer:2000du}, which is created in the side-band interval outside their $\pm50\mevcc$ mass window.
The total model is the sum of the contributions from the prompt and non-prompt signals and the combinatorial background, where each component is the product of the corresponding mass and $\log_{10}\chisqip$ distributions.
The simultaneous fits are carried out independently in each $(\pt, y^*)$ bin of the $\Dp$ or $\Ds$ mesons.
Figs.~\ref{fig:MassIPFittPA}--\ref{fig:MassIPFittAP_Ds} show the invariant mass and $\log_{10}\chisqip$ distributions, for two typical bins of $y^*$ in the forward and backward regions.

The total efficiency, $\etot$, in Eq.~\ref{eq:cross-section} is the product of the geometric acceptance, trigger, reconstruction, selection, and PID efficiencies. The geometric acceptance efficiency is estimated using simulated $pp$ events to eliminate the effect of the spatial acceptance of the LHCb detector.
The analysis uses a minimum activity trigger whose efficiency for events containing a \Dp or \Ds meson is found to be 100\%.
The reconstruction and selection efficiencies are calculated with simulated \pPb samples at 5.02$\ensuremath{\mathrm{\,Te\kern -0.1em V}}$, and corrected for known differences in tracking efficiency between data and simulation~\cite{LHCb:2014nio}.
The PID efficiency is estimated using a sample of \Dz meson decays selected from data without PID criteria~\cite{LHCb:2014set}, and collected during the same period as the \pPb sample used in the analysis.
The PID selection efficiency is calculated using the kaon and pion single-track efficiencies from calibration data, and averaging them based on the kinematic distributions observed in the simulated events in each $(\pt,y^{*})$ bin.
The reconstruction and selection efficiency and the PID efficiency are sensitive to differences between the track multiplicity distributions in \pPb data and those observed in the simulation and PID-calibration samples. This effect is corrected for by re-weighting the latter to match the distributions seen in data.

\section{Systematic uncertainties}\label{sec:syst}

The systematic uncertainties affecting the cross-section measurements are listed in Table~\ref{tab:SystematicSummary}.
The uncertainties of the forward and backward samples were evaluated separately, unless stated otherwise.
\begin{table}[tbh]
\caption{Summary of systematic and statistical uncertainties on the $D$-meson cross-section measurements ($\%$).}
\centering
\renewcommand\arraystretch{1.5}
\begin{tabular}{l|cc|ccc}
\multirow{2}{*}{Source} & \multicolumn{2}{c|}{$\Dp \to K^- \pi^+ \pi^+$} & \multicolumn{2}{c}{$\Ds \to K^- K^+ \pi^+$}\\
    & Forward    & Backward & Forward    & Backward\\
\hline
{\it Uncorrelated between bins}    &     &   &     &  \\
Prompt yield determination                  & 0.1--1.4   & 0.1--0.9 & 0--3.4   & 0--6.5\\
Simulation sample size            & 0--10   &0--8 & 1--8   &1--6\\
\hline
{\it Correlated between bins} &     &   &     &    \\
Multiplicity correction                   & 0.4--2.3   & 1.8--4.8 & 0.4--2.5   & 2.1--5.0\\
Hadronic interactions                 & 3.9   & 3.9 & 3.6   & 3.6\\
PID efficiency             & 0--18   &1--13 & 1--12   &1--17\\
Luminosity                 & 1.9   & 2.1 & 1.9   & 2.1\\
Branching fraction     & 1.6     & 1.6 & 5.8   & 5.8\\
\hline
Statistical uncertainty    &1--20    & 1--28 &3--48    & 4--52\\
\end{tabular}
\label{tab:SystematicSummary}
\end{table}

The systematic uncertainty related to the determination of the prompt signal yield has contributions both from the assumed fit model and the fixed parameters in the simultaneous fits, and from the fit method itself.
The uncertainty associated to the fit model is assigned using double CB functions for the signal component and an exponential function for the background component to fit the invariant-mass distributions, and using a Gaussian for non-prompt components to fit the $\log_{10}\chisqip$ distributions.
Parameters in the fit model are allowed to vary from their nominal values within uncertainties estimated from simulated events.
The standard deviation of the nominal and the alternative fits is taken as the uncertainty on the prompt signal yield determination. The resulting uncertainty is estimated to be less than 6.5\% for most bins.

The systematic uncertainty related to the multiplicity correction of the reconstruction and selection efficiencies is due to several contributions.
One source of uncertainty is caused by the choice of the variables used to represent the detector occupancy for weighting the distribution.
The number of charged tracks, the number of long tracks, the number of hits in the VELO, the number of hits in the IT and the OT, and the number of hits in the TT, are all considered separately. The standard deviation of the efficiencies weighted by each variable is taken as systematic uncertainty.
The effects are summed in quadrature, yielding a total uncertainty on the $\Dp$ ($\Ds$) meson multiplicity correction of $0.4-2.3\%$ ($0.4-2.5\%$) and $1.8-4.8\%$ ($2.1-5.0\%$) for the forward and backward collision samples, respectively.

The difference in tracking efficiencies between data and simulation is studied with muons. An additional uncertainty of $1.1\%$ ($1.4\%$) is assigned for each kaon (pion) present in the final state, due to the incomplete modeling of the hadronic interactions of these particles with the material of the LHCb detector~\cite{LHCb:2017yua}.
This effect is dominated by the uncertainty on knowledge of the amount of material present within the detector and hence can be assumed to be fully correlated for kaons and pions. The total uncertainty associated to this effect is $3.9\%$ ($3.6\%$) for $\Dp$ ($\Ds$) mesons.

\label{sec:PIDSys}

The finite size of the calibration sample used to determine the particle identification efficiency also contributes to the systematic uncertainty.
This contribution is evaluated by varying the pion and kaon PID efficiencies obtained from the calibration sample within their statistical uncertainties.
The uncertainty from the PID binning scheme is studied by using alternative binning schemes for the single track efficiency.
The total PID systematic uncertainty is a quadratic sum of these two sources and it ranges between 0 and 18\% depending on the kinematic region and the collision sample.

The relative uncertainty of the integrated luminosity is 
nearly $2\%$ for both forward and backward samples~\cite{LHCb:2013gmv}.
The relative uncertainty of the branching fraction is $1.6\%$~\cite{ParticleDataGroup:2022pth} and $5.8\%$~\cite{CLEO:2008hzo} for $\BF(\Dp \to K^- \pi^+ \pi^+)$ and $\BF(\Ds \to K^- K^+ \pi^+)$, respectively. 
The finite size of the simulation sample introduces uncertainties on the efficiencies which are then propagated to the cross-section measurements. This effect is negligible for the central rapidity region, while becomes significant in the region close to the boundaries of $\pt$ and $y^*$, ranging between 0 and 10\%.
The total systematic uncertainty varies gradually from 5\% in the central rapidity interval to about 20\% in the boundary interval.
\section{Results}
\label{sec:Result}
\subsection{Production cross-sections}
\begin{figure}[tbp]
\centering
\begin{minipage}[t]{0.49\textwidth}
\centering
\includegraphics[width=1.0\textwidth]{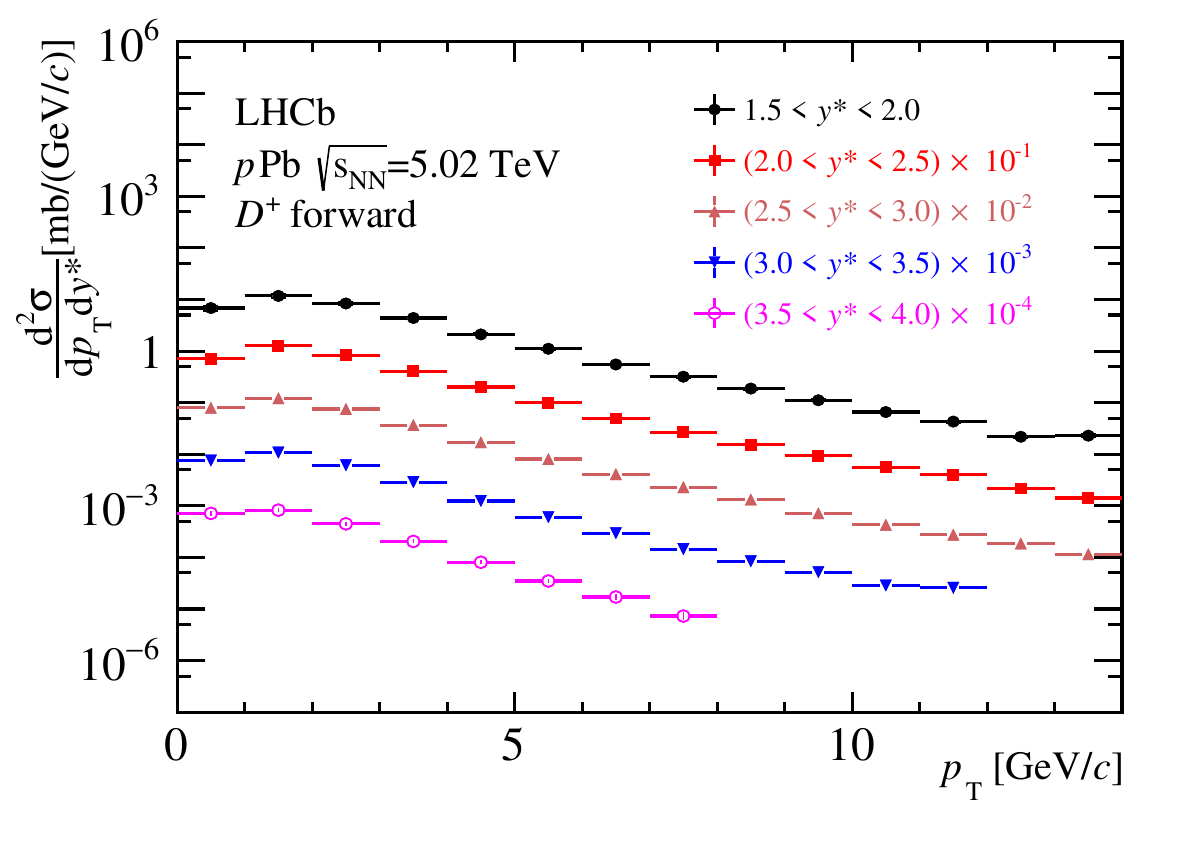}
\end{minipage}
\begin{minipage}[t]{0.49\textwidth}
\centering
\includegraphics[width=1.0\textwidth]{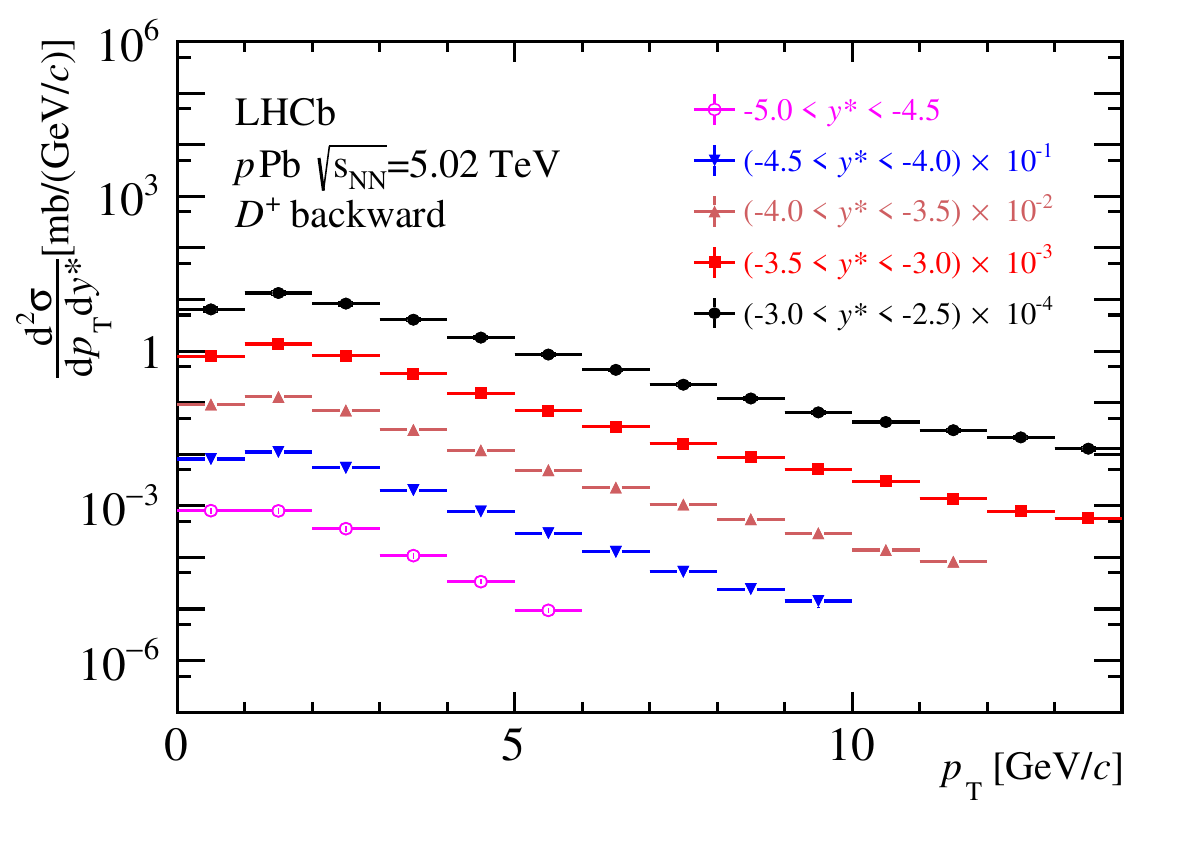}
\end{minipage}
\caption{Double-differential cross-section of prompt $\Dp$ mesons in $\pPb$ collisions for the (left) forward and (right) backward rapidities.
The error bars are the statistical uncertainty and the boxes are the systematic uncertainty, both of which are smaller than the symbol size.
The value in a particular rapidity interval is scaled by a multiplicative factor $10^{-m}$,  where the factor $m$ increase as the rapidity rises.
}
\label{fig:CrossSection2D}
\end{figure}

\begin{figure}[tbp]
\centering
\begin{minipage}[t]{0.49\textwidth}
\centering
\includegraphics[width=1.0\textwidth]{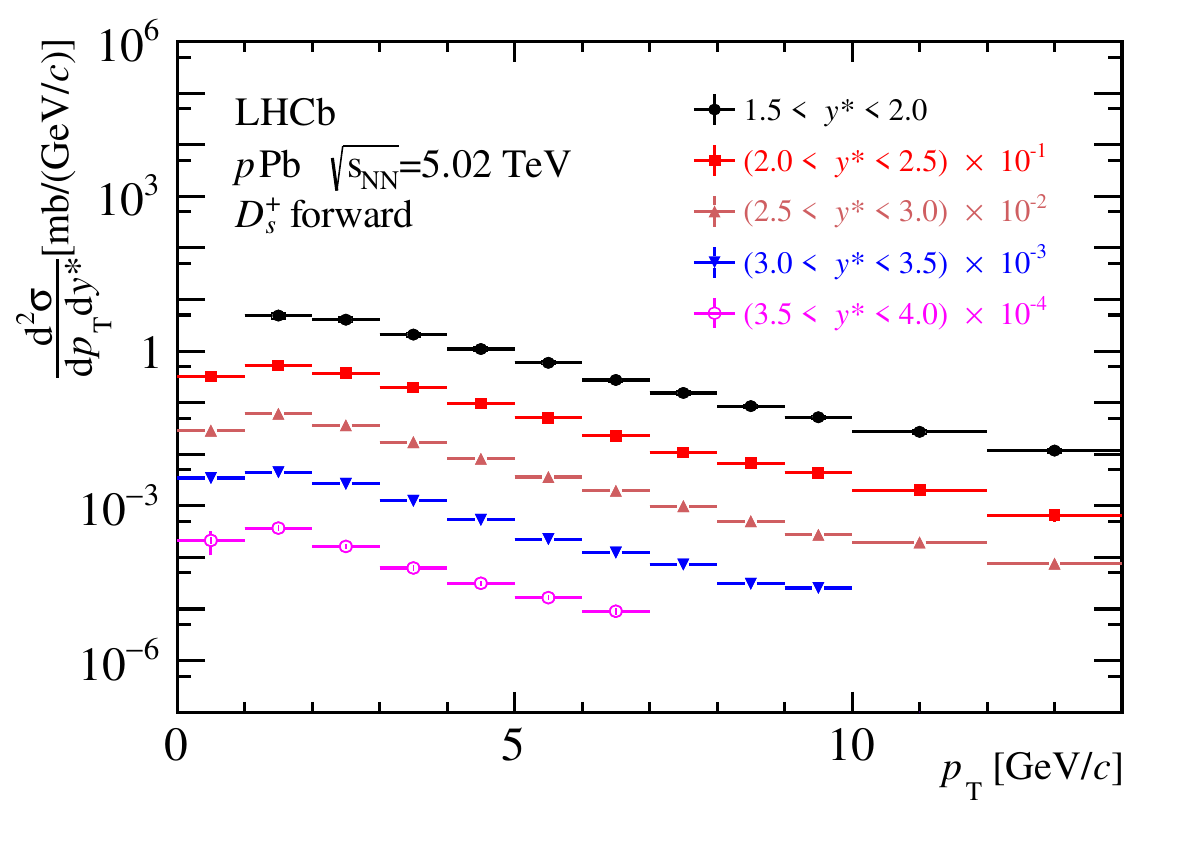}
\end{minipage}
\begin{minipage}[t]{0.49\textwidth}
\centering
\includegraphics[width=1.0\textwidth]{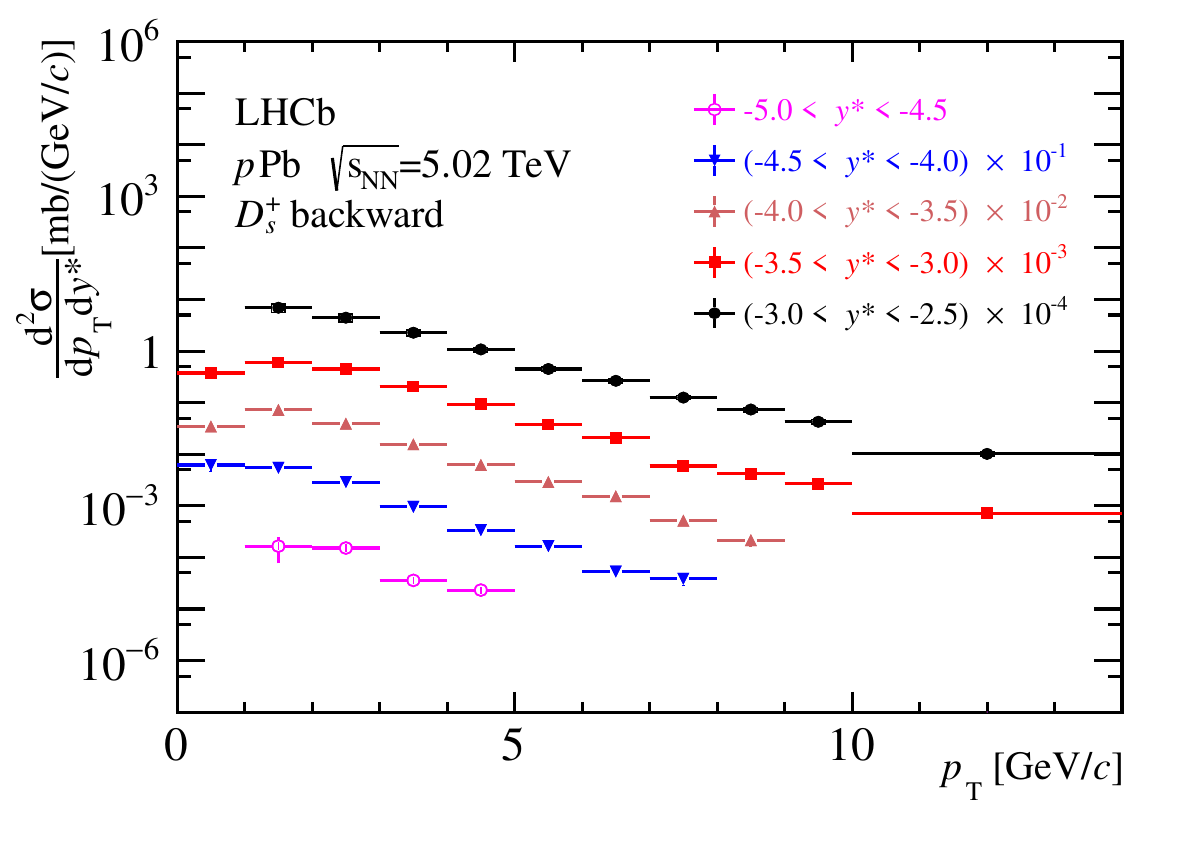}
\end{minipage}
\caption{Double-differential cross-section of prompt $\Ds$ mesons in $\pPb$ collisions for the (left) forward and (right) backward rapidities.
The error bars are the statistical uncertainty and the boxes are the systematic uncertainty, both of which are smaller than the symbol size.
The value in a particular rapidity interval is scaled by a multiplicative factor $10^{-m}$,  where the factor $m$ increase as the rapidity rises.
}
\label{fig:CrossSection2DDs}
\end{figure}

The measured values of the double-differential cross-section of prompt 
$\Dp$ and $\Ds$ mesons in proton-lead collisions for forward and backward rapidities, as a function of $\pt$ and $y^{*}$, are displayed in Figs.~\ref{fig:CrossSection2D} and \ref{fig:CrossSection2DDs} and summarized
in Tables~\ref{tab:CrossSection2Dfor}--\ref{tab:CrossSection2DDsbac} of appendix~\ref{sec:DpDs2D}. 


The one-dimensional differential prompt $\Dp$ and $\Ds$ meson cross-sections as a function of $\pt$ or $y^{*}$ are shown in Figs.~\ref{fig:CrossSection1D} and \ref{fig:CrossSection1DDs}, and are also summarized in 
appendix~\ref{sec:DpDs2D}.
The measurements are also shown as a function of $\pt$ integrated over $y^{*}$ for the common rapidity range $2.5<|y^{*}|<4.0$.

\begin{figure}[tbp]
\centering
\begin{minipage}[t]{0.49\textwidth}
\centering
\includegraphics[width=1.0\textwidth]{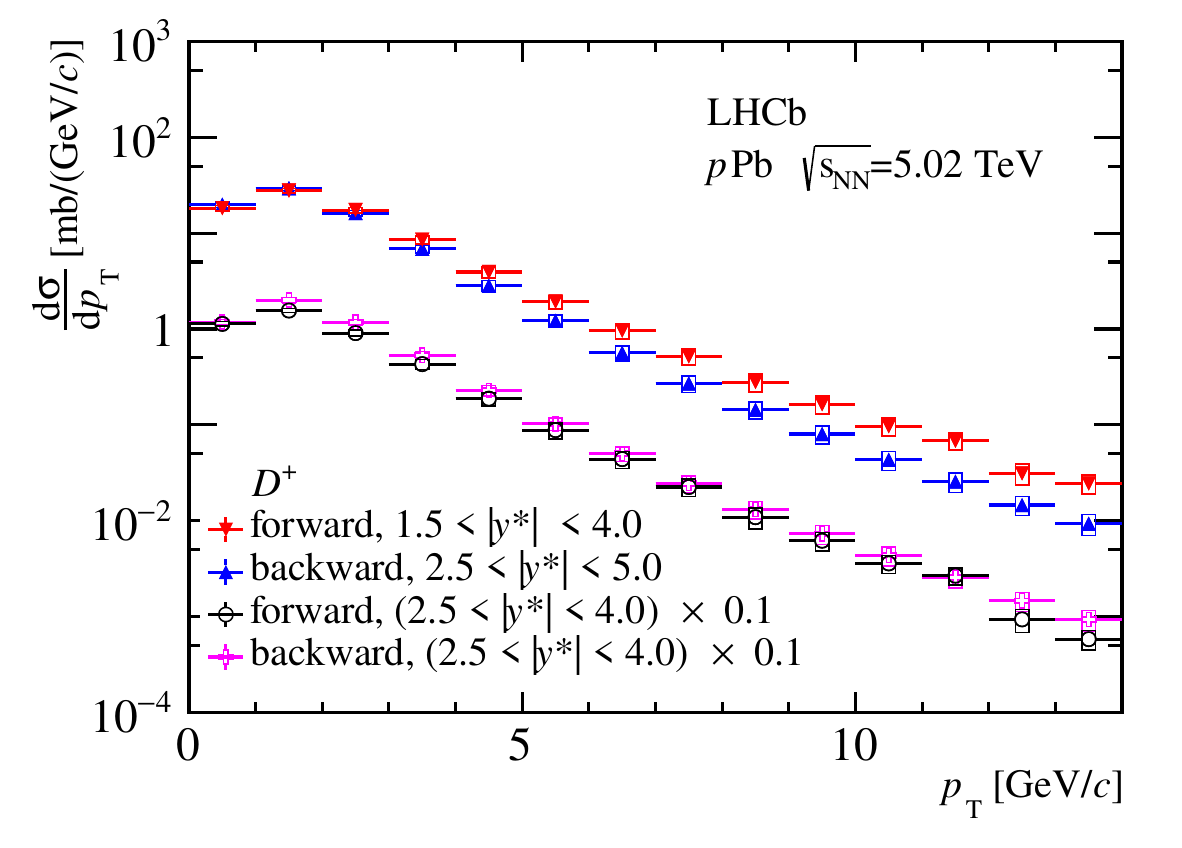}
\end{minipage}
\begin{minipage}[t]{0.49\textwidth}
\centering
\includegraphics[width=1.0\textwidth]{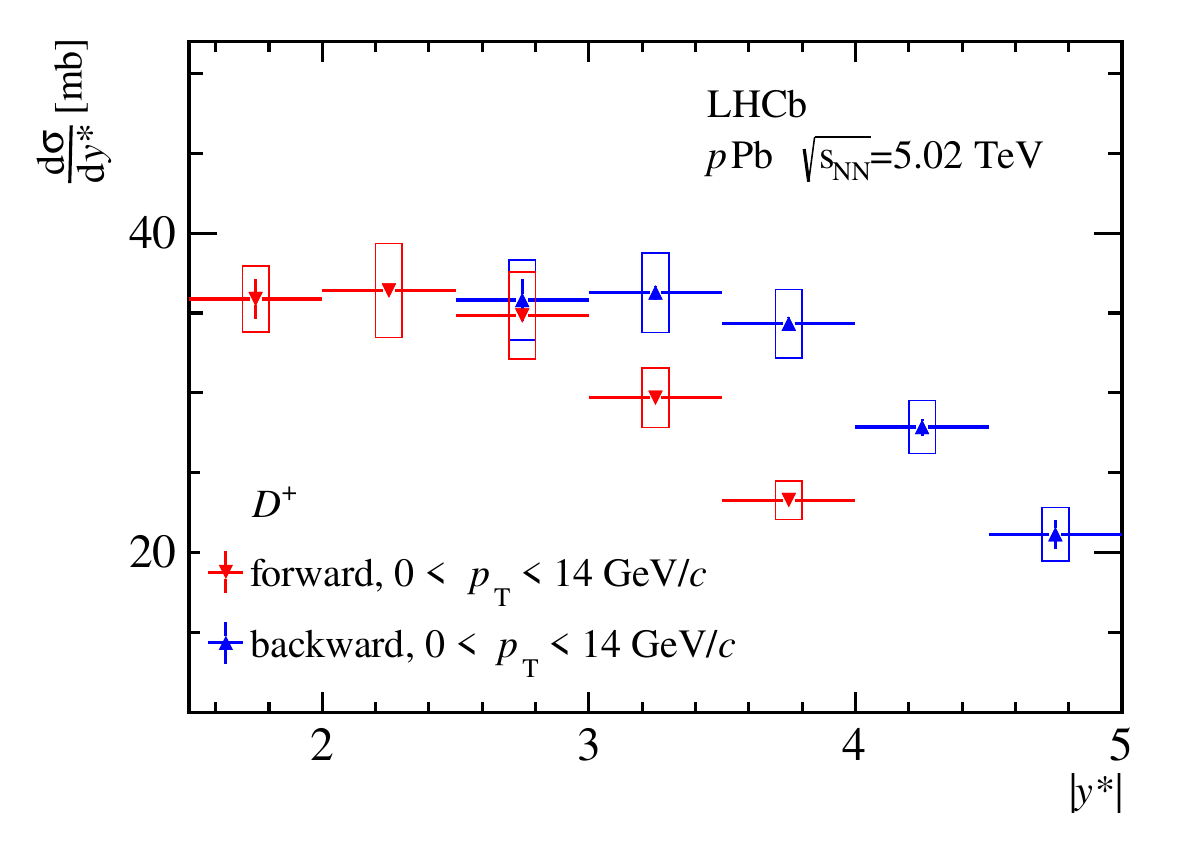}
\end{minipage}
\caption{Differential cross-section of prompt $\Dp$ meson production in $\pPb$ collisions as a function of (left) $\pt$ and (right) $y^{*}$ in the forward and backward collision samples. 
The error bars are the statistical uncertainty while the boxes are the systematic uncertainty.
}
\label{fig:CrossSection1D}
\end{figure}

\begin{figure}[tbp]
\centering
\begin{minipage}[t]{0.49\textwidth}
\centering
\includegraphics[width=1.0\textwidth]{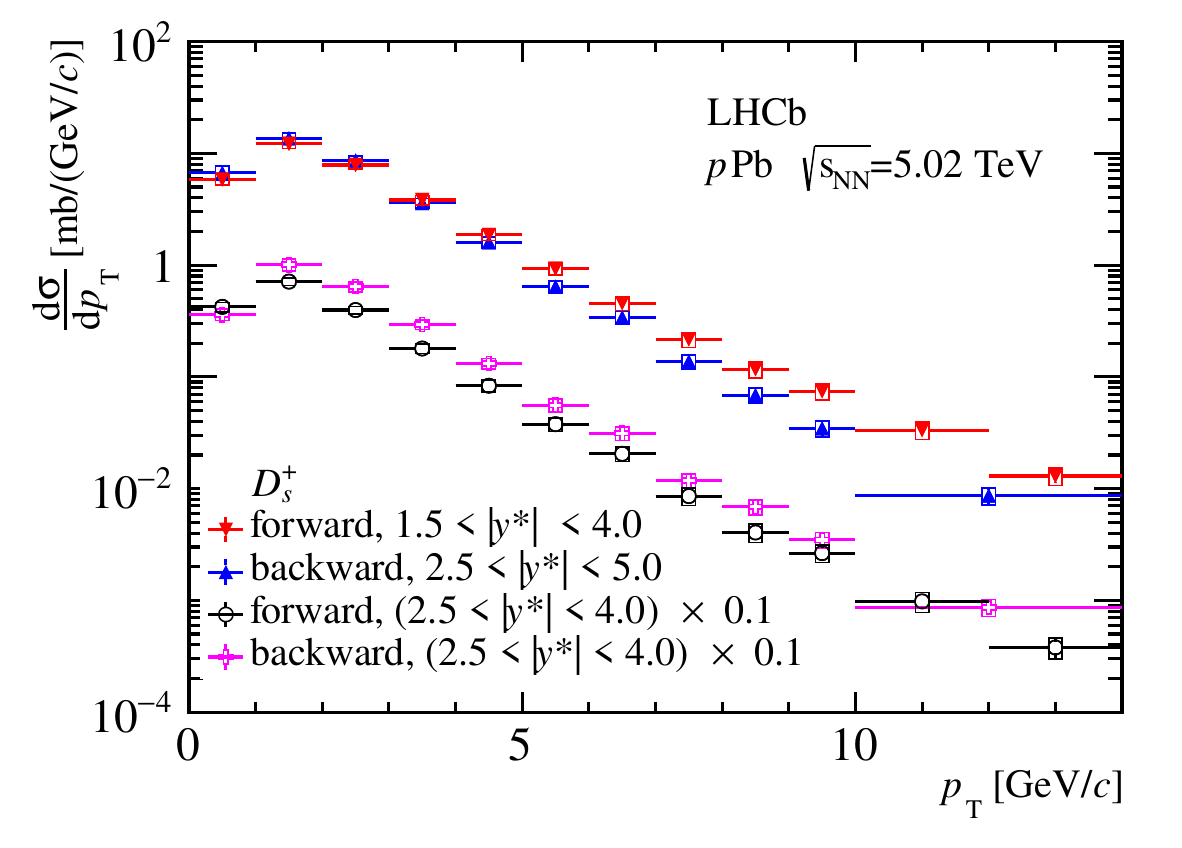}
\end{minipage}
\begin{minipage}[t]{0.49\textwidth}
\centering
\includegraphics[width=1.0\textwidth]{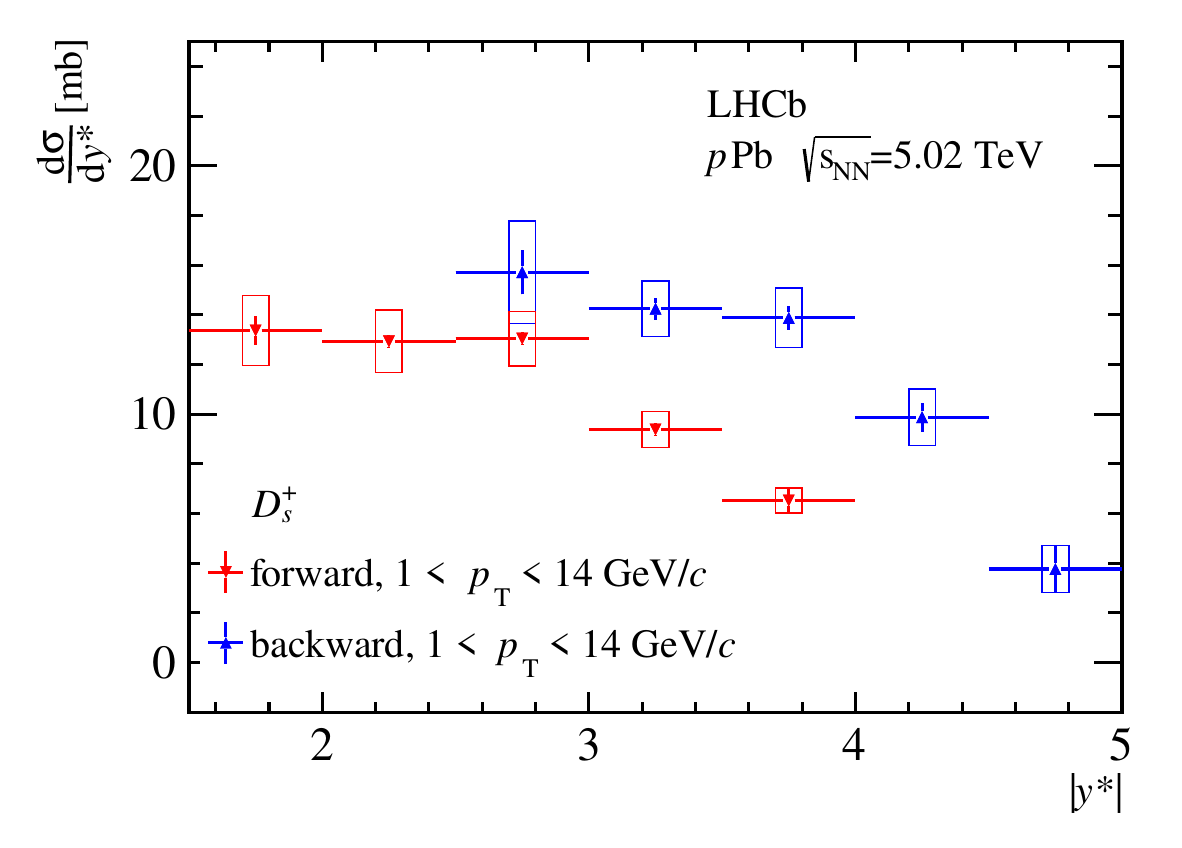}
\end{minipage}
\caption{Differential cross-section of prompt $\Ds$ meson production in $\pPb$ collisions as a function of (left) $\pt$ and (right) $y^{*}$ in the forward and backward collision samples. 
The error bars are the statistical uncertainty while the boxes are the systematic uncertainty.
}
\label{fig:CrossSection1DDs}
\end{figure}


The integrated production cross-sections of prompt $\Dp$ and $\Ds$ mesons in $\pPb$ forward collisions in the full and common fiducial regions are
\begin{equation}
\sigma(\Dp)_\mathrm{forward}(0<\pt<10\gevc, 1.5<|y^{*}|<4.0) =79.8\pm0.7\pm5.1\mbarn, \nonumber
\end{equation}
\begin{equation}
\sigma(\Dp)_\mathrm{forward}(0<\pt<10\gevc, 2.5<|y^{*}|<4.0) =43.8\pm0.2\pm2.8\mbarn, \nonumber
\end{equation}
\begin{equation}
\sigma(\Ds)_\mathrm{forward}(1<\pt<10\gevc, 1.5<|y^{*}|<4.0) =27.5\pm0.4\pm2.4\mbarn, \nonumber
\end{equation}
\begin{equation}
\sigma(\Ds)_\mathrm{forward}(1<\pt<10\gevc, 2.5<|y^{*}|<4.0) =14.4\pm0.3\pm1.1\mbarn. \nonumber
\end{equation}
The first uncertainties are statistical while the second are systematic.
The integrated production cross-sections of prompt $\Dp$ and $\Ds$ mesons in \pPb backward collisions in the full and common fiducial regions are
\begin{equation}
\sigma(\Dp)_\mathrm{backward}(0<\pt<10\gevc, 2.5<|y^{*}|<5.0) =77.6\pm0.9\pm4.9\mbarn, \nonumber
\end{equation}
\begin{equation}
\sigma(\Dp)_\mathrm{backward}(0<\pt<10\gevc, 2.5<|y^{*}|<4.0) =53.1\pm0.7\pm3.5\mbarn, \nonumber
\end{equation}
\begin{equation}
\sigma(\Ds)_\mathrm{backward}(1<\pt<10\gevc, 2.5<|y^{*}|<5.0) =28.7\pm0.8\pm3.0\mbarn, \nonumber
\end{equation}
\begin{equation}
\sigma(\Ds)_\mathrm{backward}(1<\pt<10\gevc, 2.5<|y^{*}|<4.0) =21.9\pm0.6\pm2.3\mbarn. \nonumber
\end{equation}


\subsection{Nuclear modification factors}
\begin{figure}[tbp]
\centering
\begin{minipage}[t]{0.49\textwidth}
\centering
\includegraphics[width=1.0\textwidth]{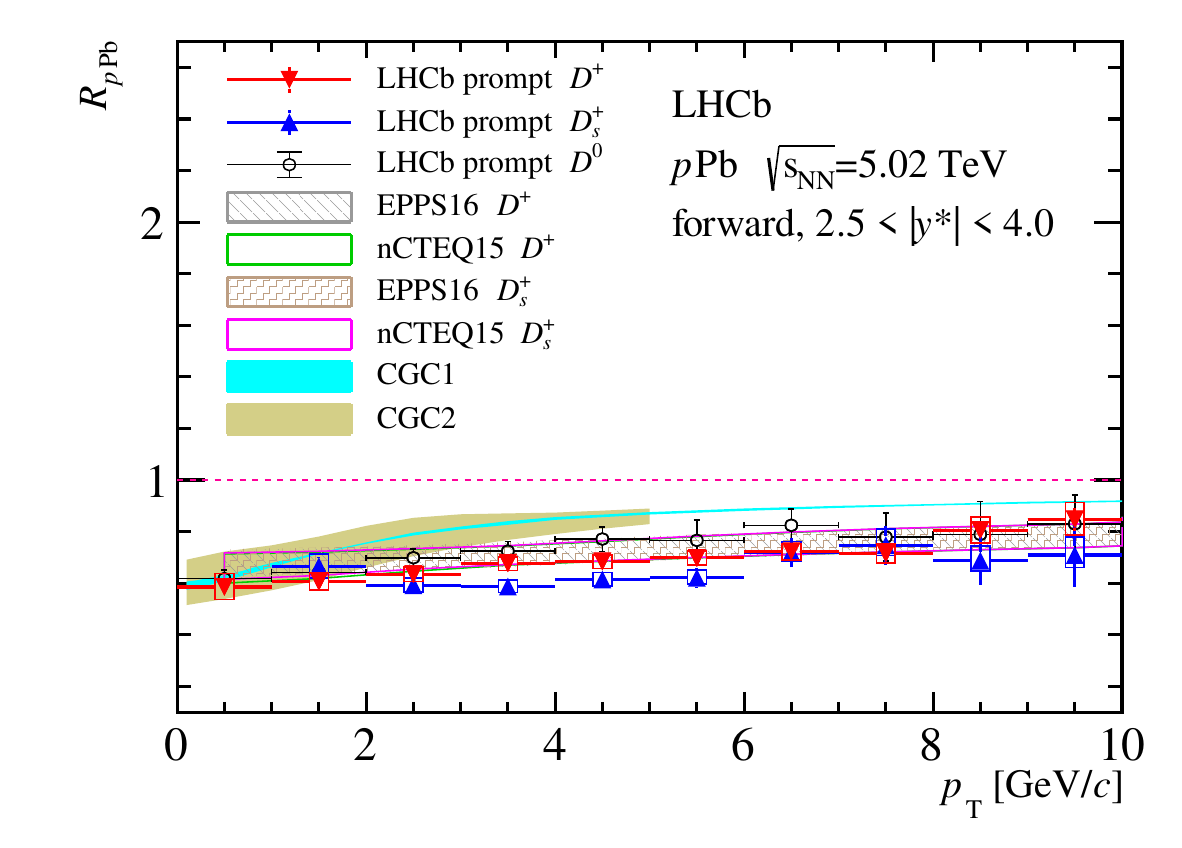}
\end{minipage}
\begin{minipage}[t]{0.49\textwidth}
\centering
\includegraphics[width=1.0\textwidth]{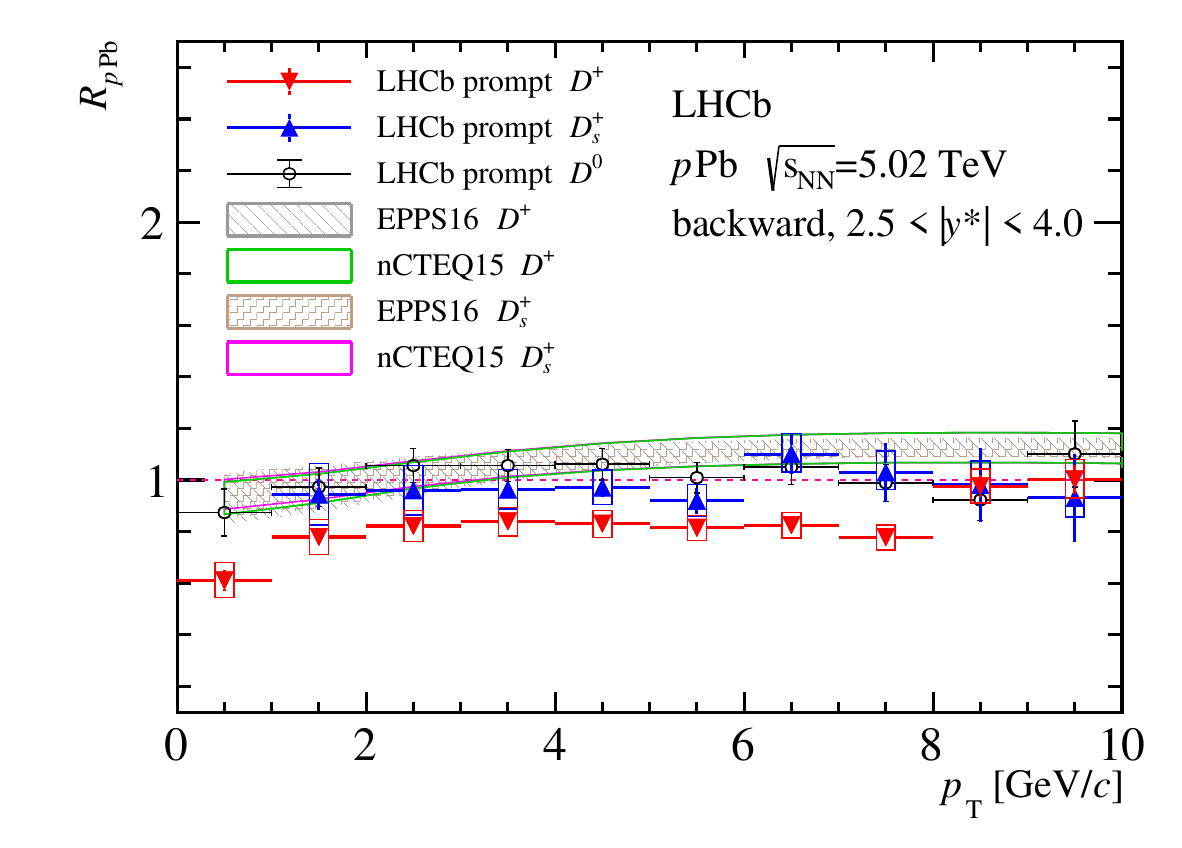}
\end{minipage}
\caption{Nuclear modification factors $R_{\pPb}$ as a function of $\pt$ for prompt $\Dp$ and $\Ds$ meson
production in the (left) forward data and (right) backward data. 
The error bars are the statistical uncertainty and the boxes are the systematic uncertainty.
The CGC~\cite{Ducloue:2015gfa,Ducloue:2016ywt,Fujii:2017rqa} predictions are only available in the forward region. 
Previous results on \Dz mesons~\cite{LHCb:2017yua} from LHCb are also shown.
}
\label{fig:RpPbPT}
\end{figure}

\begin{figure}[tbp]
\centering
\begin{minipage}[t]{0.7\textwidth}
\centering
\includegraphics[width=\textwidth]{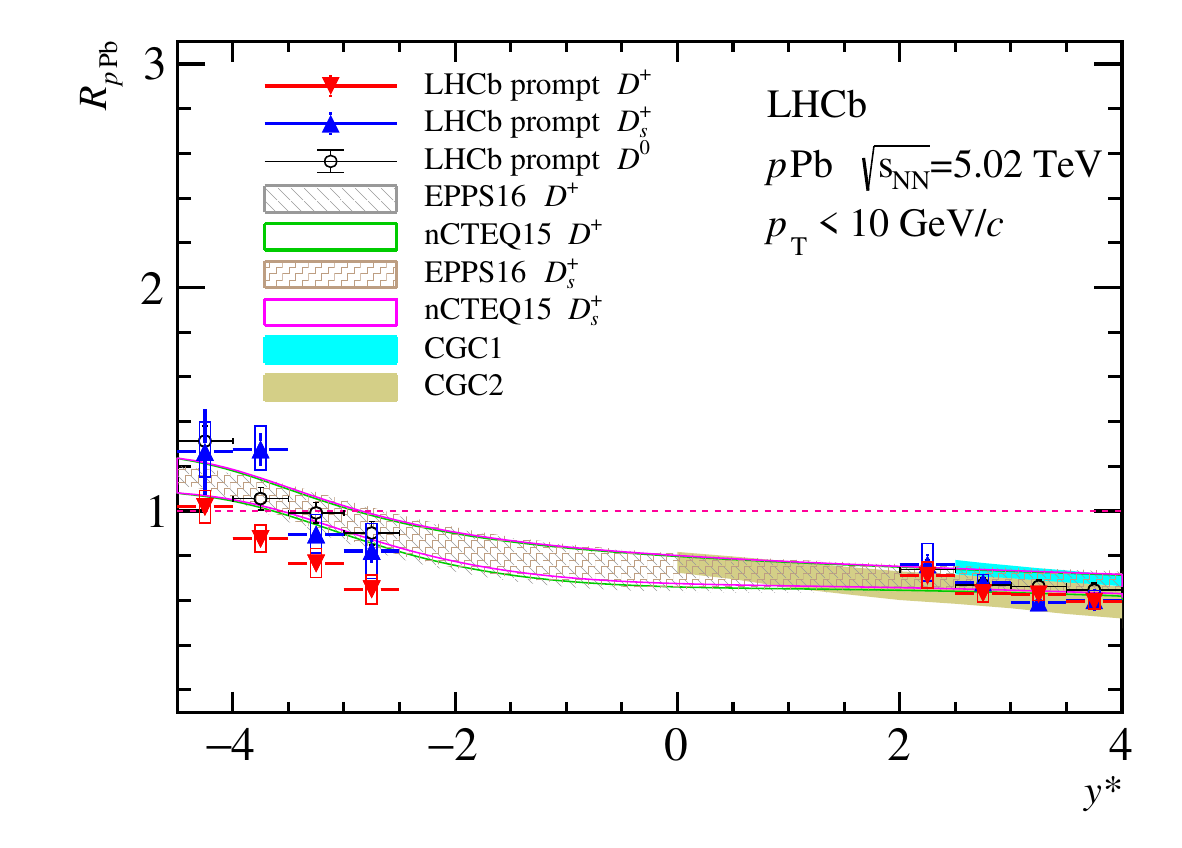}
\end{minipage}
\caption{Nuclear modification factors $R_{\pPb}$ as a function of $y^*$ for prompt $\Dp$ and $\Ds$ meson production, integrated  up to  $\pt=10\gevc$. 
The error bars are the statistical uncertainty and the boxes are the systematic uncertainty.
The CGC~\cite{Ducloue:2015gfa,Ducloue:2016ywt,Fujii:2017rqa} predictions are only available in the forward region. 
Previous results on \Dz mesons~\cite{LHCb:2017yua} from LHCb are also shown.
}
\label{fig:RpPbY}
\end{figure}

The values of the $D$-meson production cross-section in $pp$
collisions at 5.02$\ensuremath{\mathrm{\,Te\kern -0.1em V}}$, which are necessary for the measurement of the nuclear modification factor $R_{\pPb}$, are taken from Ref.~\cite{LHCb:2016ikn}.
Systematic uncertainties associated with the hadronic interaction length and branching fractions of Ref.~\cite{LHCb:2016ikn} are fully correlated with
the measurements presented here, while other uncertainties are assumed to be uncorrelated.
The nuclear modification factors for prompt $\Dp$ and $\Ds$ production, displayed in Fig.~\ref{fig:RpPbPT} in bins of $\pt$ and Fig.~\ref{fig:RpPbY} in bins of $y^*$, show a slight increase as a function of $\pt$, implying that the suppression may decrease with increasing \pt. 
The values of $R_{\pPb}$ for $\Dp$ and $\Ds$ mesons are given in 
appendix~\ref{sec:RpA}.

%

The measurements are compared with previous $\Dz$ results by the LHCb collaboration at $\snn=5.02$$\ensuremath{\mathrm{\,Te\kern -0.1em V}}$~\cite{LHCb:2017yua}, as well as with the HELAC-Onia generator~\cite{Shao:2015vga,Shao:2012iz,Lansberg:2016deg} using EPPS16 and nCTEQ15 nPDFs~\cite{Kusina:2017gkz,Kusina:2020dki}, where the nPDFs are re-weighted with the $\Dz$ cross-section~\cite{LHCb:2017yua}. 
At forward rapidity, the $\Dp$ results agree very well with the $R_\pPb$ of $\Dz$ mesons and the nPDF calculations within uncertainties. At backward rapidity, the $\Ds$ results are consistent with the $\Dz$ data and the nPDF calculations within uncertainties. However, the $\Dp$ $R_\pPb$ seems to be systematically lower than $\Dz$ and $\Ds$ mesons across the whole $\pt$ range due to the overall decrease in the $R_{\Dp/\Dz}$ ratios at backward rapidity relative to the forward in Fig.~\ref{fig:DpDzRatio}.
The \Dp $R_\pPb$ results at backward rapidity are lower than nPDF predictions with a significance of about $1-3$ standard deviations as further discussed in Sect.~\ref{sec:pratios}, suggesting possible changes in charm hadronization in \pPb collisions.

In Figs.~\ref{fig:RpPbPT} and \ref{fig:RpPbY} the measurements are also compared with predictions in the CGC framework, which include the effect of the saturation of partons at small Bjorken-$x$.
For CGC1~\cite{Ducloue:2015gfa,Ducloue:2016ywt}, the cross-section of the $D$ mesons is obtained with the optical Glauber mechanism correlating the initial state of the nucleon with that of the proton, and for CGC2~\cite{Fujii:2017rqa}, it is derived by convolving the charm-quark fragmentation function in a transverse momentum-dependent factorization framework.
The CGC models are found to be able to describe the trend of prompt $D$-meson nuclear modifications as a function of $\pt$ and of rapidity in the forward regions.

\subsection{Forward-backward ratio}

\begin{figure}[tbp]
\centering
\begin{minipage}[t]{0.49\textwidth}
\centering
\includegraphics[width=1.0\textwidth]{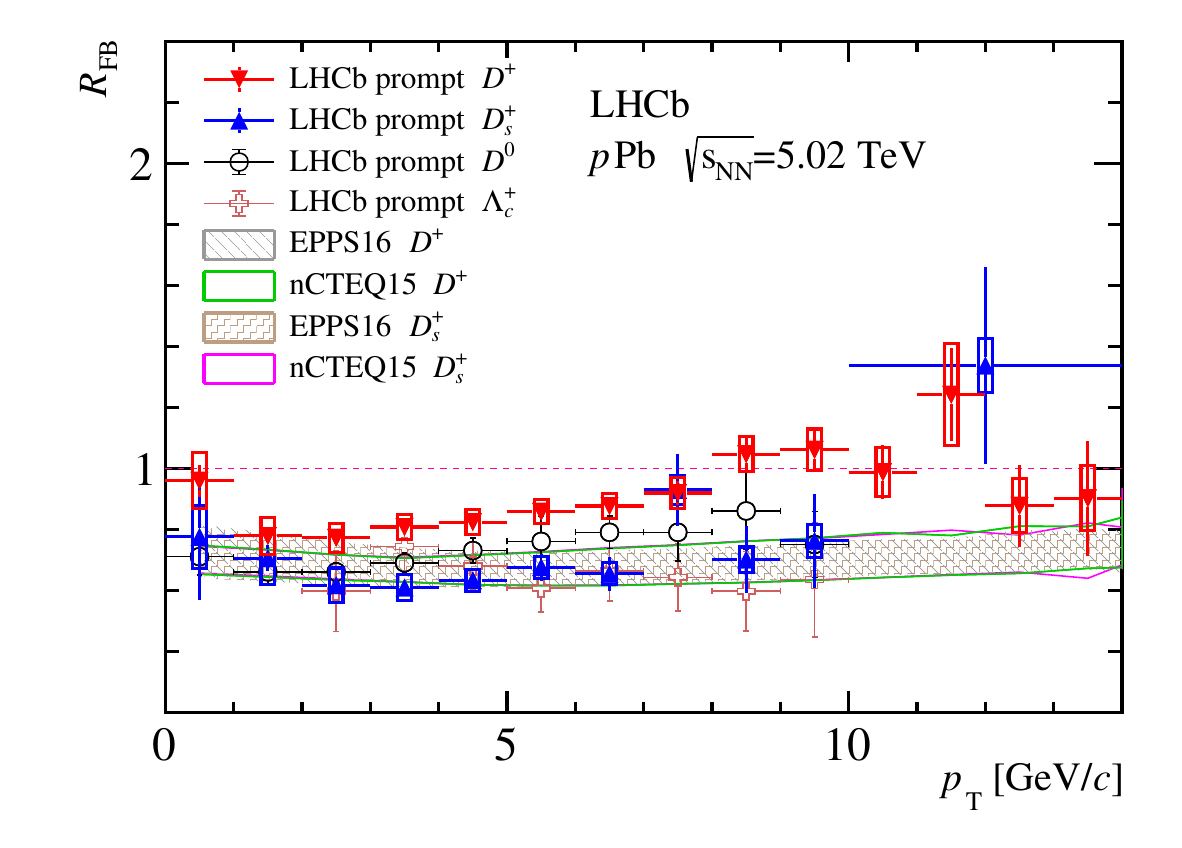}
\end{minipage}
\begin{minipage}[t]{0.49\textwidth}
\centering
\includegraphics[width=1.0\textwidth]{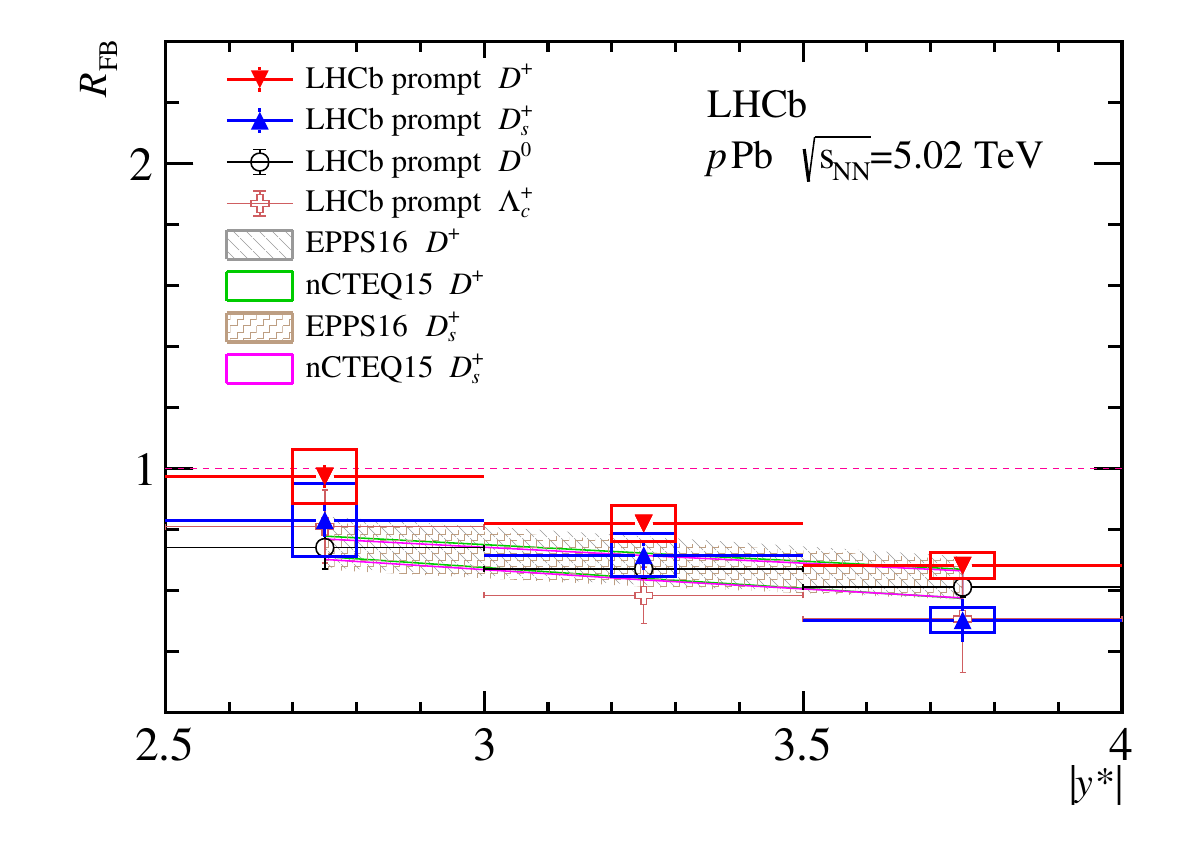}
\end{minipage}
\caption{Forward-backward ratios $R_\mathrm{FB}$ for prompt $\Dp$ and $\Ds$ meson production (left) as a function of $\pt$; (right) as a function of $y^*$. 
The error bars are the statistical uncertainty while the boxes are the systematic uncertainty. 
Previous results on \Dz~\cite{LHCb:2017yua} mesons and \Lc~\cite{LHCb:2018weo} baryons from LHCb are also shown.
} 
\label{fig:RFBResult}
\end{figure}

In the forward-backward production ratio $R_\mathrm{FB}$, the  uncertainties from common sources between the forward and backward measurements cancel out.
The uncertainties due to branching fractions and to hadronic interactions with the detector are considered fully correlated, 
while other uncertainties are conservatively treated as uncorrelated.  
The measured  $R_\mathrm{FB}$ values for \Dp and \Ds mesons are shown in Fig.~\ref{fig:RFBResult}, as a function of 
$\pt$ integrated over the range $2.5<|y^*|<4.0$,
and as a function of $y^*$ integrated up to $\pt=14\gevc$.
The $R_\mathrm{FB}$ values in different kinematic bins are tabulated in appendix~\ref{sec:RFB_DpDs}.
The $R_\mathrm{FB}$ values for \Dp and \Ds mesons are also compared to the measurements of $\Dz$ mesons~\cite{LHCb:2017yua} and \Lc baryons~\cite{LHCb:2018weo} by the LHCb collaboration at $\snn=5.02$$\ensuremath{\mathrm{\,Te\kern -0.1em V}}$ in Fig.~\ref{fig:RFBResult}.
 The $R_\mathrm{FB}$ of all open charm hadrons deviates further from unity as $|y^*|$ increases.
This behavior is consistent with the expectations from the QCD calculations, suggesting that the asymmetry becomes more pronounced in the large rapidity region.
The $R_\mathrm{FB}$ of $\Ds$ mesons as a function of $y^*$ shows reasonable agreement with $\Dz$ and \Lc results and with the EPPS16 and nCTEQ15 nPDFs~\cite{Kusina:2017gkz,Kusina:2020dki} within uncertainties.
The $R_\mathrm{FB}$ of $\Dp$ mesons as a function of $y^*$ is slightly larger than other charm hadrons and model predictions. 
At low $\pt$, the $R_\mathrm{FB}$ of $\Dp$ and $\Ds$ mesons are consistent with the $R_\mathrm{FB}$ of $\Dz$ mesons and \Lc baryons, and agree with nPDFs calculations. However, the more precise $\Dp$ data show a clear increasing trend with increasing $\pt$ and saturates at unity at high $\pt$ ($>10\gevc$) within uncertainties, which deviates from the almost $\pt$ independent nPDF calculations.
This discrepancy derives from the overall suppression of $D^+$ production as a function of \pt in the backward configuration, which is also pronounced in Fig.~\ref{fig:RpPbPT}.


\subsection{Production ratios}
\label{sec:pratios}
\begin{figure}[!tbp]
\centering
\begin{minipage}[t]{0.65\textwidth}
\centering
\includegraphics[width=1.0\textwidth]{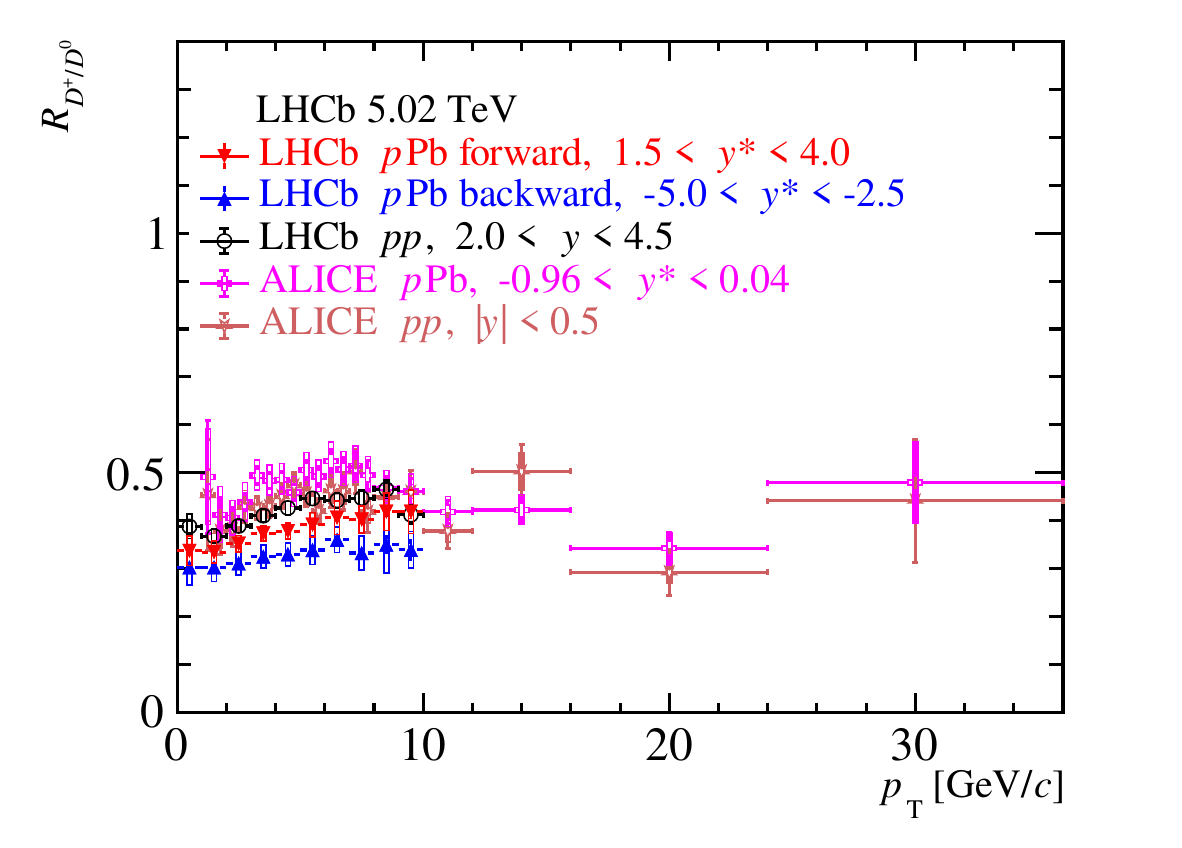}
\end{minipage}
\begin{minipage}[t]{0.65\textwidth}
\centering
\includegraphics[width=1.0\textwidth]{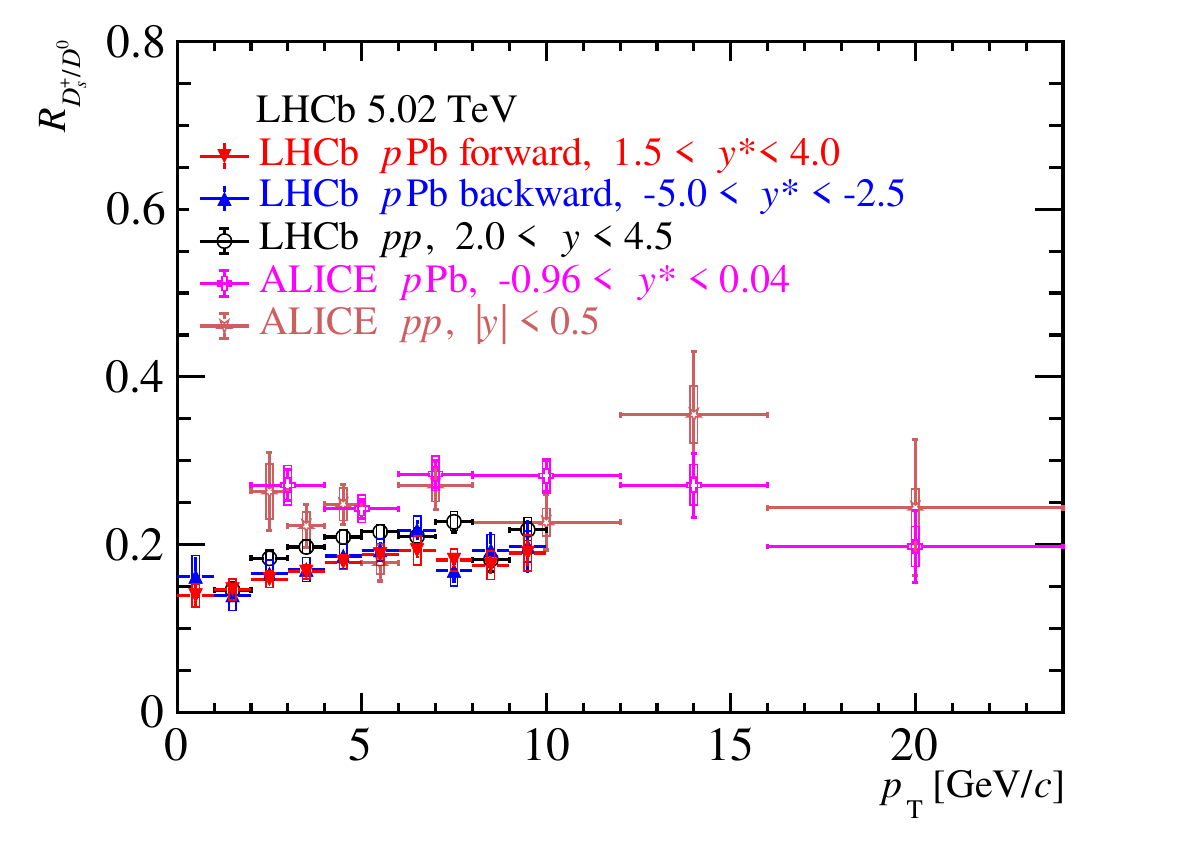}
\end{minipage}
\begin{minipage}[t]{0.65\textwidth}
\centering
\includegraphics[width=1.0\textwidth]{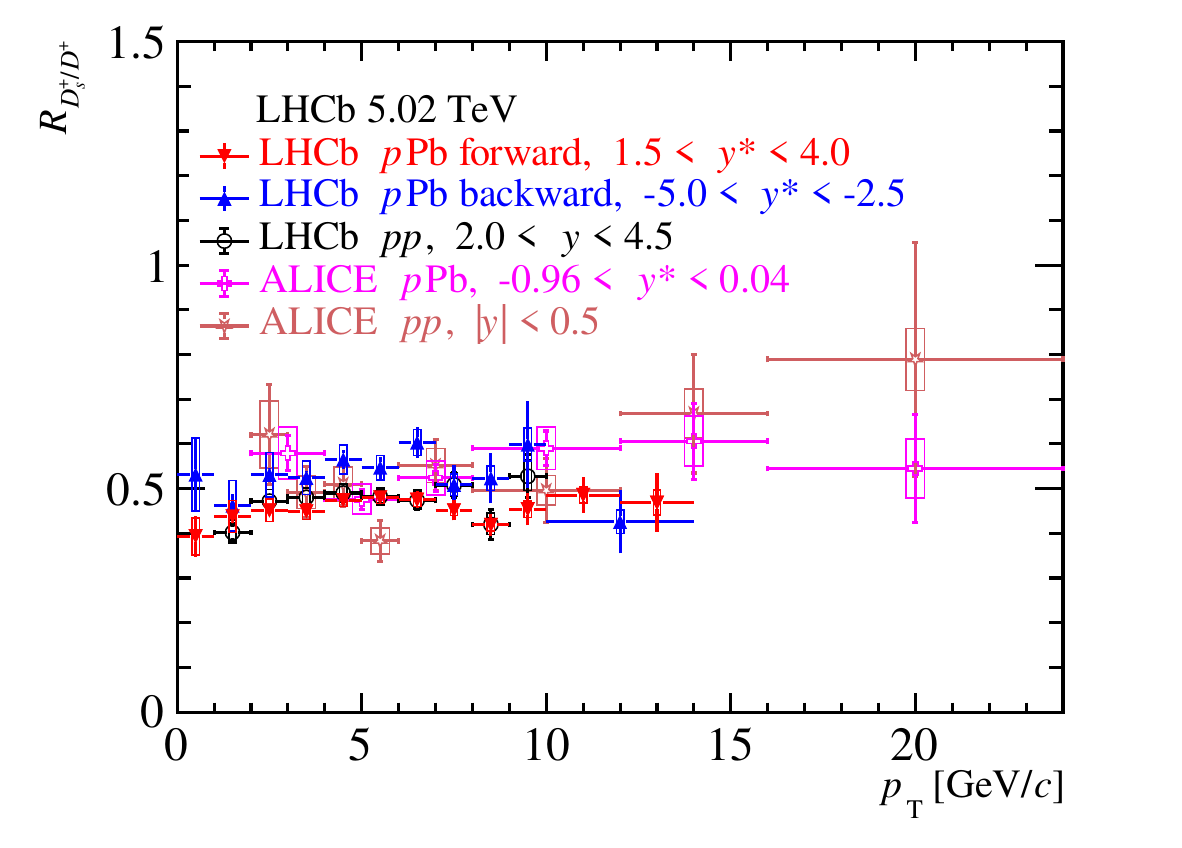}
\end{minipage}
\caption{Production ratios as a function of $\pt$ in \lhcb $\pPb$ collisions. The error bars show the statistical uncertainty while the boxes show the systematic uncertainty.
The uncertainties related to the branching fractions are not shown in the figure.
The measurements are also compared with other results of $pp$~\cite{LHCb:2016ikn,ALICE:2019nxm} and \pPb~\cite{Acharya:2019mno} collisions at the same centre-of-mass energy from \lhcb and ALICE. 
}
\label{fig:DpDzRatio}
\end{figure}

\begin{figure}[!tbp]
\centering
\begin{minipage}[t]{0.65\textwidth}
\centering
\includegraphics[width=1.0\textwidth]{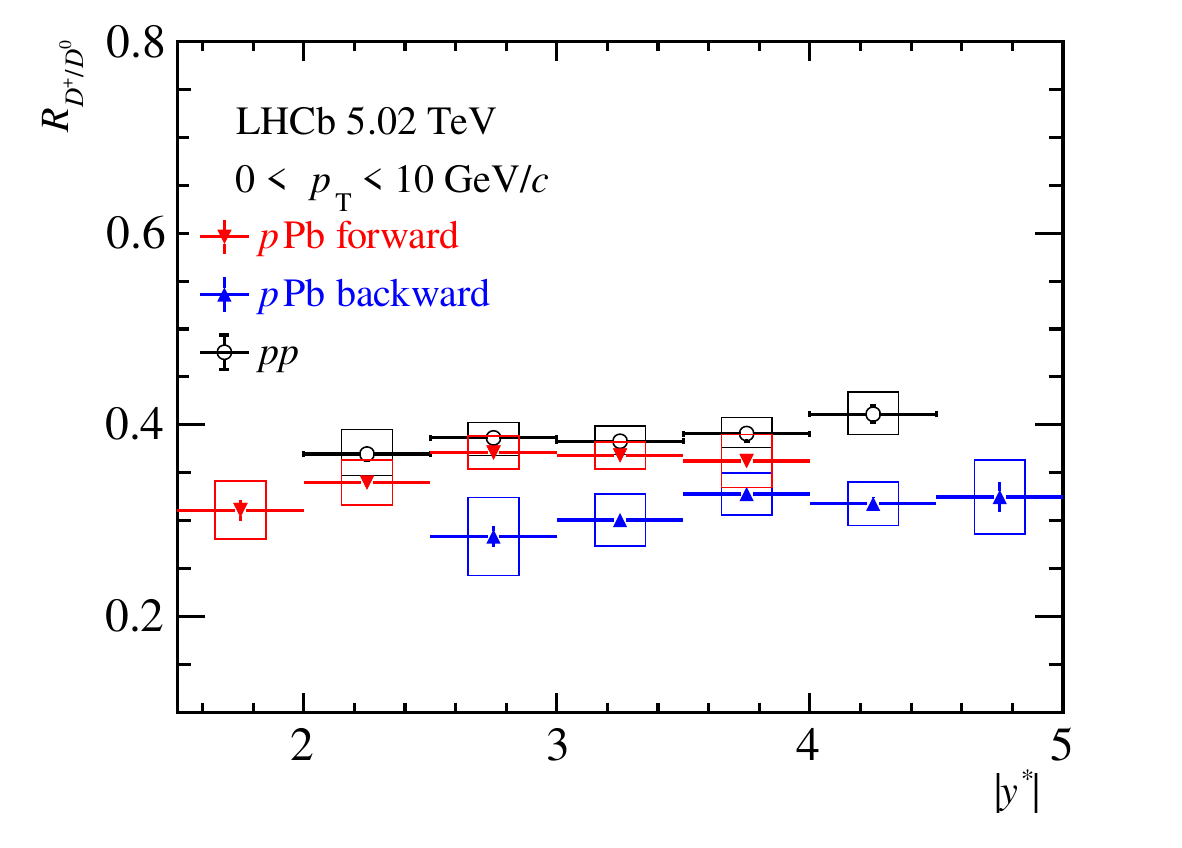}
\end{minipage}
\begin{minipage}[t]{0.65\textwidth}
\centering
\includegraphics[width=1.0\textwidth]{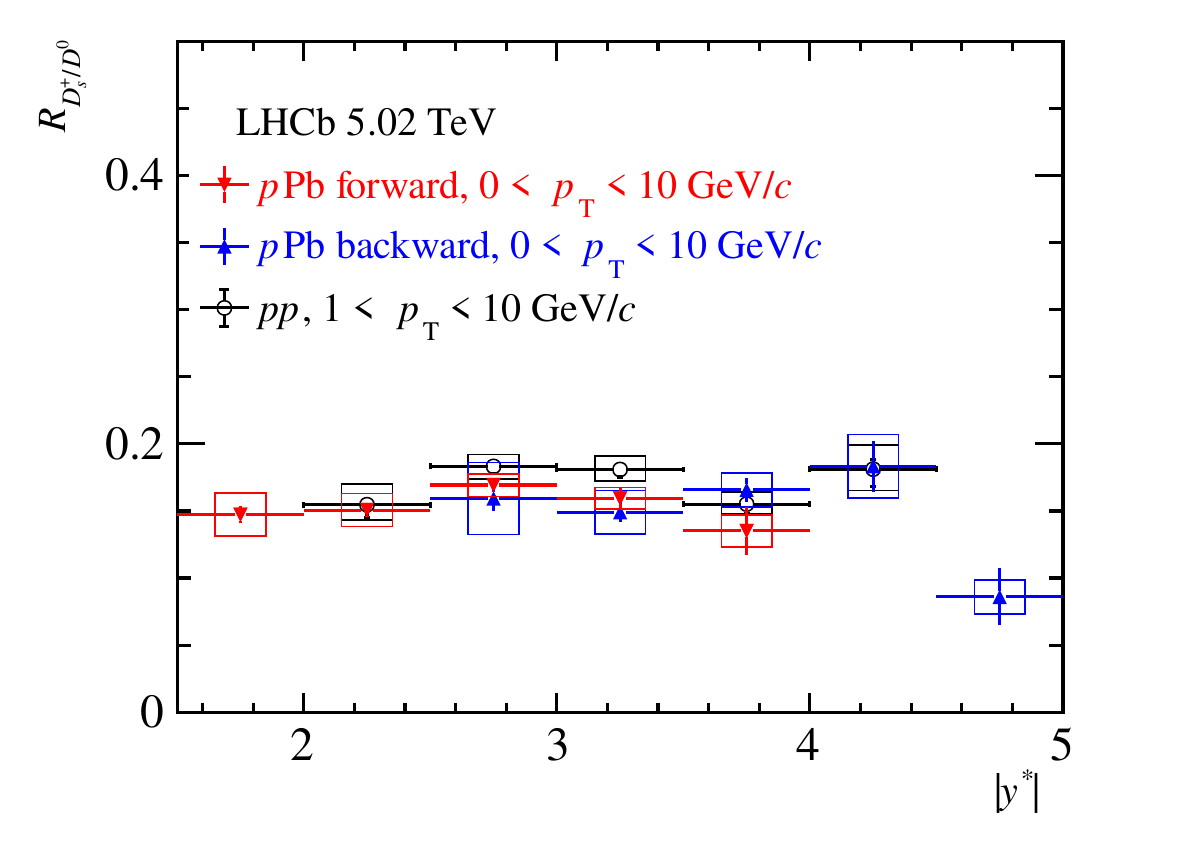}
\end{minipage}
\begin{minipage}[t]{0.65\textwidth}
\centering
\includegraphics[width=1.0\textwidth]{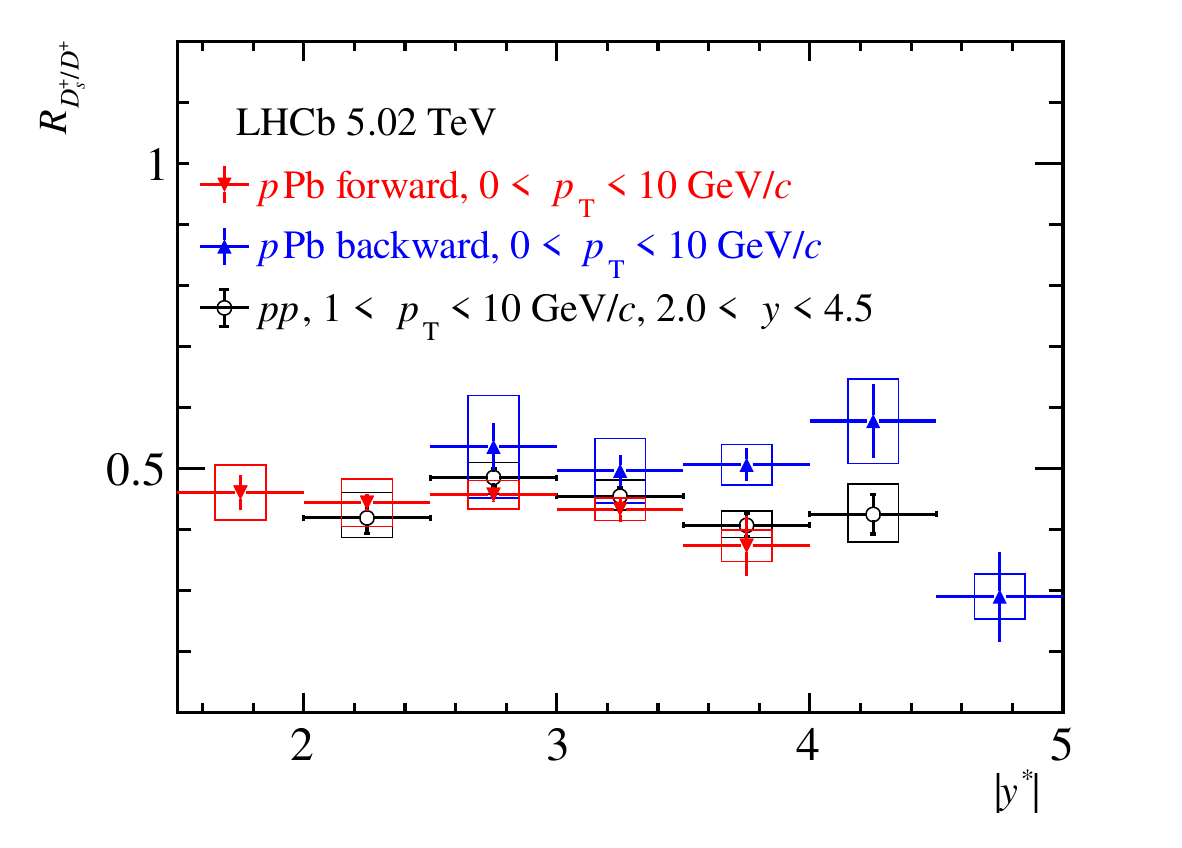}
\end{minipage}
\caption{Production ratios as a function of $y^*$ in \lhcb $\pPb$ collisions. The error bars show the statistical uncertainty while the boxes show the systematic uncertainty. The uncertainties related to the branching fractions are not shown in the figure. The measurements are also compared with the results of $pp$ collisions at the same centre-of-mass energy from \lhcb~\cite{LHCb:2016ikn}.
}
\label{fig:DpDzRatioy}
\end{figure}

The measured prompt \Ds and \Dp production cross-sections enable a calculation of $\Ds$ to $\Dp$ ratios. The LHCb measurement of $\Dz$ cross-sections in $\pPb$ collisions at $\snn=5.02$$\ensuremath{\mathrm{\,Te\kern -0.1em V}}$~\cite{LHCb:2017yua} also enables a calculation of the ratio of the production cross-sections of $\Dp$ to $\Dz$ and $\Ds$ to $\Dz$ mesons.
The uncertainty due to hadronic interactions with the detector is considered partially correlated because of different numbers of kaon and pion tracks between the numerators and denominators of $R_{\Dp/\Dz}$, $R_{\Ds/\Dz}$ or $R_{\Ds/\Dp}$. The uncertainties from the luminosity are  fully correlated, while the remaining uncertainties are uncorrelated. Figure~\ref{fig:DpDzRatio} illustrates the $R_{\Dp/\Dz}$, $R_{\Ds/\Dz}$ and $R_{\Ds/\Dp}$ ratios as a function of $\pt$ integrated over total rapidity range $1.5<y^*<4.0$ for forward and $-5.0<y^*<-2.5$ for backward. Figure~\ref{fig:DpDzRatioy} illustrates the $R_{\Dp/\Dz}$, $R_{\Ds/\Dz}$ and $R_{\Ds/\Dp}$ ratios as a function of $y^*$ integrated over the $\pt$ range up to $10\gevc$.
The measured production ratios are tabulated in appendix~\ref{sec:production_ratios}.

The results of relative cross-section ratios between $D$ mesons in \lhcb $\pPb$ collisions show a mild $\pt$ and $y^*$ dependence, which are consistent with the results from the \lhcb collaboration in $pp$~\cite{LHCb:2016ikn}, and from the ALICE collaboration in $\pPb$~\cite{Acharya:2019mno} and in $pp$~\cite{ALICE:2019nxm} collisions.
The $R_{\Ds/\Dz}$ and $R_{\Ds/\Dp}$ ratios are found not to be enhanced in neither forward nor backward rapidities in \pPb collisions at $\snn=5.02$$\ensuremath{\mathrm{\,Te\kern -0.1em V}}$. The lower $R_{\Dp/\Dz}$ ratios at backward rapidity relative to the forward region lead to differences in the results of $R_{\pPb}$ and $R_\mathrm{FB}$ for \Dp versus other $D$ mesons. On average, the multiplicity value at backward rapidity is 1.6 times higher than that at forward rapidity in terms of the backward-forward production ratio of charged particles at the same center-of-mass energy from LHCb~\cite{LHCb:2021vww}.
As some contributions of $\Dp$ and $\Dz$ mesons come from the decay of the excited charm resonance, the $D^{*+}$ meson~\cite{ParticleDataGroup:2022pth,ALICE:2021dhb}, it may be possible to further understand this phenomenon by investigating the production of $D^{*+}$ mesons in high multiplicity \pPb events.


\section{Conclusion}

The prompt $\Dp$ and $\Ds$ production cross-sections have been measured with \lhcb proton-lead collision data corresponding to an integrated luminosity of $(1.58\pm0.02)\invnb$ collected at $\snn=5.02\Tev$.
The measurement is performed in the
range of the $D$-meson transverse momentum $0<\pt<14\gevc$,
in both backward and forward collisions covering the 
rapidity ranges $1.5 < y^* < 4.0$ and $- 5.0 < y^* < - 2.5$.
This is the first measurement in this rapidity region down to zero transverse momentum
of the $\Dp$ and $\Ds$ mesons.
Nuclear modification factors and forward-backward production ratios are also measured in the same kinematic region.
A large asymmetry in the forward-backward production is observed, which is consistent with the expectations from nuclear parton distribution functions, and colour glass condensate calculations.
The $R_{\Ds/\Dz}$ and $R_{\Ds/\Dp}$ ratios show no enhancement  of the yield of strange hadrons at high transverse momentum or rapidity, and may therefore be used as a reference for experimental measurements related to investigations of the QGP in nucleus-nucleus collisions.

\clearpage
\cleardoublepage
\newpage
\section*{Acknowledgements}
%
%
\noindent We express our gratitude to our colleagues in the CERN
accelerator departments for the excellent performance of the LHC. We
thank the technical and administrative staff at the LHCb
institutes.
We acknowledge support from CERN and from the national agencies:
CAPES, CNPq, FAPERJ and FINEP (Brazil); 
MOST and NSFC (China); 
CNRS/IN2P3 (France); 
BMBF, DFG and MPG (Germany); 
INFN (Italy); 
NWO (Netherlands); 
MNiSW and NCN (Poland); 
MCID/IFA (Romania); 
MICINN (Spain); 
SNSF and SER (Switzerland); 
NASU (Ukraine); 
STFC (United Kingdom); 
DOE NP and NSF (USA).
We acknowledge the computing resources that are provided by CERN, IN2P3
(France), KIT and DESY (Germany), INFN (Italy), SURF (Netherlands),
PIC (Spain), GridPP (United Kingdom), 
CSCS (Switzerland), IFIN-HH (Romania), CBPF (Brazil),
Polish WLCG  (Poland) and NERSC (USA).
We are indebted to the communities behind the multiple open-source
software packages on which we depend.
Individual groups or members have received support from
ARC and ARDC (Australia);
Minciencias (Colombia);
AvH Foundation (Germany);
EPLANET, Marie Sk\l{}odowska-Curie Actions, ERC and NextGenerationEU (European Union);
A*MIDEX, ANR, IPhU and Labex P2IO, and R\'{e}gion Auvergne-Rh\^{o}ne-Alpes (France);
Key Research Program of Frontier Sciences of CAS, CAS PIFI, CAS CCEPP, 
Fundamental Research Funds for the Central Universities, 
and Sci. \& Tech. Program of Guangzhou (China);
GVA, XuntaGal, GENCAT, Inditex, InTalent and Prog.~Atracci\'on Talento, CM (Spain);
SRC (Sweden);
the Leverhulme Trust, the Royal Society
 and UKRI (United Kingdom).

\clearpage
\section*{Appendices}
\appendix
\section{Numerical values of the $\Dp$ and $\Ds$ mesons cross-sections}
\label{sec:DpDs2D}
Tables~\ref{tab:CrossSection2Dfor}--\ref{tab:CrossSection2DDsbac} give the numerical results for the double-differential cross-sections.
Tables~\ref{tab:CrossSection1DPTtot}--\ref{tab:CrossSection1DYDs} give the numerical results for the one-dimensional differential cross-sections.

\begin{table}[tbh]
\renewcommand\arraystretch{1.5}
\caption{Double-differential cross-section (mb) for prompt $\Dp$ mesons as functions of $\pt$ and $y^{*}$ in the forward regions. The first uncertainty is
statistical, the second is the component of the systematic uncertainty that is uncorrelated between bins and the
third is the fully correlated component.}
\centering
\scalebox{0.5}{
\begin{tabular}{cccccc}
\hline
Forward&&&&&\\
$\pt\,[\gevc]$ &$1.5<y^*<2.0$   &$2.0<y^*<2.5$    &$2.5<y^*<3.0$     &$3.0<y^*<3.5$     &$3.5<y^*<4.0$\\
\hline
$[0,1]$&$6.835\pm1.131\pm0.590\pm0.528$ &$7.069\pm0.207\pm0.104\pm0.766$    &$7.953\pm0.164\pm0.085\pm0.659$    &$7.527\pm0.185\pm0.091\pm0.393$    &$7.132\pm0.270\pm0.134\pm0.468$    \\
$[1,2]$&$11.748\pm0.481\pm0.282\pm0.848$    &$12.768\pm0.157\pm0.092\pm1.079$   &$12.198\pm0.122\pm0.069\pm0.806$   &$10.733\pm0.130\pm0.071\pm0.609$   &$8.238\pm0.187\pm0.091\pm0.487$    \\
$[2,3]$&$8.377\pm0.165\pm0.102\pm0.527$ &$8.348\pm0.067\pm0.041\pm0.671$    &$7.552\pm0.054\pm0.032\pm0.704$    &$6.096\pm0.054\pm0.034\pm0.523$    &$4.476\pm0.075\pm0.039\pm0.307$    \\
$[3,4]$&$4.383\pm0.069\pm0.044\pm0.274$ &$4.067\pm0.032\pm0.018\pm0.500$    &$3.678\pm0.027\pm0.016\pm0.539$    &$2.873\pm0.027\pm0.015\pm0.379$    &$2.040\pm0.037\pm0.019\pm0.178$    \\
$[4,5]$&$2.103\pm0.034\pm0.020\pm0.156$ &$2.015\pm0.018\pm0.010\pm0.338$    &$1.704\pm0.016\pm0.009\pm0.321$    &$1.238\pm0.015\pm0.008\pm0.216$    &$0.805\pm0.021\pm0.009\pm0.081$    \\
$[5,6]$&$1.108\pm0.020\pm0.012\pm0.099$ &$0.999\pm0.012\pm0.006\pm0.204$    &$0.808\pm0.010\pm0.005\pm0.183$    &$0.601\pm0.010\pm0.005\pm0.124$    &$0.348\pm0.015\pm0.006\pm0.033$    \\
$[6,7]$&$0.550\pm0.012\pm0.007\pm0.059$ &$0.494\pm0.008\pm0.004\pm0.115$    &$0.412\pm0.007\pm0.003\pm0.103$    &$0.294\pm0.007\pm0.003\pm0.064$    &$0.170\pm0.013\pm0.005\pm0.017$    \\
$[7,8]$&$0.318\pm0.009\pm0.005\pm0.038$ &$0.267\pm0.005\pm0.003\pm0.067$    &$0.231\pm0.005\pm0.002\pm0.061$    &$0.144\pm0.005\pm0.002\pm0.030$    &$0.073\pm0.015\pm0.005\pm0.009$    \\
$[8,9]$&$0.188\pm0.006\pm0.004\pm0.025$ &$0.154\pm0.004\pm0.002\pm0.041$    &$0.133\pm0.004\pm0.002\pm0.037$    &$0.084\pm0.004\pm0.002\pm0.016$    &$-$    \\
$[9,10]$&$0.111\pm0.005\pm0.002\pm0.017$    &$0.094\pm0.003\pm0.001\pm0.026$    &$0.072\pm0.003\pm0.001\pm0.021$    &$0.052\pm0.004\pm0.002\pm0.008$    &$-$    \\
$[10,11]$&$0.066\pm0.003\pm0.002\pm0.011$   &$0.056\pm0.002\pm0.001\pm0.016$    &$0.043\pm0.002\pm0.001\pm0.012$    &$0.028\pm0.004\pm0.002\pm0.004$    &$-$    \\
$[11,12]$&$0.043\pm0.003\pm0.001\pm0.007$   &$0.040\pm0.002\pm0.001\pm0.011$    &$0.027\pm0.002\pm0.001\pm0.007$    &$0.026\pm0.004\pm0.003\pm0.005$    &$-$    \\
$[12,13]$&$0.022\pm0.002\pm0.001\pm0.004$   &$0.022\pm0.002\pm0.001\pm0.007$    &$0.019\pm0.002\pm0.001\pm0.005$    &$-$    &$-$    \\
$[13,14]$&$0.023\pm0.002\pm0.001\pm0.004$   &$0.014\pm0.001\pm0.001\pm0.004$    &$0.012\pm0.001\pm0.001\pm0.003$    &$-$    &$-$    \\
\hline
\end{tabular}
}
\label{tab:CrossSection2Dfor}
\end{table}

\begin{table}[tbh]
\renewcommand\arraystretch{1.5}
\caption{Double-differential cross-section (mb) for prompt $\Dp$ mesons as functions of $\pt$ and $y^{*}$ in the backward regions. The first uncertainty is
statistical, the second is the component of the systematic uncertainty that is uncorrelated between bins and the
third is the fully correlated component.}
\centering
\scalebox{0.5}{
\begin{tabular}{cccccc}
\hline
Backward&&&&&\\
$\pt\,[\gevc]$ &$-3.0<y^*<-2.5$   &$-3.5<y^*<-3.0$    &$-4.0<y^*<-3.5$     &$-4.5<y^*<-4.0$     &$-5.0<y^*<-4.5$\\
\hline
$[0,1]$&$6.430\pm1.127\pm0.371\pm0.647$ &$7.921\pm0.340\pm0.115\pm0.835$    &$9.220\pm0.321\pm0.106\pm0.607$    &$8.069\pm0.438\pm0.115\pm0.571$    &$8.054\pm0.750\pm0.195\pm0.715$    \\
$[1,2]$&$13.372\pm0.641\pm0.321\pm0.966$    &$13.751\pm0.252\pm0.099\pm1.162$   &$12.931\pm0.221\pm0.073\pm0.855$   &$11.049\pm0.283\pm0.073\pm0.627$   &$7.973\pm0.497\pm0.088\pm0.471$    \\
$[2,3]$&$8.325\pm0.210\pm0.101\pm0.524$ &$8.064\pm0.102\pm0.040\pm0.649$    &$7.087\pm0.087\pm0.030\pm0.661$    &$5.434\pm0.092\pm0.030\pm0.467$    &$3.603\pm0.138\pm0.032\pm0.247$    \\
$[3,4]$&$4.058\pm0.086\pm0.040\pm0.254$ &$3.616\pm0.046\pm0.016\pm0.445$    &$2.965\pm0.039\pm0.013\pm0.435$    &$2.017\pm0.038\pm0.011\pm0.266$    &$1.081\pm0.052\pm0.010\pm0.094$    \\
$[4,5]$&$1.824\pm0.042\pm0.018\pm0.135$ &$1.533\pm0.024\pm0.008\pm0.257$    &$1.194\pm0.021\pm0.006\pm0.225$    &$0.775\pm0.020\pm0.005\pm0.135$    &$0.338\pm0.026\pm0.004\pm0.034$    \\
$[5,6]$&$0.857\pm0.024\pm0.009\pm0.077$ &$0.702\pm0.015\pm0.004\pm0.144$    &$0.492\pm0.012\pm0.003\pm0.111$    &$0.294\pm0.011\pm0.002\pm0.060$    &$0.094\pm0.017\pm0.002\pm0.009$    \\
$[6,7]$&$0.433\pm0.015\pm0.006\pm0.046$ &$0.340\pm0.009\pm0.003\pm0.079$    &$0.227\pm0.008\pm0.002\pm0.057$    &$0.128\pm0.008\pm0.001\pm0.028$    &$-$    \\
$[7,8]$&$0.223\pm0.009\pm0.003\pm0.026$ &$0.159\pm0.006\pm0.002\pm0.040$    &$0.106\pm0.005\pm0.001\pm0.028$    &$0.053\pm0.005\pm0.001\pm0.011$    &$-$    \\
$[8,9]$&$0.120\pm0.007\pm0.002\pm0.016$ &$0.087\pm0.004\pm0.001\pm0.023$    &$0.055\pm0.004\pm0.001\pm0.015$    &$0.024\pm0.004\pm0.001\pm0.005$    &$-$    \\
$[9,10]$&$0.065\pm0.005\pm0.001\pm0.010$    &$0.051\pm0.003\pm0.001\pm0.014$    &$0.030\pm0.003\pm0.001\pm0.009$    &$0.014\pm0.004\pm0.001\pm0.002$    &$-$    \\
$[10,11]$&$0.042\pm0.004\pm0.001\pm0.007$   &$0.030\pm0.003\pm0.001\pm0.009$    &$0.014\pm0.002\pm0.000\pm0.004$    &$-$    &$-$    \\
$[11,12]$&$0.029\pm0.003\pm0.001\pm0.005$   &$0.014\pm0.002\pm0.000\pm0.004$    &$0.008\pm0.001\pm0.000\pm0.002$    &$-$    &$-$    \\
$[12,13]$&$0.021\pm0.002\pm0.001\pm0.004$   &$0.008\pm0.001\pm0.000\pm0.002$    &$-$    &$-$    &$-$    \\
$[13,14]$&$0.013\pm0.002\pm0.001\pm0.002$   &$0.006\pm0.001\pm0.000\pm0.002$    &$-$    &$-$    &$-$    \\
\hline
\end{tabular}
}
\label{tab:CrossSection2Dbac}
\end{table}

\begin{table}[tbh]
\renewcommand\arraystretch{1.5}
\caption{Double-differential cross-section (mb) for prompt $\Ds$ mesons as functions of $\pt$ and $y^{*}$ in the forward regions. The first uncertainty is
statistical, the second is the component of the systematic uncertainty that is uncorrelated between bins and the
third is the fully correlated component.}
\centering
\scalebox{0.5}{
\begin{tabular}{cccccc}
\hline
Forward&&&&&\\
$\pt\,[\gevc]$ &$1.5<y^*<2.0$   &$2.0<y^*<2.5$    &$2.5<y^*<3.0$     &$3.0<y^*<3.5$     &$3.5<y^*<4.0$\\
\hline
$[0,1]$&$-$ &$3.213\pm0.403\pm0.123\pm0.441$    &$2.873\pm0.346\pm0.097\pm0.317$    &$3.454\pm0.538\pm0.147\pm0.303$    &$2.134\pm1.045\pm0.175\pm0.205$    \\
$[1,2]$&$4.885\pm0.521\pm0.170\pm0.621$ &$5.250\pm0.227\pm0.073\pm0.602$    &$6.079\pm0.219\pm0.077\pm0.505$    &$4.464\pm0.235\pm0.072\pm0.356$    &$3.699\pm0.415\pm0.110\pm0.316$    \\
$[2,3]$&$4.059\pm0.221\pm0.078\pm0.487$ &$3.738\pm0.104\pm0.034\pm0.356$    &$3.640\pm0.094\pm0.034\pm0.318$    &$2.659\pm0.098\pm0.033\pm0.217$    &$1.630\pm0.149\pm0.038\pm0.133$    \\
$[3,4]$&$2.090\pm0.093\pm0.031\pm0.211$ &$1.967\pm0.052\pm0.018\pm0.218$    &$1.717\pm0.046\pm0.017\pm0.199$    &$1.248\pm0.046\pm0.016\pm0.130$    &$0.623\pm0.065\pm0.016\pm0.054$    \\
$[4,5]$&$1.096\pm0.051\pm0.017\pm0.101$ &$0.964\pm0.030\pm0.010\pm0.133$    &$0.817\pm0.027\pm0.009\pm0.118$    &$0.534\pm0.026\pm0.008\pm0.069$    &$0.315\pm0.035\pm0.008\pm0.029$    \\
$[5,6]$&$0.592\pm0.032\pm0.011\pm0.058$ &$0.509\pm0.020\pm0.006\pm0.081$    &$0.363\pm0.017\pm0.006\pm0.059$    &$0.226\pm0.016\pm0.004\pm0.034$    &$0.166\pm0.029\pm0.006\pm0.016$    \\
$[6,7]$&$0.275\pm0.019\pm0.007\pm0.029$ &$0.227\pm0.012\pm0.004\pm0.040$    &$0.196\pm0.012\pm0.004\pm0.035$    &$0.124\pm0.011\pm0.003\pm0.020$    &$0.090\pm0.022\pm0.005\pm0.009$    \\
$[7,8]$&$0.154\pm0.013\pm0.004\pm0.018$ &$0.108\pm0.008\pm0.002\pm0.020$    &$0.099\pm0.008\pm0.002\pm0.019$    &$0.073\pm0.009\pm0.002\pm0.012$    &$-$    \\
$[8,9]$&$0.086\pm0.009\pm0.003\pm0.011$ &$0.067\pm0.006\pm0.002\pm0.014$    &$0.050\pm0.006\pm0.001\pm0.010$    &$0.031\pm0.006\pm0.001\pm0.005$    &$-$    \\
$[9,10]$&$0.052\pm0.007\pm0.002\pm0.007$    &$0.044\pm0.005\pm0.001\pm0.009$    &$0.027\pm0.004\pm0.001\pm0.006$    &$0.026\pm0.006\pm0.002\pm0.004$    &$-$    \\
$[10,12]$&$0.027\pm0.004\pm0.001\pm0.004$   &$0.020\pm0.002\pm0.001\pm0.004$    &$0.020\pm0.002\pm0.001\pm0.004$    &$-$    &$-$    \\
$[12,14]$&$0.012\pm0.002\pm0.001\pm0.002$   &$0.007\pm0.002\pm0.000\pm0.001$    &$0.008\pm0.002\pm0.000\pm0.002$    &$-$    &$-$    \\
\hline
\end{tabular}
}
\label{tab:CrossSection2DDsfor}
\end{table}

\begin{table}[tbh]
\renewcommand\arraystretch{1.5}
\caption{Double-differential cross-section (mb) for prompt $\Ds$ mesons as functions of $\pt$ and $y^{*}$ in the backward regions. The first uncertainty is
statistical, the second is the component of the systematic uncertainty that is uncorrelated between bins and the
third is the fully correlated component.}
\centering
\scalebox{0.5}{
\begin{tabular}{cccccc}
\hline
Backward&&&&&\\
$\pt\,[\gevc]$ &$-3.0<y^*<-2.5$   &$-3.5<y^*<-3.0$    &$-4.0<y^*<-3.5$     &$-4.5<y^*<-4.0$     &$-5.0<y^*<-4.5$\\
\hline
$[0,1]$&$-$ &$3.749\pm0.796\pm0.164\pm0.556$    &$3.461\pm0.780\pm0.127\pm0.351$    &$6.190\pm1.561\pm0.314\pm1.169$    &$-$    \\
$[1,2]$&$6.929\pm0.818\pm0.212\pm1.243$ &$6.020\pm0.400\pm0.086\pm0.777$    &$7.245\pm0.440\pm0.103\pm0.655$    &$5.443\pm0.549\pm0.107\pm0.678$    &$1.654\pm0.877\pm0.129\pm0.202$    \\
$[2,3]$&$4.441\pm0.305\pm0.075\pm0.723$ &$4.488\pm0.181\pm0.044\pm0.469$    &$3.931\pm0.157\pm0.039\pm0.369$    &$2.883\pm0.182\pm0.044\pm0.268$    &$1.519\pm0.265\pm0.050\pm0.205$    \\
$[3,4]$&$2.270\pm0.137\pm0.032\pm0.293$ &$2.062\pm0.082\pm0.019\pm0.235$    &$1.559\pm0.068\pm0.016\pm0.182$    &$0.955\pm0.065\pm0.014\pm0.101$    &$0.358\pm0.093\pm0.014\pm0.043$    \\
$[4,5]$&$1.077\pm0.070\pm0.017\pm0.116$ &$0.934\pm0.045\pm0.011\pm0.127$    &$0.622\pm0.036\pm0.008\pm0.087$    &$0.337\pm0.034\pm0.007\pm0.042$    &$0.232\pm0.053\pm0.011\pm0.027$    \\
$[5,6]$&$0.450\pm0.037\pm0.008\pm0.051$ &$0.377\pm0.025\pm0.005\pm0.059$    &$0.290\pm0.022\pm0.005\pm0.046$    &$0.166\pm0.020\pm0.004\pm0.024$    &$-$    \\
$[6,7]$&$0.265\pm0.026\pm0.006\pm0.032$ &$0.208\pm0.017\pm0.004\pm0.036$    &$0.154\pm0.015\pm0.004\pm0.025$    &$0.053\pm0.011\pm0.002\pm0.008$    &$-$    \\
$[7,8]$&$0.125\pm0.015\pm0.003\pm0.017$ &$0.059\pm0.008\pm0.001\pm0.011$    &$0.052\pm0.009\pm0.002\pm0.009$    &$0.038\pm0.011\pm0.002\pm0.006$    &$-$    \\
$[8,9]$&$0.074\pm0.010\pm0.003\pm0.010$ &$0.042\pm0.008\pm0.001\pm0.008$    &$0.022\pm0.006\pm0.001\pm0.004$    &$-$    &$-$    \\
$[9,10]$&$0.043\pm0.009\pm0.002\pm0.007$    &$0.027\pm0.005\pm0.001\pm0.005$    &$-$    &$-$    &$-$    \\
$[10,14]$&$0.010\pm0.002\pm0.000\pm0.002$   &$0.007\pm0.001\pm0.000\pm0.002$    &$-$    &$-$    &$-$    \\
\hline
\end{tabular}
}
\label{tab:CrossSection2DDsbac}
\end{table}

\begin{table}[btp]
\renewcommand\arraystretch{1.5}
\caption{Differential cross-section for prompt $\Dp$ mesons as a function of $\pt$ in the total forward and backward rapidity regions, respectively. The first uncertainty is
statistical, the second is the component of the systematic uncertainty that is uncorrelated between bins and the
third is the fully correlated component. 
}
\centering
\scalebox{0.88}{
\begin{tabular}{ccc}
\hline
Forward\\
$\pt\,[\gevc]$&$y^*$&$\frac{\deriv \sigma}{\deriv \pt}$ [\mbarn/(\gevc)]\\
\hline
$[0,1]$&$[1.5,4.0]$&$18.258\pm0.603\pm0.313\pm1.180$\\
$[1,2]$&$[1.5,4.0]$&$27.842\pm0.284\pm0.163\pm1.510$\\
$[2,3]$&$[1.5,4.0]$&$17.424\pm0.104\pm0.063\pm1.103$\\
$[3,4]$&$[1.5,4.0]$&$8.521\pm0.047\pm0.028\pm0.840$\\
$[4,5]$&$[1.5,4.0]$&$3.933\pm0.025\pm0.014\pm0.526$\\
$[5,6]$&$[1.5,4.0]$&$1.932\pm0.015\pm0.008\pm0.310$\\
$[6,7]$&$[1.5,4.0]$&$0.960\pm0.011\pm0.005\pm0.171$\\
$[7,8]$&$[1.5,4.0]$&$0.517\pm0.010\pm0.004\pm0.097$\\
$[8,9]$&$[1.5,3.5]$&$0.279\pm0.005\pm0.002\pm0.059$\\
$[9,10]$&$[1.5,3.5]$&$0.164\pm0.004\pm0.002\pm0.034$\\
$[10,11]$&$[1.5,3.5]$&$0.097\pm0.003\pm0.001\pm0.021$\\
$[11,12]$&$[1.5,3.5]$&$0.068\pm0.003\pm0.002\pm0.014$\\
$[12,13]$&$[1.5,3.0]$&$0.031\pm0.001\pm0.001\pm0.008$\\
$[13,14]$&$[1.5,3.0]$&$0.024\pm0.001\pm0.001\pm0.005$\\
\hline
Backward\\
$\pt\,[\gevc]$&$y^*$&$\frac{\deriv \sigma}{\deriv \pt}$ [\mbarn/(\gevc)]\\
\hline
$[0,1]$&$[-5.0,-2.5]$&$19.847\pm0.749\pm0.231\pm1.276$\\
$[1,2]$&$[-5.0,-2.5]$&$29.538\pm0.461\pm0.169\pm1.705$\\
$[2,3]$&$[-5.0,-2.5]$&$16.257\pm0.150\pm0.058\pm1.070$\\
$[3,4]$&$[-5.0,-2.5]$&$6.869\pm0.062\pm0.024\pm0.663$\\
$[4,5]$&$[-5.0,-2.5]$&$2.832\pm0.031\pm0.011\pm0.365$\\
$[5,6]$&$[-5.0,-2.5]$&$1.219\pm0.018\pm0.006\pm0.187$\\
$[6,7]$&$[-4.5,-2.5]$&$0.564\pm0.010\pm0.004\pm0.100$\\
$[7,8]$&$[-4.5,-2.5]$&$0.270\pm0.007\pm0.002\pm0.052$\\
$[8,9]$&$[-4.5,-2.5]$&$0.143\pm0.005\pm0.002\pm0.029$\\
$[9,10]$&$[-4.5,-2.5]$&$0.080\pm0.004\pm0.001\pm0.017$\\
$[10,11]$&$[-4.0,-2.5]$&$0.043\pm0.002\pm0.001\pm0.010$\\
$[11,12]$&$[-4.0,-2.5]$&$0.026\pm0.002\pm0.001\pm0.006$\\
$[12,13]$&$[-3.5,-2.5]$&$0.015\pm0.001\pm0.000\pm0.003$\\
$[13,14]$&$[-3.5,-2.5]$&$0.009\pm0.001\pm0.000\pm0.002$\\
\hline
\end{tabular}
}
\label{tab:CrossSection1DPTtot}
\end{table}

\begin{table}[btp]
\renewcommand\arraystretch{1.5}
\caption{Differential cross-section for prompt $\Ds$ mesons as a function of $\pt$ in the total forward and backward rapidity regions, respectively. The first uncertainty is
statistical, the second is the component of the systematic uncertainty that is uncorrelated between bins and the third is the fully correlated component. 
}
\centering
\scalebox{1.0}{
\begin{tabular}{ccc}
\hline
Forward\\
$\pt\,[\gevc]$&$y^*$&$\frac{\deriv \sigma}{\deriv \pt}$ [\mbarn/(\gevc)]\\
\hline
$[0,1]$&$[2.0,4.0]$&$5.837\pm0.645\pm0.139\pm0.591$\\
$[1,2]$&$[1.5,4.0]$&$12.189\pm0.387\pm0.120\pm1.087$\\
$[2,3]$&$[1.5,4.0]$&$7.864\pm0.158\pm0.052\pm0.649$\\
$[3,4]$&$[1.5,4.0]$&$3.823\pm0.070\pm0.023\pm0.360$\\
$[4,5]$&$[1.5,4.0]$&$1.863\pm0.039\pm0.012\pm0.209$\\
$[5,6]$&$[1.5,4.0]$&$0.928\pm0.026\pm0.008\pm0.117$\\
$[6,7]$&$[1.5,4.0]$&$0.456\pm0.018\pm0.005\pm0.063$\\
$[7,8]$&$[1.5,3.5]$&$0.217\pm0.010\pm0.003\pm0.033$\\
$[8,9]$&$[1.5,3.5]$&$0.117\pm0.007\pm0.002\pm0.019$\\
$[9,10]$&$[1.5,3.5]$&$0.074\pm0.006\pm0.002\pm0.012$\\
$[10,12]$&$[1.5,3.0]$&$0.033\pm0.002\pm0.001\pm0.006$\\
$[12,14]$&$[1.5,3.0]$&$0.013\pm0.002\pm0.000\pm0.002$\\
\hline
Backward\\
$\pt\,[\gevc]$&$y^*$&$\frac{\deriv \sigma}{\deriv \pt}$ [\mbarn/(\gevc)]\\
\hline
$[0,1]$&$[-4.5,-3.0]$&$6.700\pm0.959\pm0.188\pm0.966$\\
$[1,2]$&$[-5.0,-2.5]$&$13.646\pm0.723\pm0.151\pm1.587$\\
$[2,3]$&$[-5.0,-2.5]$&$8.631\pm0.252\pm0.058\pm0.850$\\
$[3,4]$&$[-5.0,-2.5]$&$3.602\pm0.104\pm0.023\pm0.355$\\
$[4,5]$&$[-5.0,-2.5]$&$1.601\pm0.055\pm0.013\pm0.175$\\
$[5,6]$&$[-4.5,-2.5]$&$0.641\pm0.027\pm0.006\pm0.083$\\
$[6,7]$&$[-4.5,-2.5]$&$0.340\pm0.018\pm0.004\pm0.047$\\
$[7,8]$&$[-4.5,-2.5]$&$0.137\pm0.011\pm0.002\pm0.020$\\
$[8,9]$&$[-4.0,-2.5]$&$0.069\pm0.007\pm0.002\pm0.010$\\
$[9,10]$&$[-3.5,-2.5]$&$0.035\pm0.005\pm0.001\pm0.006$\\
$[10,14]$&$[-3.5,-2.5]$&$0.009\pm0.001\pm0.000\pm0.001$\\
\hline
\end{tabular}
}
\label{tab:CrossSection1DPTtotDs}
\end{table}

\begin{table}[btp]
\renewcommand\arraystretch{1.5}
\caption{Differential cross-section for prompt $\Dp$ mesons as a function of $\pt$ in the common forward and backward regions, respectively. The first uncertainty is
statistical, the second is the component of the systematic uncertainty that is uncorrelated between bins and the
third is the fully correlated component. 
}
\centering
\scalebox{0.88}{
\begin{tabular}{ccc}
\hline
Forward\\
$\pt\,[\gevc]$&$y^*$&$\frac{\deriv \sigma}{\deriv \pt}$ [\mbarn/(\gevc)]\\
\hline
$[0,1]$&$[2.5,4.0]$&$11.306\pm0.183\pm0.091\pm0.604$\\
$[1,2]$&$[2.5,4.0]$&$15.584\pm0.129\pm0.067\pm0.725$\\
$[2,3]$&$[2.5,4.0]$&$9.062\pm0.054\pm0.030\pm0.669$\\
$[3,4]$&$[2.5,4.0]$&$4.295\pm0.026\pm0.014\pm0.518$\\
$[4,5]$&$[2.5,4.0]$&$1.874\pm0.015\pm0.007\pm0.300$\\
$[5,6]$&$[2.5,4.0]$&$0.879\pm0.010\pm0.005\pm0.166$\\
$[6,7]$&$[2.5,4.0]$&$0.438\pm0.008\pm0.003\pm0.087$\\
$[7,8]$&$[2.5,4.0]$&$0.224\pm0.008\pm0.003\pm0.046$\\
$[8,9]$&$[2.5,3.5]$&$0.108\pm0.003\pm0.001\pm0.026$\\
$[9,10]$&$[2.5,3.5]$&$0.062\pm0.002\pm0.001\pm0.014$\\
$[10,11]$&$[2.5,3.5]$&$0.036\pm0.002\pm0.001\pm0.008$\\
$[11,12]$&$[2.5,3.5]$&$0.027\pm0.002\pm0.001\pm0.005$\\
$[12,13]$&$[2.5,3.0]$&$0.009\pm0.001\pm0.000\pm0.002$\\
$[13,14]$&$[2.5,3.0]$&$0.006\pm0.001\pm0.000\pm0.001$\\
\hline
Backward\\
$\pt\,[\gevc]$&$y^*$&$\frac{\deriv \sigma}{\deriv \pt}$ [\mbarn/(\gevc)]\\
\hline
$[0,1]$&$[-4.0,-3.0]$&$11.786\pm0.610\pm0.201\pm0.800$\\
$[1,2]$&$[-4.0,-2.5]$&$20.026\pm0.362\pm0.146\pm1.242$\\
$[2,3]$&$[-4.0,-2.5]$&$11.738\pm0.125\pm0.051\pm0.755$\\
$[3,4]$&$[-4.0,-2.5]$&$5.320\pm0.053\pm0.021\pm0.505$\\
$[4,5]$&$[-4.0,-2.5]$&$2.275\pm0.026\pm0.010\pm0.292$\\
$[5,6]$&$[-4.0,-2.5]$&$1.026\pm0.015\pm0.005\pm0.157$\\
$[6,7]$&$[-4.0,-2.5]$&$0.500\pm0.010\pm0.003\pm0.088$\\
$[7,8]$&$[-4.0,-2.5]$&$0.244\pm0.006\pm0.002\pm0.047$\\
$[8,9]$&$[-4.0,-2.5]$&$0.131\pm0.004\pm0.001\pm0.027$\\
$[9,10]$&$[-4.0,-2.5]$&$0.073\pm0.003\pm0.001\pm0.016$\\
$[10,11]$&$[-4.0,-2.5]$&$0.043\pm0.002\pm0.001\pm0.010$\\
$[11,12]$&$[-4.0,-2.5]$&$0.026\pm0.002\pm0.001\pm0.006$\\
$[12,13]$&$[-3.5,-2.5]$&$0.015\pm0.001\pm0.000\pm0.003$\\
$[13,14]$&$[-3.5,-2.5]$&$0.009\pm0.001\pm0.000\pm0.002$\\
\hline
\end{tabular}
}
\label{tab:CrossSection1DPTcom}
\end{table}

\begin{table}[btp]
\renewcommand\arraystretch{1.5}
\caption{Differential cross-section for prompt $\Ds$ mesons as a function of $\pt$ in the common forward and backward regions, respectively. The first uncertainty is
statistical, the second is the component of the systematic uncertainty that is uncorrelated between bins and the third is the fully correlated component. 
}
\centering
\scalebox{1.0}{
\begin{tabular}{ccc}
\hline
Forward\\
$\pt\,[\gevc]$&$y^*$&$\frac{\deriv \sigma}{\deriv \pt}$ [\mbarn/(\gevc)]\\
\hline
$[0,1]$&$[2.5,4.0]$&$4.230\pm0.613\pm0.124\pm0.382$\\
$[1,2]$&$[2.5,4.0]$&$7.121\pm0.262\pm0.076\pm0.530$\\
$[2,3]$&$[2.5,4.0]$&$3.965\pm0.101\pm0.030\pm0.296$\\
$[3,4]$&$[2.5,4.0]$&$1.794\pm0.046\pm0.014\pm0.177$\\
$[4,5]$&$[2.5,4.0]$&$0.833\pm0.026\pm0.007\pm0.103$\\
$[5,6]$&$[2.5,4.0]$&$0.377\pm0.018\pm0.005\pm0.052$\\
$[6,7]$&$[2.5,4.0]$&$0.205\pm0.014\pm0.003\pm0.030$\\
$[7,8]$&$[2.5,3.5]$&$0.086\pm0.006\pm0.002\pm0.015$\\
$[8,9]$&$[2.5,3.5]$&$0.041\pm0.004\pm0.001\pm0.007$\\
$[9,10]$&$[2.5,3.5]$&$0.026\pm0.004\pm0.001\pm0.005$\\
$[10,12]$&$[2.5,3.0]$&$0.010\pm0.001\pm0.000\pm0.002$\\
$[12,14]$&$[2.5,3.0]$&$0.004\pm0.001\pm0.000\pm0.001$\\
\hline
Backward\\
$\pt\,[\gevc]$&$y^*$&$\frac{\deriv \sigma}{\deriv \pt}$ [\mbarn/(\gevc)]\\
\hline
$[0,1]$&$[-4.0,-3.0]$&$3.605\pm0.558\pm0.104\pm0.409$\\
$[1,2]$&$[-4.0,-2.5]$&$10.097\pm0.506\pm0.125\pm1.186$\\
$[2,3]$&$[-4.0,-2.5]$&$6.430\pm0.194\pm0.048\pm0.652$\\
$[3,4]$&$[-4.0,-2.5]$&$2.945\pm0.087\pm0.020\pm0.294$\\
$[4,5]$&$[-4.0,-2.5]$&$1.316\pm0.045\pm0.011\pm0.146$\\
$[5,6]$&$[-4.0,-2.5]$&$0.558\pm0.025\pm0.005\pm0.072$\\
$[6,7]$&$[-4.0,-2.5]$&$0.314\pm0.017\pm0.004\pm0.043$\\
$[7,8]$&$[-4.0,-2.5]$&$0.118\pm0.010\pm0.002\pm0.017$\\
$[8,9]$&$[-4.0,-2.5]$&$0.069\pm0.007\pm0.002\pm0.010$\\
$[9,10]$&$[-3.5,-2.5]$&$0.035\pm0.005\pm0.001\pm0.006$\\
$[10,14]$&$[-3.5,-2.5]$&$0.009\pm0.001\pm0.000\pm0.001$\\
\hline
\end{tabular}
}
\label{tab:CrossSection1DPTcomDs}
\end{table}

\begin{table}[btp]
\renewcommand\arraystretch{1.5}
\caption{Differential cross-section for prompt $\Dp$ mesons as a function of $|y^{*}|$ integrated over $1<\pt<14\gevc$ for the forward and backward regions, respectively. The first uncertainty is statistical, the second is systematic.
}
\centering
\begin{tabular}{cc}
\hline
Forward &\\
$y^*$&$\frac{\deriv \sigma}{\deriv y^*}$ [\mbarn]\\
\hline
$[1.5,2.0]$&$35.87\pm1.24\pm2.08$\\
$[2.0,2.5]$&$36.41\pm0.27\pm2.95$\\
$[2.5,3.0]$&$34.84\pm0.21\pm2.73$\\
$[3.0,3.5]$&$29.70\pm0.23\pm1.86$\\
$[3.5,4.0]$&$23.28\pm0.34\pm1.21$\\
\hline
Backward &\\
$y^*$&$\frac{\deriv \sigma}{\deriv y^*}$ [\mbarn]\\
\hline
$[-3.0,-2.5]$&$35.81\pm1.32\pm2.49$\\
$[-3.5,-3.0]$&$36.28\pm0.44\pm2.49$\\
$[-4.0,-3.5]$&$34.33\pm0.40\pm2.14$\\
$[-4.5,-4.0]$&$27.86\pm0.53\pm1.65$\\
$[-5.0,-4.5]$&$21.14\pm0.91\pm1.67$\\
\hline
\end{tabular}
\label{tab:CrossSection1DY}
\end{table}

\begin{table}[btp]
\renewcommand\arraystretch{1.5}
\caption{Differential cross-section for prompt $\Ds$ mesons as a function of $|y^{*}|$ integrated over $1<\pt<14\gevc$ for the forward and backward regions, respectively. The first uncertainty is statistical, the second is systematic.
}
\centering
\begin{tabular}{cc}
\hline
Forward &\\
$y^*$&$\frac{\deriv \sigma}{\deriv y^*}$ [\mbarn]\\
\hline
$[1.5,2.0]$&$13.37\pm0.58\pm1.42$\\
$[2.0,2.5]$&$12.93\pm0.26\pm1.26$\\
$[2.5,3.0]$&$13.04\pm0.25\pm1.10$\\
$[3.0,3.5]$&$9.38\pm0.26\pm0.73$\\
$[3.5,4.0]$&$6.52\pm0.45\pm0.51$\\
\hline
Backward &\\
$y^*$&$\frac{\deriv \sigma}{\deriv y^*}$ [\mbarn]\\
\hline
$[-3.0,-2.5]$&$15.71\pm0.89\pm2.07$\\
$[-3.5,-3.0]$&$14.25\pm0.45\pm1.12$\\
$[-4.0,-3.5]$&$13.88\pm0.47\pm1.20$\\
$[-4.5,-4.0]$&$9.88\pm0.58\pm1.14$\\
$[-5.0,-4.5]$&$3.76\pm0.92\pm0.94$\\
\hline
\end{tabular}
\label{tab:CrossSection1DYDs}
\end{table}
\clearpage

\section{Nuclear modification factor $R_{\pPb}$}
\label{sec:RpA}
Tables~\ref{tab:RpPbResult}--\ref{tab:RpPbResult_Ds} give the numerical results for the  nuclear modification factor as a function of $\pt$.
Tables~\ref{tab:RpPbResultY}--\ref{tab:RpPbResultY_Ds} give the numerical results for the  nuclear modification factor as a function of $y^*$.
\begin{table}[tbh]
\renewcommand\arraystretch{1.5}
\caption{Nuclear modification factor $R_{\pPb}$ for prompt $\Dp$ meson production in different $\pt$ intervals, for the forward and backward rapidity regions.
The first uncertainty is statistical, the second is systematic.
}
\centering
\begin{tabular}{ccc}
\hline
$\pt\,[\gevc]$ &Forward &Backward \\
\hline
$[0,1]$&$0.585_{-\ 0.029-\ 0.048}^{+\ 0.029+\ 0.052}$	&$0.610_{-\ 0.042-\ 0.066}^{+\ 0.042+\ 0.070}$\\
$[1,2]$&$0.606_{-\ 0.006-\ 0.032}^{+\ 0.006+\ 0.033}$	&$0.779_{-\ 0.015-\ 0.067}^{+\ 0.015+\ 0.069}$\\
$[2,3]$&$0.634_{-\ 0.005-\ 0.028}^{+\ 0.005+\ 0.028}$	&$0.822_{-\ 0.009-\ 0.061}^{+\ 0.009+\ 0.060}$\\
$[3,4]$&$0.678_{-\ 0.005-\ 0.029}^{+\ 0.005+\ 0.024}$	&$0.839_{-\ 0.009-\ 0.057}^{+\ 0.009+\ 0.053}$\\
$[4,5]$&$0.684_{-\ 0.007-\ 0.027}^{+\ 0.007+\ 0.025}$	&$0.831_{-\ 0.011-\ 0.053}^{+\ 0.011+\ 0.051}$\\
$[5,6]$&$0.698_{-\ 0.010-\ 0.028}^{+\ 0.010+\ 0.026}$	&$0.814_{-\ 0.013-\ 0.050}^{+\ 0.013+\ 0.048}$\\
$[6,7]$&$0.723_{-\ 0.015-\ 0.032}^{+\ 0.015+\ 0.031}$	&$0.825_{-\ 0.018-\ 0.050}^{+\ 0.018+\ 0.050}$\\
$[7,8]$&$0.715_{-\ 0.028-\ 0.037}^{+\ 0.028+\ 0.036}$	&$0.778_{-\ 0.023-\ 0.048}^{+\ 0.023+\ 0.048}$\\
$[8,9]$&$0.806_{-\ 0.030-\ 0.050}^{+\ 0.030+\ 0.052}$	&$0.975_{-\ 0.043-\ 0.065}^{+\ 0.043+\ 0.066}$\\
$[9,10]$&$0.847_{-\ 0.045-\ 0.060}^{+\ 0.045+\ 0.065}$	&$1.002_{-\ 0.056-\ 0.073}^{+\ 0.056+\ 0.077}$\\
\hline
\end{tabular}
\label{tab:RpPbResult}
\end{table}

\begin{table}[tbh]
\renewcommand\arraystretch{1.5}
\caption{Nuclear modification factor $R_{\pPb}$ for prompt $\Ds$ meson production in different $\pt$ intervals, for the forward and backward rapidity regions.
The first uncertainty is statistical, the second is systematic.
}
\centering
\begin{tabular}{ccc}
\hline
$\pt\,[\gevc]$ &Forward &Backward \\
\hline
$[1,2]$&$0.665_{-\ 0.036-\ 0.044}^{+\ 0.036+\ 0.047}$	&$0.944_{-\ 0.060-\ 0.119}^{+\ 0.060+\ 0.121}$\\
$[2,3]$&$0.592_{-\ 0.019-\ 0.027}^{+\ 0.019+\ 0.028}$	&$0.960_{-\ 0.034-\ 0.096}^{+\ 0.034+\ 0.097}$\\
$[3,4]$&$0.587_{-\ 0.019-\ 0.024}^{+\ 0.019+\ 0.026}$	&$0.964_{-\ 0.034-\ 0.075}^{+\ 0.034+\ 0.077}$\\
$[4,5]$&$0.614_{-\ 0.022-\ 0.025}^{+\ 0.022+\ 0.027}$	&$0.971_{-\ 0.038-\ 0.066}^{+\ 0.038+\ 0.068}$\\
$[5,6]$&$0.621_{-\ 0.036-\ 0.025}^{+\ 0.036+\ 0.030}$	&$0.920_{-\ 0.051-\ 0.059}^{+\ 0.051+\ 0.064}$\\
$[6,7]$&$0.720_{-\ 0.057-\ 0.034}^{+\ 0.057+\ 0.043}$	&$1.101_{-\ 0.078-\ 0.069}^{+\ 0.078+\ 0.080}$\\
$[7,8]$&$0.748_{-\ 0.076-\ 0.039}^{+\ 0.076+\ 0.062}$	&$1.030_{-\ 0.116-\ 0.066}^{+\ 0.116+\ 0.083}$\\
$[8,9]$&$0.687_{-\ 0.092-\ 0.039}^{+\ 0.092+\ 0.058}$	&$0.981_{-\ 0.145-\ 0.070}^{+\ 0.145+\ 0.093}$\\
$[9,10]$&$0.709_{-\ 0.123-\ 0.048}^{+\ 0.123+\ 0.070}$	&$0.931_{-\ 0.170-\ 0.075}^{+\ 0.170+\ 0.099}$\\
\hline
\end{tabular}
\label{tab:RpPbResult_Ds}
\end{table}

\begin{table}[btp]
\renewcommand\arraystretch{1.5}
\caption{Nuclear modification factor $R_{\pPb}$ for prompt $\Dp$ meson production in different $y^*$ intervals, integrated up to $\pt=10\gevc$.
The first uncertainty is statistical, the second is systematic.
}
\centering
\begin{tabular}{cc}
\hline
$y^{*}$ &$R_{\pPb}$ \\
\hline
$[-4.5,-4.0]$&$1.018_{-\ 0.031-\ 0.071}^{+\ 0.031+\ 0.074}$	\\
$[-4.0,-3.5]$&$0.875_{-\ 0.022-\ 0.059}^{+\ 0.022+\ 0.062}$	\\
$[-3.5,-3.0]$&$0.765_{-\ 0.015-\ 0.062}^{+\ 0.015+\ 0.066}$	\\
$[-3.0,-2.5]$&$0.649_{-\ 0.029-\ 0.065}^{+\ 0.029+\ 0.065}$	\\
$[2.0,2.5]$&$0.710_{-\ 0.013-\ 0.056}^{+\ 0.013+\ 0.062}$	\\
$[2.5,3.0]$&$0.632_{-\ 0.017-\ 0.039}^{+\ 0.017+\ 0.038}$	\\
$[3.0,3.5]$&$0.626_{-\ 0.011-\ 0.027}^{+\ 0.011+\ 0.032}$	\\
$[3.5,4.0]$&$0.594_{-\ 0.016-\ 0.033}^{+\ 0.016+\ 0.036}$	\\
\hline

\end{tabular}
\label{tab:RpPbResultY}
\end{table}

\begin{table}[btp]
\renewcommand\arraystretch{1.5}
\caption{Nuclear modification factor $R_{\pPb}$ for prompt $\Ds$ meson production in different $y^*$ intervals, integrated up to  $\pt=10\gevc$.
The first uncertainty is statistical, the second is systematic.
}
\centering
\begin{tabular}{cc}
\hline
$y^{*}$ &$R_{\pPb}$ \\
\hline
$[-4.5,-4.0]$&$1.264_{-\ 0.191-\ 0.111}^{+\ 0.191+\ 0.134}$	\\
$[-4.0,-3.5]$&$1.275_{-\ 0.073-\ 0.093}^{+\ 0.073+\ 0.104}$	\\
$[-3.5,-3.0]$&$0.895_{-\ 0.039-\ 0.084}^{+\ 0.039+\ 0.089}$	\\
$[-3.0,-2.5]$&$0.821_{-\ 0.052-\ 0.124}^{+\ 0.052+\ 0.123}$	\\
$[2.0,2.5]$&$0.760_{-\ 0.048-\ 0.073}^{+\ 0.048+\ 0.093}$	\\
$[2.5,3.0]$&$0.680_{-\ 0.024-\ 0.038}^{+\ 0.024+\ 0.036}$	\\
$[3.0,3.5]$&$0.591_{-\ 0.024-\ 0.030}^{+\ 0.024+\ 0.036}$	\\
$[3.5,4.0]$&$0.601_{-\ 0.050-\ 0.038}^{+\ 0.050+\ 0.044}$	\\
\hline
\end{tabular}
\label{tab:RpPbResultY_Ds}
\end{table}
\clearpage

\section{Forward-backward production ratios of $\Dp$ and $\Ds$}
\label{sec:RFB_DpDs}
Tables~\ref{tab:RFBResult}--\ref{tab:RFBResult_Ds} give the numerical results for the forward-backward production ratios.
\begin{table}[tbh]
\renewcommand\arraystretch{1.5}
\caption{Forward-backward production ratios of $\Dp$ mesons as a function of $\pt$, integrated over the common rapidity range $2.5<|y^*|<4.0$; and as a function of $y^*$, integrated over $0<\pt<14\gevc$. The first uncertainty is statistical, the second is systematic.}
\centering
\begin{tabular}{cc}
\hline
\pt [\gevc] & $R_{\mathrm{FB}}$ \\
\hline
$[0,1]$ &$0.959\pm 0.052 \pm 0.092$ \\
$[1,2]$ &$0.778\pm 0.015 \pm 0.062$ \\
$[2,3]$ &$0.772\pm 0.009 \pm 0.048$ \\
$[3,4]$ &$0.807\pm 0.009 \pm 0.042$ \\
$[4,5]$ &$0.823\pm 0.012 \pm 0.040$ \\
$[5,6]$ &$0.857\pm 0.016 \pm 0.039$ \\
$[6,7]$ &$0.876\pm 0.023 \pm 0.042$ \\
$[7,8]$ &$0.919\pm 0.041 \pm 0.051$ \\
$[8,9]$ &$1.046\pm 0.052 \pm 0.057$ \\
$[9,10]$ &$1.061\pm 0.071 \pm 0.068$ \\
$[10,11]$ &$0.987\pm 0.088 \pm 0.079$ \\
$[11,12]$ &$1.241\pm 0.152 \pm 0.167$ \\
$[12,13]$ &$0.877\pm 0.135 \pm 0.089$ \\
$[13,14]$ &$0.902\pm 0.189 \pm 0.107$ \\
\hline
$|y^{*}|$ & $R_{\mathrm{FB}}$ \\
\hline
$[2.5,3.0]$ &$0.973\pm 0.036 \pm 0.089$ \\
$[3.0,3.5]$ &$0.819\pm 0.012 \pm 0.058$ \\
$[3.5,4.0]$ &$0.680\pm 0.013 \pm 0.043$ \\
\hline
\end{tabular}
\label{tab:RFBResult}
\end{table}

\begin{table}[tbh]
\renewcommand\arraystretch{1.5}
\caption{Forward-backward production ratios of $\Ds$ mesons as a function of $\pt$, integrated over the common rapidity range $2.5<|y^*|<4.0$; and as a function of $y^*$, integrated over $0<\pt<14\gevc$. The first uncertainty is statistical, the second is systematic.}
\centering
\begin{tabular}{cc}
\hline
\pt [\gevc] & $R_{\mathrm{FB}}$ \\
\hline
$[0,1]$ &$0.775\pm 0.208 \pm 0.104$ \\
$[1,2]$ &$0.705\pm 0.044 \pm 0.086$ \\
$[2,3]$ &$0.617\pm 0.024 \pm 0.056$ \\
$[3,4]$ &$0.609\pm 0.024 \pm 0.042$ \\
$[4,5]$ &$0.633\pm 0.029 \pm 0.036$ \\
$[5,6]$ &$0.675\pm 0.045 \pm 0.036$ \\
$[6,7]$ &$0.654\pm 0.057 \pm 0.036$ \\
$[7,8]$ &$0.929\pm 0.119 \pm 0.047$ \\
$[8,9]$ &$0.701\pm 0.111 \pm 0.041$ \\
$[9,10]$ &$0.762\pm 0.155 \pm 0.054$ \\
$[10,14]$ &$1.337\pm 0.323 \pm 0.089$ \\
\hline
$|y^{*}|$ & $R_{\mathrm{FB}}$ \\
\hline
$[2.5,3.0]$ &$0.830\pm 0.054 \pm 0.119$ \\
$[3.0,3.5]$ &$0.715\pm 0.049 \pm 0.071$ \\
$[3.5,4.0]$ &$0.502\pm 0.071 \pm 0.041$ \\
\hline
\end{tabular}
\label{tab:RFBResult_Ds}
\end{table}
\clearpage

\section{Production ratios between $\Dp$, $\Ds$ and $\Dz$}
\label{sec:production_ratios}
Tables~\ref{tab:1DDpDzRatio}--\ref{tab:1DDsDpRatio} give the numerical results for the cross-section ratios between mesons as a function of $\pt$.
Tables~\ref{tab:y1DDpDzRatio}--\ref{tab:y1DDsDpRatio} give the numerical results for the cross-section ratios between mesons as a function of $y^*$.
\begin{table}[tbh]
\renewcommand\arraystretch{1.5}
\caption{Measured of $R_{\Dp/\Dz}$ as a function of $\pt$ in \lhcb $\pPb$ collisions at 5.02$\ensuremath{\mathrm{\,Te\kern -0.1em V}}$ in the forward and backward regions, integrated over $2.5<|y^*|<4.0$. The first uncertainty is statistical, the second is systematic.}
\centering
\begin{tabular}{ccc}
\hline
$\pt\,[\gevc]$ &Forward &Backward \\
\hline
$[0,1]$&$0.337\pm0.011\pm0.031$	&$0.302\pm0.012\pm0.036$\\
$[1,2]$&$0.334\pm0.004\pm0.023$	&$0.302\pm0.005\pm0.029$\\
$[2,3]$&$0.351\pm0.002\pm0.017$	&$0.311\pm0.003\pm0.024$\\
$[3,4]$&$0.373\pm0.002\pm0.016$	&$0.325\pm0.004\pm0.024$\\
$[4,5]$&$0.378\pm0.003\pm0.017$	&$0.330\pm0.005\pm0.024$\\
$[5,6]$&$0.391\pm0.005\pm0.026$	&$0.339\pm0.009\pm0.030$\\
$[6,7]$&$0.406\pm0.009\pm0.035$	&$0.360\pm0.009\pm0.027$\\
$[7,8]$&$0.402\pm0.010\pm0.028$	&$0.332\pm0.012\pm0.035$\\
$[8,9]$&$0.418\pm0.010\pm0.039$	&$0.350\pm0.020\pm0.060$\\
$[9,10]$&$0.418\pm0.015\pm0.046$	&$0.339\pm0.023\pm0.038$\\
\hline
\end{tabular}
\label{tab:1DDpDzRatio}
\end{table}
\begin{table}[tbh]
\renewcommand\arraystretch{1.5}
\caption{Measured of $R_{\Ds/\Dz}$ ratio as a function of $\pt$ in \lhcb $\pPb$ collisions at 5.02$\ensuremath{\mathrm{\,Te\kern -0.1em V}}$ in the forward and backward regions, integrated over $2.5<|y^*|<4.0$. The first uncertainty is statistical, the second is systematic.}
\centering
\begin{tabular}{ccc}
\hline
$\pt\,[\gevc]$ &Forward &Backward \\
\hline
$[0,1]$&$0.139\pm0.015\pm0.014$	&$0.162\pm0.023\pm0.025$\\
$[1,2]$&$0.146\pm0.005\pm0.012$	&$0.140\pm0.007\pm0.018$\\
$[2,3]$&$0.159\pm0.003\pm0.010$	&$0.165\pm0.005\pm0.016$\\
$[3,4]$&$0.167\pm0.003\pm0.008$	&$0.170\pm0.005\pm0.014$\\
$[4,5]$&$0.179\pm0.004\pm0.009$	&$0.186\pm0.007\pm0.015$\\
$[5,6]$&$0.188\pm0.006\pm0.013$	&$0.193\pm0.009\pm0.013$\\
$[6,7]$&$0.193\pm0.008\pm0.017$	&$0.217\pm0.012\pm0.017$\\
$[7,8]$&$0.181\pm0.008\pm0.013$	&$0.169\pm0.015\pm0.018$\\
$[8,9]$&$0.175\pm0.011\pm0.017$	&$0.193\pm0.022\pm0.019$\\
$[9,10]$&$0.190\pm0.015\pm0.022$	&$0.198\pm0.031\pm0.019$\\
\hline
\end{tabular}
\label{tab:1DDsDzRatio}
\end{table}

\begin{table}[tbh]
\renewcommand\arraystretch{1.5}
\caption{Measured of $R_{\Ds/\Dp}$ ratio as a function of $\pt$ in \lhcb $\pPb$ collisions at 5.02$\ensuremath{\mathrm{\,Te\kern -0.1em V}}$ in the forward and backward regions, integrated over $2.5<|y^*|<4.0$. The first uncertainty is statistical, the second is systematic.}
\centering
\begin{tabular}{ccc}
\hline
$\pt\,[\gevc]$ &Forward &Backward \\
\hline
$[0,1]$&$0.393\pm0.046\pm0.041$	&$0.532\pm0.082\pm0.081$\\
$[1,2]$&$0.438\pm0.015\pm0.036$	&$0.462\pm0.026\pm0.057$\\
$[2,3]$&$0.451\pm0.009\pm0.025$	&$0.531\pm0.016\pm0.050$\\
$[3,4]$&$0.449\pm0.009\pm0.018$	&$0.524\pm0.016\pm0.038$\\
$[4,5]$&$0.474\pm0.010\pm0.014$	&$0.565\pm0.020\pm0.033$\\
$[5,6]$&$0.480\pm0.014\pm0.013$	&$0.547\pm0.025\pm0.027$\\
$[6,7]$&$0.475\pm0.019\pm0.015$	&$0.603\pm0.034\pm0.029$\\
$[7,8]$&$0.451\pm0.022\pm0.012$	&$0.509\pm0.044\pm0.024$\\
$[8,9]$&$0.419\pm0.025\pm0.014$	&$0.523\pm0.056\pm0.028$\\
$[9,10]$&$0.454\pm0.036\pm0.018$	&$0.598\pm0.098\pm0.037$\\
$[10,12]$&$0.486\pm0.040\pm0.018$	&$-$\\
$[12,14]$&$0.469\pm0.065\pm0.028$	&$-$\\
$[10,14]$&$-$&$0.426\pm0.070\pm0.026$\\
\hline
\end{tabular}
\label{tab:1DDsDpRatio}
\end{table}

\begin{table}[tbh]
\renewcommand\arraystretch{1.5}
\caption{Measured of $R_{\Dp/\Dz}$ as a function of $|y^*|$ in \lhcb $\pPb$ collisions at 5.02$\ensuremath{\mathrm{\,Te\kern -0.1em V}}$ in the forward and backward regions, integrated over $0<\pt<10\gevc$. The first uncertainty is statistical, the second is systematic.}
\centering
\begin{tabular}{ccc}
\hline
$\lvert y^{*}|$ &Forward &Backward \\
\hline
$[1.5,2.0]$&$0.311\pm0.011\pm0.030$	&$-$	\\
$[2.0,2.5]$&$0.340\pm0.003\pm0.024$	&$-$	\\
$[2.5,3.0]$&$0.371\pm0.003\pm0.017$	&$0.283\pm0.011\pm0.041$\\
$[3.0,3.5]$&$0.368\pm0.003\pm0.014$	&$0.301\pm0.004\pm0.027$\\
$[3.5,4.0]$&$0.362\pm0.006\pm0.028$	&$0.328\pm0.004\pm0.022$\\
$[4.0,4.5]$&$-$	&$0.317\pm0.007\pm0.023$\\
$[4.5,5.0]$&$-$	&$0.325\pm0.016\pm0.039$\\
\hline
\end{tabular}
\label{tab:y1DDpDzRatio}
\end{table}

\begin{table}[tbh]
\renewcommand\arraystretch{1.5}
\caption{Measured of $R_{\Ds/\Dz}$ as a function of $|y^*|$ in \lhcb $\pPb$ collisions at 5.02$\ensuremath{\mathrm{\,Te\kern -0.1em V}}$ in the forward and backward regions, integrated over $0<\pt<10\gevc$. The first uncertainty is statistical, the second is systematic.}
\centering
\begin{tabular}{ccc}
\hline
$\lvert y^{*}|$ &Forward &Backward \\
\hline
$[1.5,2.0]$&$0.147\pm0.006\pm0.016$	&$-$	\\
$[2.0,2.5]$&$0.151\pm0.005\pm0.012$	&$-$	\\
$[2.5,3.0]$&$0.169\pm0.005\pm0.008$	&$0.159\pm0.009\pm0.027$\\
$[3.0,3.5]$&$0.159\pm0.007\pm0.008$	&$0.149\pm0.008\pm0.016$\\
$[3.5,4.0]$&$0.135\pm0.018\pm0.012$	&$0.166\pm0.009\pm0.013$\\
$[4.0,4.5]$&$-$	&$0.183\pm0.019\pm0.024$\\
$[4.5,5.0]$&$-$	&$0.086\pm0.021\pm0.013$\\
\hline
\end{tabular}
\label{tab:y1DDsDzRatio}
\end{table}

\begin{table}[tbh]
\renewcommand\arraystretch{1.5}  
\caption{Measured $R_{\Ds/\Dp}$ as a function of $|y^*|$ in \lhcb $\pPb$ collisions at 5.02$\ensuremath{\mathrm{\,Te\kern -0.1em V}}$ in the forward and backward regions, integrated over $0<\pt<10\gevc$. The first uncertainty is statistical, the second is systematic.}
\centering
\begin{tabular}{ccc}
\hline
$\lvert y^{*}|$ &Forward &Backward \\
\hline
$[1.5,2.0]$&$0.460\pm0.028\pm0.045$	&$-$	\\
$[2.0,2.5]$&$0.444\pm0.014\pm0.039$	&$-$	\\
$[2.5,3.0]$&$0.457\pm0.013\pm0.024$	&$0.535\pm0.039\pm0.084$\\
$[3.0,3.5]$&$0.433\pm0.020\pm0.018$	&$0.496\pm0.026\pm0.053$\\
$[3.5,4.0]$&$0.373\pm0.049\pm0.026$	&$0.506\pm0.027\pm0.034$\\
$[4.0,4.5]$&$-$	&$0.578\pm0.061\pm0.069$\\
$[4.5,5.0]$&$-$	&$0.290\pm0.074\pm0.037$\\
\hline
\end{tabular}
\label{tab:y1DDsDpRatio}
\end{table}
\clearpage


\addcontentsline{toc}{section}{References}
\bibliographystyle{LHCb}


\ifx\mcitethebibliography\mciteundefinedmacro
\PackageError{LHCb.bst}{mciteplus.sty has not been loaded}
{This bibstyle requires the use of the mciteplus package.}\fi
\providecommand{\href}[2]{#2}

\newpage

\centerline
{\large\bf LHCb collaboration}
\begin
{flushleft}
\small
R.~Aaij$^{32}$\lhcborcid{0000-0003-0533-1952},
A.S.W.~Abdelmotteleb$^{51}$\lhcborcid{0000-0001-7905-0542},
C.~Abellan~Beteta$^{45}$,
F.~Abudin{\'e}n$^{51}$\lhcborcid{0000-0002-6737-3528},
T.~Ackernley$^{55}$\lhcborcid{0000-0002-5951-3498},
B.~Adeva$^{41}$\lhcborcid{0000-0001-9756-3712},
M.~Adinolfi$^{49}$\lhcborcid{0000-0002-1326-1264},
P.~Adlarson$^{77}$\lhcborcid{0000-0001-6280-3851},
H.~Afsharnia$^{9}$,
C.~Agapopoulou$^{43}$\lhcborcid{0000-0002-2368-0147}
% ...

{\footnotesize \it

% ...
$^{9}$Universit{\'e} Clermont Auvergne, CNRS/IN2P3, LPC, Clermont-Ferrand, France\\
% ...
$^{32}$Nikhef National Institute for Subatomic Physics, Amsterdam, Netherlands\\
% ...
$^{41}$Instituto Galego de F{\'\i}sica de Altas Enerx{\'\i}as (IGFAE), Universidade de Santiago de Compostela, Santiago de Compostela, Spain\\
% ...
$^{43}$European Organization for Nuclear Research (CERN), Geneva, Switzerland\\
% ...
$^{45}$Physik-Institut, Universit{\"a}t Z{\"u}rich, Z{\"u}rich, Switzerland\\
% ...
$^{49}$H.H. Wills Physics Laboratory, University of Bristol, Bristol, United Kingdom\\
% ...
$^{51}$Department of Physics, University of Warwick, Coventry, United Kingdom\\
% ...
$^{54}$School of Physics and Astronomy, University of Glasgow, Glasgow, United Kingdom\\
$^{55}$Oliver Lodge Laboratory, University of Liverpool, Liverpool, United Kingdom\\
% ...
$^{77}$Department of Physics and Astronomy, Uppsala University, Uppsala, Sweden, associated to $^{54}$\\
% ...
}
\end{flushleft}

\end{document}